\documentclass[aps,pra,superscriptaddress,twocolumn]{revtex4-2}
\usepackage{graphicx,amsmath,amssymb,amsfonts,latexsym,color,dcolumn,bm}

\newcommand{\beq}{\begin{equation}}
\newcommand{\eeq}{\end{equation}}
\newcommand{\bea}{\begin{eqnarray}}
\newcommand{\eea}{\end{eqnarray}}

\providecommand{\bra}[1]{\langle #1 \rvert}
\providecommand{\ket}[1]{\lvert #1 \rangle}
\providecommand{\braket}[2]{\langle #1 \rvert #2 \rangle}

\usepackage{amsfonts,amssymb}
\usepackage{dsfont}
\usepackage{physics}
\usepackage{tocvsec2}

\usepackage{appendix}
\usepackage{stmaryrd} 

\usepackage{tikz}

\usepackage{soul} 

\usepackage{amsmath}
\usepackage{bbold}
\usepackage{hyperref}
\hypersetup{
     colorlinks=true,      
    linkcolor=blue,        
    citecolor=blue,        
    filecolor=magenta, 
    urlcolor=black          
}

\usepackage{comment}

\DeclareMathOperator*{\sumint}{%
\mathchoice%
  {\ooalign{$\displaystyle\sum$\cr\hidewidth$\displaystyle\int$\hidewidth\cr}}
  {\ooalign{\raisebox{.14\height}{\scalebox{.7}{$\textstyle\sum$}}\cr\hidewidth$\textstyle\int$\hidewidth\cr}}
  {\ooalign{\raisebox{.2\height}{\scalebox{.6}{$\scriptstyle\sum$}}\cr$\scriptstyle\int$\cr}}
  {\ooalign{\raisebox{.2\height}{\scalebox{.6}{$\scriptstyle\sum$}}\cr$\scriptstyle\int$\cr}}
}

\usepackage{braket} 

\usepackage{soul} 

\begin{document}
\author{Mohamed Meguebel}
\email{mohamed.meguebel@telecom-paris.fr}
\affiliation{Telecom Paris, Institut Polytechnique de Paris, 19 Place Marguerite Perey, 91120 Palaiseau, France}
\author{Maxime Federico}
\affiliation{Telecom Paris, Institut Polytechnique de Paris, 19 Place Marguerite Perey, 91120 Palaiseau, France}
\affiliation{Laboratoire Interdisciplinaire Carnot de Bourgogne, UMR 6303 CNRS - Université de Bourgogne Franche-Comté, BP47870, 21078 Dijon, France}
\author{Louis Garbe}
\affiliation{Technical University of Munich, TUM School of Natural Sciences,
Physics Department, 85748 Garching, Germany}
\affiliation{Walther-Meißner-Institut, Bayerische Akademie der Wissenschaften, 85748 Garching, Germany and
Munich Center for Quantum Science and Technology (MCQST), 80799 Munich, Germany}
\author{Nadia Belabas}
\affiliation{Centre for Nanosciences and Nanotechnology, Université Paris-Saclay, 10 Bd Thomas Gobert, 91120 Palaiseau, France}
\author{Nicolas Fabre}
\email{nicolas.fabre@telecom-paris.fr}
\affiliation{Telecom Paris, Institut Polytechnique de Paris, 19 Place Marguerite Perey, 91120 Palaiseau, France}

\date{\today}
\begin{abstract}
    We propose an adiabatic-elimination formalism in the dispersive regime based on a transition-centric perturbation theory. The perturbative expansion is recast into a diagrammatic framework, while adiabatic elimination is implemented through controlled projections onto transition subspaces. Our approach applies systematically at arbitrary perturbation order, and is suited to multilevel systems and multiple qubits in both cavity and waveguide quantum electrodynamics. It ultimately enables the explicit construction of effective higher-order Hamiltonians while bypassing important limitations of existing techniques, thereby providing a practical toolbox for multiphoton processes in the dispersive regime.
\end{abstract}
\vskip2pc 

\title{Effective Hamiltonians in Cavity and Waveguide QED from Transition-Operator Diagrammatic Perturbation Theory}
\maketitle

\section{Introduction}
Adiabatic elimination is an instrumental approximation in light-matter interaction theory, employed to eliminate rapidly evolving degrees of freedom in the presence of a clear timescale separation. Adiabatic elimination is most often implemented in one of two settings: an isolated, closed-system, \textit{bath-free} framework or an open-system, \textit{bath-traced} framework. In the bath-traced approach~\cite{Generalopenquantumsystembreuer2002theory,Adiabaticeliminationsubspacefinkelstein2020adiabatic,Adiabaticeliminationprojectorappraochgonzalez2024tutorial,AdiabaticeliminationHeisenbergRouchonle2023heisenberg}, the fast degrees of freedom of an external bath are traced over -- using for instance projection-operator techniques --, producing a reduced master equation that incorporates dissipation. In the bath-free approach, standard in atomic physics~\cite{adiabaticeliminationatomicphysicspuri1988quantum,ArticleracineRamanpassagegerry1990dynamics,adiabaticeliminationatomicphysicswu1996effective,adiabaticeliminationatomicphysicsalexanian1995unitary,AdiabaticeliminationLambdasystembrion2007adiabatic,Adiabaticeliminationbeyondpaulisch2014beyond,adiabaticeliminationatomicphysicslarson2021jaynes,Adiabaticeliminationmultiphotonmaity2024adiabatic}, states that are far off resonance are eliminated in the dispersive regime~\cite{Interacregimespedernales2015quantum}, that is when detunings are much larger than the coupling strengths. This procedure yields effective Hamiltonian corrections such as energy shifts or mediated couplings between initially uncoupled atomic levels. This closed-system avenue is the bedrock of a plethora of atomic physics phenomena, most notably the two-photon Raman transitions~\cite{Ramantwophotonlinskens1996two}. \\
Numerous articles have investigated the theoretical inner workings of bath-free adiabatic elimination procedure. Brion \textit{et al.}~\cite{AdiabaticeliminationLambdasystembrion2007adiabatic} examine the subtleties of rotating-frame selection and apply the Feshbach projection-operator method, combined with Green’s functions, to rigorously eliminate the excited state in a three-level $\Lambda$ system. In~\cite{Adiabaticeliminationbeyondpaulisch2014beyond}, Paulisch \textit{et al.} analyze a more general multi-level atomic system and clarify the implicit orders of the Markovian approximation underlying earlier adiabatic-elimination calculations. In both~\cite{AdiabaticeliminationLambdasystembrion2007adiabatic, Adiabaticeliminationbeyondpaulisch2014beyond}, the adiabatic elimination is performed in state-vector pictures and in a semi-classical model. In~\cite{adiabaticeliminationatomicphysicspuri1988quantum,ArticleracineRamanpassagegerry1990dynamics, adiabaticeliminationatomicphysicswu1996effective,adiabaticeliminationatomicphysicsalexanian1995unitary,adiabaticeliminationatomicphysicslarson2021jaynes,Adiabaticeliminationmultiphotonmaity2024adiabatic} on the other hand, the electromagnetic field is quantized and the derivation is done in the Heisenberg picture. In \cite{adiabaticeliminationatomicphysicsalexanian1995unitary} in particular, the authors employ Schrieffer-Wolff perturbation theory~\cite{SchriefferWolffbravyi2011schrieffer}, which is rooted in a non-trivial unitary transformation that block-diagonalizes the Hamiltonian to decouple well-separated subspaces. Nonetheless, the work in~\cite{ArticleracineRamanpassagegerry1990dynamics, adiabaticeliminationatomicphysicswu1996effective,adiabaticeliminationatomicphysicsalexanian1995unitary,adiabaticeliminationatomicphysicslarson2021jaynes,Adiabaticeliminationmultiphotonmaity2024adiabatic} consists in manipulating disjoint transition operators instead of treating light-matter transition operators collectively. More generally, in hybrid light-matter systems, adiabatic elimination is often applied separately to the light and matter degrees of freedom, an approximation that suffices when the field remains essentially classical but breaks down once the electromagnetic field is fully quantized. Puri \textit{et al.}~\cite{adiabaticeliminationatomicphysicspuri1988quantum} did address the physical effects that emerge when evolving joint transition operators, for example, spontaneous photon-number-dependent Stark shifts. However, \cite{adiabaticeliminationatomicphysicspuri1988quantum,ArticleracineRamanpassagegerry1990dynamics,adiabaticeliminationatomicphysicswu1996effective,adiabaticeliminationatomicphysicsalexanian1995unitary,adiabaticeliminationatomicphysicslarson2021jaynes,Adiabaticeliminationmultiphotonmaity2024adiabatic} are restricted to single- or two-mode cavities and do not offer a systematic adiabatic-elimination scheme at all orders and for an arbitrary number of frequency modes. \\ In~\cite{Jamesmethodjames2007effective} James \textit{et al.} propose another adiabatic elimination procedure -- although not referred to as such -- with the derivation of effective Hamiltonians for closed-system dynamics through time-averaged unitary evolution. In~\cite{Jamesmethodgamel2010time}, they extend their method to account for remarkable effective losses in the Liouvillian density-matrix evolution from discarded fast-evolving interactions. Nevertheless, their method hinges on an abrupt \textit{all-or-nothing} low-pass filtering of high-frequency contributions and does not transparently show the physical interactions leading to the effective Hamiltonian nor provides a prescription for handling frequency continua. \\
Diagrammatic methods have also been explored, not specifically for adiabatic elimination, but to analyze the perturbative structure of the underlying physical processes. In particular, Feynman diagrams -- ubiquitous in perturbation theory -- play a central role in quantum field theory~\cite{QFTbogolubov2002quantum,QDTzee2010quantum}. A somewhat identical diagrammatic representation developed by Bordé~\cite{Doublesidedborde1983density} called double-sided Feynman diagrams has been employed in nonlinear quantum optics and spectroscopy when the electromagnetic field is not quantized~\cite{DoublesidedFeynmandiagramsboyd2008nonlinear,DoublesidedFeynmandiagramshache2016optique}. This representation is based upon the time-evolution of matter (\textit{e.g.} atoms, molecules) density matrix where losses to an outer system, \textit{e.g.} a bath, may be added phenomenologically. In a similar fashion, the quantized electromagnetic field can be incorporated into the density matrix formalism through both explicit double-sided Feynman diagrams and their compact counterpart: nonlinear-spectroscopy closed-time path loop (CTPL) diagrams~\cite{CloseloopMukamelmarx2008nonlinear,CloseloopMukamelroslyak2009unified,DoublesidedFeynmandiagramsmukamel2010ultrafast,DoublesidedFeynmandiagramsRevModPhys.88.045008}. Originally formulated by Marx \textit{et al.}~\cite{CloseloopMukamelmarx2008nonlinear}, these diagrams are conceptually close to Schwinger-Keldysh contour~\cite{Keldyshmarques2004time} and compactly represents several doubled-sided Feynman diagrams. There are other diagrammatic techniques utilized in nonlinear quantum optics~\cite{DoublesidedFeynmandiagramsvergari2025diagrammatic}, such as Liouville pathways~\cite{DoublesidedFeynmandiagramsmukamel1995principles}. The benefit of these diagrammatic methods is to provide a clear bookkeeping of the physical processes occurring at each order. \\ \\
In this work, we propose a framework in which adiabatic elimination is formulated at the level of transitions, exploiting the fact that transition operators are eigenoperators of the free evolution in Liouville space. Performing a perturbation theory of the \textit{joint light-matter} (JLM) transition operators in the dispersive regime, we construct JLM diagrams to organize and systematically bookkeep the perturbative expansion within a unified operator-oriented formalism. This approach enables a transparent identification of resonant and off-resonant transitions and a controlled adiabatic elimination of the latter, thereby bringing diagrammatic calculus and adiabatic elimination together; a strategy that is, to our knowledge, novel in the adiabatic-elimination literature. Our framework provides a systematic route to effective Hamiltonians at arbitrary perturbative order for multilevel and multiple-qubit systems coupled to bosonic modes, both in cavity QED and in waveguide QED. To the best of our knowledge, such a general derivation of higher-order effective Hamiltonians has not previously been considered in the context of waveguide QED, despite the current interest in the field. Our approach thus offers a novel avenue for the Hamiltonian-engineering challenges characteristic of advanced quantum-information protocols. This toolbox is specifically tailored for the light-matter-interaction and quantum-optics communities working to implement physical systems suitable for quantum-information tasks, for instance non-Gaussian operations with higher orders of squeezing~\cite{Highersqueezingsutherland2021universal}. \\ \\
This work is organized as follows. In Sec.~\ref{Section Closed-system diagrammatic adiabatic elimination of joint light-matter operators}, we first define the JLM transition operators before deriving an operator-based Heisenberg-picture perturbation theory~\cite{PerturbationtheoryHeisenbergackerhalt1975heisenberg,PerturbationtheoryHeisenbergfranson2002perturbation,PerturbationtheoryHeisenbergfernandez2003perturbation} which we translate into JLM diagrams with their subsequent construction rules. In Sec.~\ref{Section Examples}, we exploit this JLM transition operator perturbation theory up to the second order in the interaction Hamiltonian. We recover the AC quantum Stark shift~\cite{QuantumACStarkshiftblais2004cavity}; the effective qubit-qubit coupling in the dispersive Tavis-Cummings model and discuss the contribution of counter-rotating terms from the Dicke model~\cite{dicke1954coherence,JCmodellarson2021jaynes}; and derive the effective dynamics of a three-level system coupled to a continuum, yielding two-photon Raman transitions and Stark-shift contributions mediated by the continuum modes. This manuscript is closely connected to a companion article~\cite{CompanionarticlePRL}, which elevates the present transition-centric framework to a broader conceptual shift in which transitions, rather than states, constitute the fundamental objects of the dynamics. In particular, it highlights the persistence of polaritonic hybridization in the Jaynes-Cummings model across both dispersive and resonant regimes.
\section{Joint light-matter transition operator-diagrammatic adiabatic elimination}
\label{Section Closed-system diagrammatic adiabatic elimination of joint light-matter operators}
\subsection{Joint light-matter transition operator-perturbation theory}
\subsubsection{Joint light-matter transition operators}
In this subsection, we introduce the JLM transition operators that underpins the transition-centered formalism. Consider the closed system composed of a $N$-level system $\{\ket{i}\}_{i \in \llbracket 1,N \rrbracket}$ corresponding to matter which is coupled to the electromagnetic field. The free Hamiltonian reads ($\hbar = 1$)
\begin{equation}
    \hat H_{\text{free}} = \sum_{i=1}^N\omega_i\ket{i}\bra{i}\otimes \mathbb {1}_{\omega} + \mathbb {1}_{\alpha}\otimes \sumint_{\boldsymbol \sigma}\sumint_{\omega}\omega\hat a_{\boldsymbol \sigma}^\dagger(\omega)\hat a_{\boldsymbol \sigma}(\omega),
\end{equation}
where no energy ordering is assumed for the matter states. The $\sumint$ shorthand notation denotes the unified "sum-integral" operator acting on both discrete and continuous indices, \textit{e.g}. $\sumint_{\omega}$ reduces to $ \sum_{\omega}$ when only a discrete frequency set is present. $\mathbb{1}_\omega$ and $\mathbb{1}_\alpha$ denote, respectively, the identity in the matter Hilbert space and in the quantized electromagnetic field Fock space. These two identity operators will be dropped for readability unless indicated otherwise. The index $\boldsymbol{\sigma}$ accounts for auxiliary discrete and/or continuous degrees of freedom, such as the polarization. The bosonic canonical commutation relation can be expressed as 
\begin{equation}
\label{Equation bosonic commutation relation}
    [\hat a_{\boldsymbol \sigma}(\omega),\hat a_{\boldsymbol{\sigma}'}^\dagger(\omega')] = \delta(\boldsymbol \sigma - \boldsymbol \sigma')\delta(\omega-\omega').
\end{equation} 
The general interaction Hamiltonian is taken to be 
\begin{equation}
\label{Equation general interaction Hamiltonian}
    \hat H_{\text{int}} = \sum_{i=1}^N\sum_{j\neq i}\sumint_{\boldsymbol \sigma}\sumint_{\omega} g_{i\boldsymbol \sigma}^{j}(\omega)\ket{j}\bra{i}\otimes \hat a_{\boldsymbol \sigma}(\omega) + \text{h.c},
\end{equation}
that is all states may be connected through photon absorption or emission events, with a coupling term $g_{i\boldsymbol \sigma}^j(\omega)$ for the $\ket{i}\rightarrow \ket{j}$ transition by absorption of a photon at frequency $\omega$ and auxiliary degrees of freedom $\boldsymbol \sigma$. Both $\hat H_{\text{free}}$ and $\hat H_{\text{int}}$ are time-independent. Thus far, the interaction Hamiltonian Eq.~\eqref{Equation general interaction Hamiltonian} describes interaction processes both within and beyond the rotating wave approximation (RWA), \textit{i.e.} terms of the form $\ket{j}\bra{i}\otimes \hat a_{\boldsymbol \sigma}(\omega)$ and $\ket{j}\bra{i}\otimes \hat a_{\boldsymbol \sigma}^\dagger(\omega)$ are \textit{a priori} included. \\
We now define the one-photon JLM transition operator $\hat \xi_{i\boldsymbol \sigma}^{j}(\omega) \equiv \ket{j}\bra{i}\otimes \hat a_{\boldsymbol \sigma}(\omega)$ which depicts the matter transition $\ket{i}\rightarrow \ket{j}$ by annihilation of a photon at frequency $\omega$ and auxiliary degrees of freedom $\boldsymbol \sigma$. The Hermitian conjugate $\hat \xi_{i\boldsymbol \sigma}^{j\dagger}(\omega) \equiv \hat \xi_{j}^{i\boldsymbol \sigma}(\omega)$ equivalently corresponds to the matter transition $\ket{j}\rightarrow \ket{i}$ by emission of a photon at frequency $\omega$ and auxiliary degrees of freedom $ \boldsymbol \sigma$. \\
Note that we also use throughout this manuscript the shorthand expression of the interaction Hamiltonian Eq.~\eqref{Equation general interaction Hamiltonian} 
\begin{equation}
\label{Equation compact general interaction Hamiltonian}
\hat H_{\text{int}} = \sum_{m=1}^{M/2}\hat \xi_m+\text{h.c}
\end{equation}
where $M$ denotes the number of matter transitions, with the associated photon processes absorbed into the operators $\hat \xi_m$. Each operator comes with its Hermitian conjugate, therefore $M$ is even. Each $\hat \xi_m$ is then of the following form
\begin{equation}
    \hat \xi_m \equiv \hat \xi_{i}^j = \sumint_{\boldsymbol \sigma}\sumint_{\omega} g_{i\boldsymbol \sigma}^{j}(\omega)\ket{j}\bra{i}\otimes \hat a_{\boldsymbol \sigma}(\omega).
\end{equation}
Eq.~\eqref{Equation compact general interaction Hamiltonian} will be particularly useful for combinatorial and abstract reasoning. \\
In the Heisenberg picture, the JLM transition operators evolve as
\begin{equation}
\label{Equation equation of motion in Heisenberg picture}
    \frac{d\hat \xi_{i\boldsymbol \sigma}^{j}(\omega,t)}{dt}=-i\mathcal L\hat \xi_{i\boldsymbol \sigma}^{j}(\omega,t),
\end{equation}
where $\mathcal L$ is the Liouvillian superoperator defined as $\mathcal L\hat \xi = [\hat \xi,\hat H]$ for any element $\hat \xi$ of the Liouville space~\cite{Liouvillespacegyamfi2020fundamentals}, that is the space of operators acting on the Hilbert space. The Liouvillian $\mathcal L$ may be broken down into a free Liouvillian $\mathcal L_{\text{free}}$ and an interaction Liouvillian $\mathcal L_{\text{int}}$ acting on an element $\hat \xi$ of the Liouville space as $\mathcal L_{\text{free}}\hat \xi = [\hat \xi, \hat H_{\text{free}}] $ and $\mathcal L_{\text{int}}\hat \xi = [\hat \xi, \hat H_{\text{int}}]$. The interaction term contribution can be expressed in terms of commutators of one-photon JLM transition operators 
\begin{equation}
\label{Equation joint operators interaction Hamiltonian commutator}
\begin{split}
    \mathcal L_{\text{int}}\hat \xi_{i\boldsymbol\sigma}^j(\omega,t) &= \sum_{k=1}^N\sum_{l\neq k}\sumint_{\boldsymbol \sigma'}\sumint_{\omega'}\\
    &\bigg(g_{k\boldsymbol \sigma'}^l(\omega')[\hat \xi_{i\boldsymbol\sigma}^j(\omega,t), \hat \xi_{k\boldsymbol\sigma'}^l(\omega',t)]\\
    &+g_{k\boldsymbol \sigma'}^{l*}(\omega')[\hat \xi_{i\boldsymbol\sigma}^j(\omega,t), \hat \xi_{l}^{k\boldsymbol\sigma'}(\omega',t)]\bigg).
\end{split}
\end{equation}
The interaction Liouvillian thus couples distinct one-photon JLM transition operators, thereby generating new matter transitions through the absorption or emission of an additional photon. In that regard, the first commutator in Eq.~\eqref{Equation joint operators interaction Hamiltonian commutator} is composed of two-photon JLM transition operators where two photons are absorbed while the second commutator contains two-photon JLM transition operators where one photon is first absorbed (respectively, emitted) before another one is emitted (respectively, absorbed). These operators have the general form 
\begin{align}
    &\hat \xi_{k\boldsymbol\sigma'\boldsymbol\sigma}^{j}(\omega',\omega,t) = \big(\ket{j}\bra{k}\otimes \hat a_{\boldsymbol\sigma'}(\omega')\hat a_{\boldsymbol \sigma}(\omega)\big)(t) \\ 
    &\hat \xi_{k\boldsymbol\sigma}^{j\boldsymbol \sigma'}(\omega',\omega,t) = \big(\ket{j}\bra{k}\otimes \hat a_{\boldsymbol\sigma'}^\dagger(\omega')\hat a_{\boldsymbol \sigma}(\omega)\big)(t),
\end{align}
where the process ordering is indicated by the ordering $(\omega',\omega,t)$ of the frequency parameter corresponding to each photon, read from the right to the left. For the absorption-absorption two-photon JLM transition operators for example, $\hat \xi_{k\boldsymbol\sigma'\boldsymbol\sigma}^{j}(\omega',\omega,t)$ should be understood as having transitioned from $\ket{k}$ to $\ket{j}$ by first absorbing a photon at frequency $\omega$ and auxiliary degrees of freedom $\boldsymbol \sigma$ and then another photon at frequency $\omega'$ and auxiliary degrees of freedom $\boldsymbol \sigma'$. Likewise, $\hat \xi_{k\boldsymbol\sigma}^{j\boldsymbol \sigma'}(\omega',\omega,t)$ depicts a transition from $\ket{k}$ to $\ket{j}$ by first absorbing a photon at frequency $\omega$ and auxiliary degrees of freedom $\boldsymbol \sigma$ and then emitting another photon at frequency $\omega'$ and auxiliary degrees of freedom $\boldsymbol \sigma'$. For absorption-absorption and emission-emission processes, the ordering is irrelevant according to the bosonic commutation relation Eq.~\eqref{Equation bosonic commutation relation} whereas for emission-absorption and absorption-emission the ordering might influence the light-matter dynamics. Note that these higher-order JLM transition operators can in principle be of the form $\ket{k}\bra{k}\otimes \dots$ with "$\dots$" indicating a combination of bosonic operators. These terms will amount to diagonal energy-renormalization terms.
\subsubsection{Time-dependent perturbation theory in the Heisenberg-Dirac picture}
\label{Subsubsection Time-dependent perturbation theory in the Heisenberg-Dirac picture}
We now expand the one-photon JLM transition operator in different orders of coupling terms $g_{i\boldsymbol \sigma}^{j}(\omega)$: $\hat \xi_{i\boldsymbol \sigma}^j(\omega,t) = \sum_{n=0}^{+\infty} \hat \xi_{i\boldsymbol \sigma}^{j(n)}(\omega,t)$, where $ \hat \xi_{i\boldsymbol \sigma}^{j(n)}(\omega,t)$ is multiplied by a product of $n$ coupling terms. The interaction Hamiltonian $\hat H_{\text{int}}$ Eq.~\eqref{Equation general interaction Hamiltonian} is thus implicitly a first-order term containing zeroth-order unperturbed JLM transition operators. Reinjecting the perturbative expansion in the equation of motion Eq.~\eqref{Equation equation of motion in Heisenberg picture} and identifying the different orders, the one-photon JLM transition operators' time-evolution can be iteratively expressed for each order as
\begin{align}
    &\frac{d\hat \xi_{i\boldsymbol\sigma}^{j(n)}(\omega,t)}{dt} = -i\mathcal L_{\text{free}}\hat \xi_{i\boldsymbol\sigma}^{j(n)}(\omega,t) -i\mathcal L_{\text{int}}\hat \xi_{i\boldsymbol\sigma}^{j(n-1)}(\omega,t) \\
     &\frac{d\hat \xi_{i\boldsymbol\sigma}^{j(0)}(\omega,t)}{dt} = -i\mathcal L_{\text{free}}\hat \xi_{i\boldsymbol\sigma}^{j(0)}(\omega,t),
\end{align}
for $n\geq 1$. Hence, one recovers a standard perturbation-theory result: the $n$th order contribution evolves via its free propagator -- which preserves the perturbation order -- plus an interaction-induced source term $\mathcal L_{\text{int}}$ that promotes the $(n-1)$th-order contribution to $n$th order. Let us now perform the following gauge transformation
\begin{equation}
\label{Equation interaction picture JLM transition operator}
    \hat \xi_{i\boldsymbol\sigma}^{j}(\omega,t) = e^{-i\mathcal L_{\text{free}}(t-t_0)}\hat \xi_{i\boldsymbol\sigma}^{j(I)}(\omega,t),
\end{equation}
This unitary transformation can be proven -- using Baker-Campbell-Hausdorff formula~\cite{Generalsakurai2020modern,MathematicalQBakerformulahall2013quantum} -- to be equal to the standard $\hat \xi \mapsto e^{i\hat H_{\text{free}}(t-t_0)}\hat \xi e^{-i\hat H_{\text{free}}(t-t_0)}$ interaction picture map, but here acting on a Heisenberg-picture operator. Accordingly, $\hat \xi_{i\boldsymbol\sigma}^{j(I)}(\omega,t)$ may be referred to as an interaction picture JLM transition operator. The time $t_0$ then labels the time when the interaction kicks off, ensuring that at the initial time $\hat \xi_{i\boldsymbol \sigma}^j(\omega,t_0)$ is equal to the zeroth-order term in the perturbative expansion $\sum_{n=0}^{+\infty}\hat \xi_{i\boldsymbol \sigma}^{j(n)}(\omega,t)$. \\
Expressing the transformation Eq.~\eqref{Equation interaction picture JLM transition operator} with the commutator is motivated by the fact that the class of operators $\hat \xi$ -- that is the JLM transition operator -- are eigenoperators~\cite{DoublesidedFeynmandiagramsmukamel1995principles,Generalopenquantumsystembreuer2002theory} of the free Liouvillian $\mathcal L_{\text{free}}$ for all orders. More precisely, the eigenvalues of $\mathcal L_{\text{free}}$ quantifies the detuning $\Delta_{\hat \xi}$ associated with a given light-matter transition operator $\hat \xi$; $\mathcal L_{\text{free}}\hat \xi = \Delta_{\hat \xi} \hat \xi$. The detunings are to be computed as $\sum_{l=0}^n(-1)^{c_l}\omega_l-(\omega_j-\omega_i)_n$ where $c_l=0$ for an absorbed photon and $c_l=1$ for an emitted photon and where $(\omega_j-\omega_i)_n$ denotes the energy difference between the two levels involved in the $n$th-order transition. We further introduce regularization parameters $\theta_l > 0$ by promoting each individual detuning to the complex plane, $\delta_l \rightarrow \delta_l-i\theta_l$. This prescription regularizes the expressions and prevents potential divergences, particularly in the presence of continua, see Sec.~\ref{Section Three-level systems in a continuum}. The regularization terms may also be interpreted as incorporating external loss channels. In the absence of such dissipation, the regulators $\theta_l$ are taken to zero at the end of the calculation so that no additional \textit{ad hoc} physical parameters are introduced.  \\
Defining the time-evolution superoperator map $\mathcal{U}_I(t,t_0)$ which acts on the interaction picture JLM transition operator according to 
$\hat\xi_{i\boldsymbol \sigma}^{j(I)}(\omega,t) = \mathcal{U}_I(t,t_0)\hat\xi_{i\boldsymbol \sigma}^{j(I)}(\omega,t_0)$,
and injecting this last expression in the equation of motion yields 
\begin{equation}
\label{Equation of motion for the superoperator map}
    \frac{d\;\mathcal{U}_I(t,t_0)}{dt} = -i\mathcal L_{\text{int}}(t)\;\mathcal{U}_I(t,t_0),
\end{equation}
with $\mathcal L_{\text{int}}(t) \equiv e^{i\mathcal L_{\text{free}}(t-t_0)}\mathcal L_{\text{int}}e^{-i\mathcal L_{\text{free}}(t-t_0)}$. Formally integrating Eq.~\eqref{Equation of motion for the superoperator map} iteratively results in a Dyson series~\cite{Generalsakurai2020modern} for the time-evolution superoperator 
\begin{equation}
\label{Equation time-evolution superoperator in the interaction picture}
     \mathcal{U}_I(t,t_0) = \mathcal T_{\leftarrow} \exp\left(-i\int_{t_0}^td\tau\;\mathcal L_{\text{int}}(\tau) \right),
\end{equation}
where $\mathcal T_{\leftarrow}$ is the time-ordering operator. Eq.~\eqref{Equation time-evolution superoperator in the interaction picture} is akin to the standard Dyson series in the Schrödinger interaction picture. Considering the expansion of the interaction picture JLM transition operators $\hat\xi_{i\boldsymbol \sigma}^{j(I)}(\omega,t) = \mathcal{U}_I(t,t_0)\hat\xi_{i\boldsymbol \sigma}^{j(I)}(\omega,t_0)$ in orders of the coupling terms $\hat \xi_{i\boldsymbol\sigma}^j(\omega,t) = \sum_{n=0}^{+\infty}\hat \xi_{i\boldsymbol\sigma}^{j(n)}(\omega,t)$,
one can identify the Dyson series contributions 
\begin{widetext}
\begin{equation}
\label{Equation perturbation theory expansion in Heisenberg-interaction picture}
\begin{split}
    \hat \xi_{i\boldsymbol\sigma}^{j(0)}(\omega,t) &= e^{-i\mathcal L_{\text{free}}(t-t_0)}\hat \xi_{i\boldsymbol\sigma}^{j}(\omega) \\ 
    \hat \xi_{i\boldsymbol\sigma}^{j(1)}(\omega,t) &= -i\int_{t_0}^td\tau\; e^{-i\mathcal L_{\text{free}}(t-\tau)}\mathcal L_{\text{int}}e^{-i\mathcal L_{\text{free}}(\tau-t_0)}\hat \xi_{i\boldsymbol\sigma}^{j}(\omega) \\ 
    \hat \xi_{i\boldsymbol\sigma}^{j(2)}(\omega,t) &= (-i)^2\int_{t_0}^t\int_{t_0}^\tau d\tau d\tau'\; e^{-i\mathcal L_{\text{free}}(t-\tau)}\mathcal L_{\text{int}}e^{-i\mathcal L_{\text{free}}(\tau-\tau')}\mathcal L_{\text{int}}e^{-i\mathcal L_{\text{free}}(\tau'-t_0)}\hat \xi_{i\boldsymbol\sigma}^{j}(\omega) \\ 
    &\;\;\vdots,
\end{split}
\end{equation}
\end{widetext}
where we have used the fact that at $t=t_0$ all pictures are equivalent $\hat \xi_{i\boldsymbol\sigma}^{j(I)(0)}(\omega,t_0) = \hat \xi_{i\boldsymbol\sigma}^{j(0)}(\omega,t_0) = \hat \xi_{i\boldsymbol\sigma}^{j}(\omega)$ and toggled back to the Heisenberg picture. The different $n$th-order contributions from Eq.~\eqref{Equation perturbation theory expansion in Heisenberg-interaction picture}, when reinjected into the interaction Hamiltonian Eq.~\eqref{Equation general interaction Hamiltonian}, lead to corrections $\Delta \hat H_{\text{int}}^{(n)}(t)$ at order $n\geq 1$, which contain the $n$th-order JLM transition operators, each weighted by a time-dependent factor $W_n(t)$ in $\Delta \hat H_{\text{int}}^{(n)}(t)$, which we discuss in depth in Sec.~\ref{Section JLM diagram method}. \\ 
Eq.~\eqref{Equation perturbation theory expansion in Heisenberg-interaction picture} is a standard form in the Schrödinger interaction picture perturbation theory wherein there are successions of free propagations and time-local interactions. Interestingly enough however, $N$-photon processes are associated with a $(N-1)$th order of the perturbation theory, \textit{e.g.} for a four-photon process, one ought to compute the third-order contribution. In a sense, we exploit that the interaction Hamiltonian is, by definition, first order -- composed of one-photon JLM transition operators -- to generate $N$-photon JLM transition operators by proceeding to the $(N-1)$th order in the perturbative expansion. Furthermore, the perturbation expansion Eq.~\eqref{Equation perturbation theory expansion in Heisenberg-interaction picture} is expressed in the standard Heisenberg picture which is uniquely defined as opposed to the interaction picture. Indeed, the latter depends on how the Hamiltonian is broken down into a free part and an interaction part. The gauge transformation Eq.~\eqref{Equation interaction picture JLM transition operator} is then a bookkeeping procedure that rewrites the same operator dynamics in a familiar Schrödinger-representation-like form commonly performed in terms of either vector states or density matrices. Another practical advantage of this transition-oriented description is that spectral labels, detunings and the multiple timescales associated with virtual processes remain explicit throughout the expansion. This formulation thus circumvents the selection of a non-unique frame transformation required in state-centric approaches, such as in Brion \textit{et al.}~\cite{AdiabaticeliminationLambdasystembrion2007adiabatic} or Paulish \textit{et al.}~\cite{Adiabaticeliminationbeyondpaulisch2014beyond}. 
\subsubsection{Resolvent method and and adiabatic elimination projection in Liouville space}
\begin{figure}
    \centering
    \includegraphics[width=0.9\linewidth]{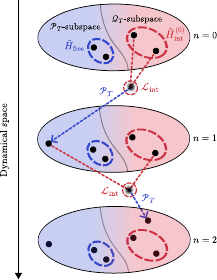}
    \caption{Schematics illustrating the dispersive perturbative expansion of JLM transition operators and the associated adiabatic elimination procedure. Each point represents a JLM transition operator. At zeroth order, the JLM transition operators constituting the interaction Hamiltonian $\hat H_{\text{int}}^{(0)} = \hat H_{\text{int}}$ Eq.~\eqref{Equation compact general interaction Hamiltonian} lie in the $\mathcal Q_T$-subspace in the dispersive regime. The interaction Liouvillian $\mathcal L_{\text{int}}$ Eq.~\eqref{Equation joint operators interaction Hamiltonian commutator} -- that is the commutator with the interaction Hamiltonian -- then couples them in the perturbative expansion, generating new transition operators whose cumulative detunings given by the free Liouvillian $\mathcal L_{\text{free}}$ determine whether they are projected with Eq.~\eqref{Equation projection time-averaging map} onto the $\mathcal P_T$-subspace or onto the $\mathcal Q_T$-subspace at first order, thereby implementing the adiabatic elimination step. The newly generated operator may then couple to a zeroth-order JLM transition operator, inducing further transitions and progressively building higher-order contributions, and so forth. In the present figure, the first-order generated operator is projected onto the resonant-transition $\mathcal P_T$-subspace, whereas the second-order generated operator is projected onto the off-resonant-transition $\mathcal Q_T$-subspace. }
    \label{Figure successive perturbation-projection in transition-operator space}
\end{figure}
The Dyson-series contributions in Eq.~\eqref{Equation perturbation theory expansion in Heisenberg-interaction picture} can be treated by exploiting the fact that all JLM transition operators are eigenoperators of the free Liouvillian. This property permits the nested time integrals in the Dyson expansion to be resumed into convolution products, which are conveniently handled in Laplace space where time-nonlocal integrals become local expressions. The detailed derivation is provided in Appendix~\ref{Appendix Joint light-matter transition operators' weights calculation}.
Equivalently, one may directly toggle to the Laplace space at the stage of the evolution operator $\mathcal U(t,t_0)=e^{-i\mathcal L(t-t_0)}$ in the Heisenberg picture instead of the interaction picture, resulting in the resolvent $\mathcal G(s)$ of the Liouvillian $\mathcal L$ 
\begin{equation}
    \mathcal G(s) = \frac{1}{s+i\mathcal L},\; \Re(s) > 0,
\end{equation}
which can be fathomed as the evolution operator in Laplace space. The dispersive perturbation theory now rises by expanding the resolvent $\mathcal G(s)$ by means of a Neumann series about the ratio $\mathcal L_{\text{int}}\left(s+i\mathcal L_{\text{free}}\right)^{-1}$
\begin{equation}
\label{Equation Neumann series of the resolvent of the Liouvillian}
    \mathcal G(s) = \left(s+i\mathcal L_{\text{free}}\right)^{-1}\sum_{n=0}^{+\infty}(-i)^n\left[\mathcal L_{\text{int}}\left(s+i\mathcal L_{\text{free}}\right)^{-1} \right]^{n}.
\end{equation}
Indeed, the $n$-order term $\mathcal G^{(n)}(s)$ of the series Eq.~\eqref{Equation Neumann series of the resolvent of the Liouvillian} corresponds to a product of $\mathcal L_{\text{int}}\left(s+i\mathcal L_{\text{free}}\right)^{-1}$ ratios $n$ times, with the operator norms $||\mathcal L_{\text{int}}||$ containing the coupling terms contributions and $||\mathcal L_{\text{free}}||$ the detuning contributions. As a consequence, the perturbative expansion $\hat \xi_{i\boldsymbol \sigma}^j(\omega,t) = \sum_{n=0}^{+\infty} \hat \xi_{i\boldsymbol \sigma}^{j(n)}(\omega,t)$ is not governed simply by products of $n$ coupling terms, but rather by products of $n$ coupling-to-detuning ratios, convenient for a dispersive regime description. The series Eq.~\eqref{Equation Neumann series of the resolvent of the Liouvillian} mirrors the pattern free evolution $\rightarrow$ interaction $\rightarrow$ free evolution that appears in the nested time integrals of the Dyson expansion Eq.~\eqref{Equation perturbation theory expansion in Heisenberg-interaction picture} but casts it as an algebraic form that is straightforward to manipulate because the JLM transition operators on which $\mathcal G(s)$ acts are eigenoperators of the free Liouvillian. Further computational details are given in Appendix~\ref{Appendix Computation with the resolvent} where we derive the JLM transition operators contributions to the interaction Hamiltonian at arbitrary perturbation order. \\
The resolvent method~\cite{AdiabaticeliminationLambdasystembrion2007adiabatic,cohen2024atom} is particularly useful in perturbative problems that involve splitting the state Hilbert space into a physically relevant and complementary subspace of respective projectors $\mathcal P$ and $\mathcal Q= 1-\mathcal P$. In the present framework, where we seek effective dynamics in the dispersive regime, a suitable choice of irrelevant subspace is the set of JLM transition operators $\hat \xi$ whose free-Liouvillian eigenvalues $\Delta_{\hat \xi}$ are large with respect to a timescale $T$,
\begin{equation}
\label{Equation relevant resonant subspace}
\mathcal Q_T\text{-subspace} = \{\hat \xi, \; |\Delta_{\hat \xi}| \gg 1/T \},
\end{equation}
\textit{i.e.} the light-matter transitions that are effectively off-resonant over the interval $T$. The projection onto the complementary relevant subspace is hence defined through the time-averaging map
\begin{equation}
\label{Equation projection time-averaging map}
\mathcal P_T \hat \xi = \frac{1}{T}\int_{0}^{T} dt\, \hat \xi(t).
\end{equation}
Operators are retained as long as their detuning does not generate rapid oscillations over $T$, that is $|\Delta_{\hat \xi}|T \not\gg 1$, whereas strongly detuned operators average out onto the complementary $\mathcal Q_T$-subspace. In this way, the spectral structure of $\mathcal L_{\text{free}}$, through its detuning eigenvalues, induces a canonical projection splitting of Liouville space. The perturbative resolvent expansion Eq.~\eqref{Equation Neumann series of the resolvent of the Liouvillian} then generates excursions between these spectral sectors, see Fig.~\ref{Figure successive perturbation-projection in transition-operator space}. As the perturbative order increases, the Liouville space explored by the expansion progressively enlarges and uncovers additional sectors of both $\mathcal P_T$ and $\mathcal Q_T$. The only selection criterion throughout is the resonance condition on the timescale $T$, which determines which light-matter processes are effectively regarded as off-resonant. We stress that the associated map should be understood as a physically motivated elimination prescription rather than as a strict mathematical projector. \\
Within this formalism, the kernel $\ker \mathcal L_{\text{free}}$ of the free Liouvillian prior to the interaction plays a distinguished role. At zeroth perturbative order $(n=0)$, this kernel coincides with the operators composing the free Hamiltonian itself and thus provides the spectral reference relative to which transition operators acquire their detunings $\Delta_{\hat \xi}$. From this perspective, the dispersive perturbative expansion and subsequent adiabatic elimination can be viewed as a systematic renormalization of this reference kernel, governed by the detunings furnished by the same reference kernel. \\
In light of all of the above, in the present formalism, the perturbative expansion Eq.~\eqref{Equation Neumann series of the resolvent of the Liouvillian} of the resolvent $\mathcal G(s)$ provides the formulation of the dynamics in the \textit{dispersive regime} while the restriction to the resonant subspace $\mathcal P_T$-subspace Eq.~\eqref{Equation relevant resonant subspace} is equivalent to implementing an \textit{adiabatic elimination procedure}. \\ 
In line with refinements of James' formalism~\cite{Jamesmethodgamel2010time}, effective dissipation arising from the discarded rapidly oscillating transition sectors can also be obtained at the operator level. For simplicity however, we neglect these dissipative corrections in the present work.
\subsection{JLM diagram method}
\label{Section JLM diagram method}
\subsubsection{JLM diagram construction rules}
\begin{figure}
    \centering
    \includegraphics[width=0.9\linewidth]{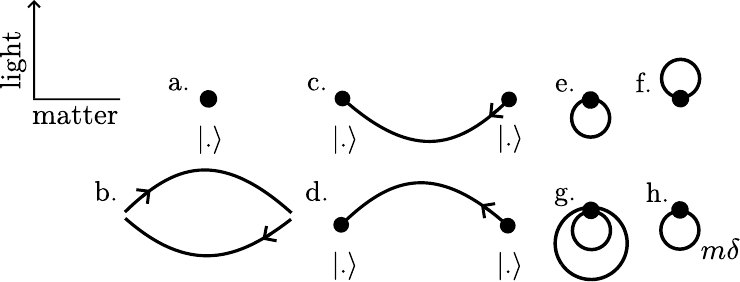}
    \caption{JLM diagrams elementary blocks. States of matter are represented by point vertices labeled by the corresponding ket $\ket{\cdot}$ (panel a.). Photon emission and absorption events are indicated by arrows (panel b.). Photon-induced matter transitions are indicated by arrows. Absorption events are drawn below the matter axis defined by the matter-state vertices (panel c.), while emission events are drawn above it (panel d.). Free evolution of a JLM transition operator is represented by a free-evolution loop returning to the same matter vertex. When the free evolution follows an absorption (emission) event, the loop is drawn below (above) the matter axis (panels e. and f.). Multiple free-evolution loops occurring at the same matter vertex may be stacked (panel g.). For higher-order processes in which several one-photon detunings accumulate at a given vertex, the corresponding detunings $\delta$ and their multiplicities $m$ may be annotated on the free-evolution loop (panel h.).}
    \label{Figure JLM diagrams building blocks}
\end{figure}

\begin{figure}
    \centering
    \includegraphics[width=0.7\linewidth]{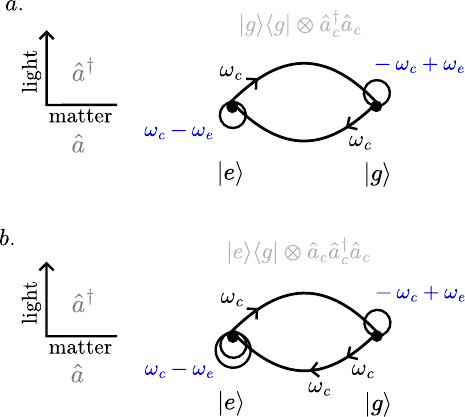}
    \caption{Example of a first-order (panel a.) and second-order (panel b.) JLM diagram starting from $\hat \xi_{g\boldsymbol \sigma_c}^e = \ket{e}\bra{g}\otimes \hat a_c$ and resulting in $\hat \xi_{g \boldsymbol \sigma_c}^{g \boldsymbol \sigma_c}$ and $\hat \xi_{g\boldsymbol \sigma_c \boldsymbol \sigma_c}^{e\boldsymbol \sigma_c}$, respectively. Following rule (R1), the matter vertices (see panel a. from Fig.~\ref{Figure JLM diagrams building blocks}) are placed along the so-called matter axis and light-driven transitions are represented by arrows (see panel b. from Fig.~\ref{Figure JLM diagrams building blocks}) connecting matter states together. An arrow indicates a photon annihilation or creation event (see panels c. and d. from Fig.~\ref{Figure JLM diagrams building blocks}) depending on whether it is located below or above the matter axis, respectively. Light–matter axes are included for clarity. The light axis is oriented upward to emphasize photon creation and annihilation events, while the matter axis carries no preferred orientation since the matter levels are not necessarily energy ordered. Note that two arrowheads are placed on the $\ket{e}\bra{g}\otimes \hat a_c$ transition branch to indicate that this transition occurs twice. The loop attached to each matter vertex implements rule (R2): following each interaction event, the system undergoes free evolution at the cumulative detuning (see panels e., f. and g. from Fig.~\ref{Figure JLM diagrams building blocks}) associated with the sequence of interaction arrows leading to that matter state. In practice, this amounts to summing the detunings (here indicated in blue) of all the preceding transitions. For instance, in the panel a. and b., the diagram corresponds to an operator evolving at zero and $\omega_c-\omega_e$ frequency, respectively. According to rule (R3), the diagram in the panel a. is circularly invariant and thus Hermitian, while the diagram in panel b. is not and its Hermitian conjugate should be added to the interaction-Hamiltonian correction $\Delta \hat H_{\text{int}}^{n\leq 2}$; whose different contributions are calculated using rule (R4).}
    \label{Figure example of double-sided Feynman diagrams for three-level system in RWA third-order}
\end{figure}

\begin{figure}
    \centering
    \includegraphics[width=0.65\linewidth]{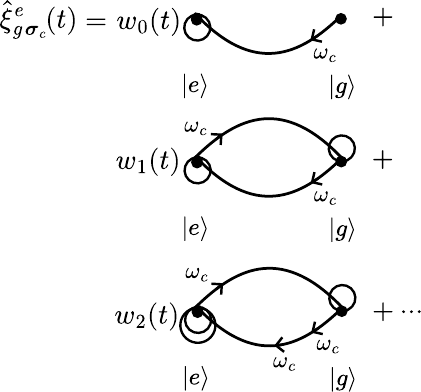}
    \caption{Perturbative expansion up to second order of the $\hat \xi_{g\boldsymbol \sigma_c}^e(t)$ JLM transition operator depicted in terms of JLM diagrams. Each diagram has a time-dependent weight $w_n(t)$ which depends on the one-photon detunings through its time-dependent part $v_n(t)$ Eq.~\eqref{Equation general expression of the time-dependent loop diagram} and which is a product of $n$ one-photon coupling terms. }
    \label{Figure perturbative expansion illustrative example}
\end{figure}

As previously discussed, from Eq.~\eqref{Equation perturbation theory expansion in Heisenberg-interaction picture} and Eq.~\eqref{Equation Neumann series of the resolvent of the Liouvillian}, one can observe that each perturbative order $n\in \mathbb N$ of the JLM transition operators is acted upon $n$ times by the interaction Liouvillian $\mathcal L_{\text{int}}$. The resulting commutator applies $\hat H_{\text{int}}$ either on the right or left side of the JLM transition operators, and each application increases the perturbative order by one. The different perturbative-expansion contributions $\hat \xi_{i\boldsymbol \sigma}^{j(n)}(\omega,t)$ in $\hat \xi_{i\boldsymbol \sigma}^j(\omega,t) = \sum_{n=0}^{+\infty}\hat \xi_{i\boldsymbol \sigma}^{j(n)}(\omega,t)$ can therefore be tracked based on how many times the interaction Hamiltonian $\hat H_{\text{int}}$ has been applied and on which side of the JLM transition operators. This perturbative bookkeeping gives rise to JLM diagrams, which explicitly represent successive interaction events leading to a $n$th-order $\hat \xi_{i\boldsymbol \sigma}^{j(n)}(\omega,t)$ contribution in the perturbative expansion. Because JLM transition operators jointly encode light and matter degrees of freedom, the diagrams must make both explicit. We therefore construct them such that light dynamics is tracked along one axis and matter dynamics along the other. For a $n$th-order JLM transition operator, the JLM diagram construction rules are set to be the following
\begin{itemize}
    \item (R1) Place the matter states on vertices. The matter states do not have to be energy ordered. One-photon JLM transition operators appear as strands linking vertices, each labeled by the associated photon frequency and, if relevant, auxiliary degrees of freedom. JLM-transition-operator strand drawn below (respectively, above) the axis carrying the matter-state vertices signals a photon absorption (respectively, emission) event, see Fig.~\ref{Figure JLM diagrams building blocks}. 
\end{itemize}
The two-dimensional diagrammatic representation introduced by rule (R1) exploits the interaction Hamiltonian’s formulation in terms of photon creation and annihilation operators, allowing transitions to be depicted directly as creation and annihilation events rather than in terms of photon states. This transition representation is hence particularly advantageous for processes involving photons from different modes, as in waveguide QED.
\begin{itemize}
    \item (R2) Between the $v$ and $v+1$ JLM-transition-operator strands ($v\in \llbracket 0,n \rrbracket$), the $v$-order JLM transition operator undergoes free evolution and accumulates a phase $e^{-i\Delta_v t}$ at the regularized cumulative detunings $\Delta_v = \sum_{l=0}^v\left[(-1)^{c_l}\omega_l-i\theta_l\right]-(\omega_j-\omega_i)_l$, where $c_l=0$ for an absorbed photon, $c_l = 1$ for an emitted photon, $(\omega_j-\omega_i)_l$ denotes the energy difference between the two levels involves in the $l$-order transition, and $\theta_l$ the regularization terms. During the free evolution, the matter state is left unchanged. The free evolution is thus represented by a free-evolution loop that starts and ends at the matter-state vertex, see Fig.~\ref{Figure JLM diagrams building blocks}. 
\end{itemize}
An $n$th-order JLM diagram therefore encodes the $n+1$ detunings associated with the corresponding $(n+1)$-photon transition which dictates the multiple timescales. 
\begin{itemize}
    \item (R3) For a $n$th-order JLM diagram, the Hermitian conjugate is obtained by reading each JLM strand backwards. If the $n$th-order JLM diagram is circularly invariant, the corresponding $n$th-order JLM transition operator is already Hermitian.
    \item (R4) An $n$-th order JLM diagram has $w_n(t)$ coefficient in the $n$th-order interaction-Hamiltonian correction $\Delta \hat H_{\text{int}}^{(n)}(t)$ resulting from the JLM transition operators perturbative expansion. The time-dependent part $v_n(t)$ of $w_n(t)$ -- \textit{dropping the (time-dependent) coupling terms for readability} -- can be expressed as
\begin{equation}
\begin{split}
\label{Equation general expression of the time-dependent loop diagram}
   v_n(t) &=
   (-1)^{N_L} \sum_{l=0}^n e^{-i\Delta_l t}e^{-\Theta_l t} \\
   &\times \prod_{\substack{k=0 \\ k\neq l}}^n\frac{1}{\Delta_l-\Delta_k+i(\Theta_k-\Theta_l)},
 \end{split}
\end{equation}
where $\Delta_l$ and $\Theta_l$ denote the cumulative detunings and cumulative regularization terms at $l$th order, respectively. The extra $(-1)^{N_L}$ accounts for the $N_L$ operators acting on the left of the zeroth-order JLM transition operator and is due to the commutator applied by the interaction Liouvillian $\mathcal L_{\text{int}}$.
\end{itemize}
Eq.~\eqref{Equation general expression of the time-dependent loop diagram} is free of divergences because the partial-fraction (residue-derivative) decomposition is carried out with finite regulators $\theta_l > 0$, which keep all regularized cumulative detunings $\Delta_k-i\Theta_k$ distinct and thus prevent singular denominators. The computation is given in Appendix~\ref{Appendix Joint light-matter transition operators' weights calculation} and Appendix~\ref{Appendix Computation with the resolvent}, using nested convolution products and directly the resolvent, respectively. When some cumulative detunings coincide in the unregularized limit, \textit{i.e.} $\Delta_l = \Delta_k$ for certain $l\neq k$, the limits $\Theta_l,\Theta_k \rightarrow 0^+$ must be taken carefully, as considered in the Appendix~\ref{Appendix Dealing with cumulative-detuning degeneracies}. \\
We illustrate the previous construction rules with the Jaynes-Cummings model~\cite{JCmodellarson2021jaynes}, wherein the interaction Hamiltonian reads 
\begin{equation}
    \hat H_{\text{int}} = g_{g\boldsymbol \sigma_c}^e\hat \xi_{g\boldsymbol \sigma_c}^e + \text{h.c},
\end{equation}
where $\hat \xi_{g\boldsymbol \sigma_c}^e = \ket{e}\bra{g}\otimes \hat a_c$, corresponding the transition from the two-level atomic ground state $\ket{g}$ to the two-level atomic excited state $\ket{e}$ by absorption of a photon of frequency $\omega_c$, and where $g_{g\boldsymbol \sigma_c}^e$ is the associated coupling term. Let us construct in Fig.~\ref{Figure example of double-sided Feynman diagrams for three-level system in RWA third-order} the first-order (panel a.) and the second-order (panel b.) JLM diagrams starting from the $\hat \xi_{g\boldsymbol \sigma_c}^e$ one-photon JLM transition operator at zeroth-order and leading to the two-photon and three-photon JLM transition operators $\hat \xi_{g \boldsymbol \sigma_c}^{g \boldsymbol \sigma_c}$ and $\hat \xi_{g\boldsymbol \sigma_c \boldsymbol \sigma_c}^{e\boldsymbol \sigma_c}$, respectively.
The perturbative expansion of $\hat \xi_{g\boldsymbol \sigma_c}^e(t) = \sum_{n=0}^{+\infty}\hat \xi_{g\boldsymbol \sigma_c}^{e(n)}(t)$ is represented by means of JLM diagrams in Fig.~\ref{Figure perturbative expansion illustrative example}. In the companion article~\cite{CompanionarticlePRL}, we argue that the present diagrammatic representation goes beyond systematic construction rules and emerges because of the very definition of the JLM transition operators, in particular their spectral properties. 
\subsubsection{Process and time ordering}
\label{Subsubsection Process and time ordering}
A central aspect of the present approach is the distinction between two different notions of ordering. The \textit{process ordering} and the \textit{perturbative time ordering}. The process ordering refers to the sequence of elementary transitions defining a given physical pathway, whereas the perturbative ordering specifies the order in which interaction operators are applied within the perturbative expansion. An $n$th-order JLM transition operator is represented by an ordered string of zeroth-order transition operators, $\hat{\Xi}^{(n)} = \hat{\xi}_n \cdots \hat{\xi}_1 \hat{\xi}_0$, which defines the process ordering. This string is to be read from right to left. The physical process corresponds to first applying the perturbed operator $\hat{\xi}_0$, followed by $\hat{\xi}_1$, then $\hat{\xi}_2$, and so on. The ordering $\hat \xi_n\dots \hat \xi_1\hat \xi_0$ is fixed by the physical sequence of transitions and is independent of the perturbative construction. \\
On the other hand, the perturbative time ordering arises from the perturbation theory of zeroth-order JLM transition operators, which is built from nested commutators with the interaction Hamiltonian, itself composed of zeroth-order JLM transition operators. As a consequence, at each perturbative step, JLM transition operators may act either on the left or on the right of the previously constructed operator. If $n-r$ operators are applied on the left of the zeroth-order operator and $r$ on its right, then the $n-r$ operators are time ordered from right to left and the $r$ operators on the right are ordered from left to right, while no time ordering is imposed between operators acting on opposite sides of the zeroth-order operator. For example, consider the second-order operator $\hat \Xi^{(2)} = \hat \xi_k \hat \xi_j \hat \xi_i$. If $\hat \xi_i$ is chosen as the zeroth-order operator, then $\hat \xi_j$ acts first at first order on its left, followed by $\hat \xi_k$ at second order. Conversely, if $\hat \xi_k$ is taken as the zeroth-order operator, then $\hat \xi_j$ first acts at first order on its right, before $\hat \xi_i$ is applied at second order. Finally, if $\hat \xi_j$ is selected as the zeroth-order operator, the perturbative construction admits two equivalent orderings: either $\hat \xi_i$ acts first on the right at first order followed by $\hat \xi_k$ on the left at second order, or alternatively $\hat \xi_k$ may first act on the left at first order before $\hat \xi_i$ is applied on the right at second order. \\
This apparent violation of causality is an artefact of perturbative expansions. The different time orderings are encoded within the nested time integrations of Eq.~\eqref{Equation perturbation theory expansion in Heisenberg-interaction picture}; after integration, only the ordering corresponding to the physical process survives. The weight $w_n(t)$ of a $n$th-order JLM diagram $\hat \Xi^{(n)}$ obtained from the perturbative expansion of a zeroth-order operator drawn from the interaction Hamiltonian Eq.~\eqref{Equation compact general interaction Hamiltonian} should account for the different possible time-ordering contributions: $\sum_{\text{time orderings}}w_n(t) \leftarrow w_n(t)$. For instance, the second-order operator $\hat \Xi^{(2)} = \hat \xi_k \hat \xi_j\hat \xi_i$ is associated with four possible perturbative time orderings.
\subsubsection{Perturbator and perturbed object}
\label{Section Perturbator and perturbed object}
\begin{figure}
    \centering
    \includegraphics[width=0.9\linewidth]{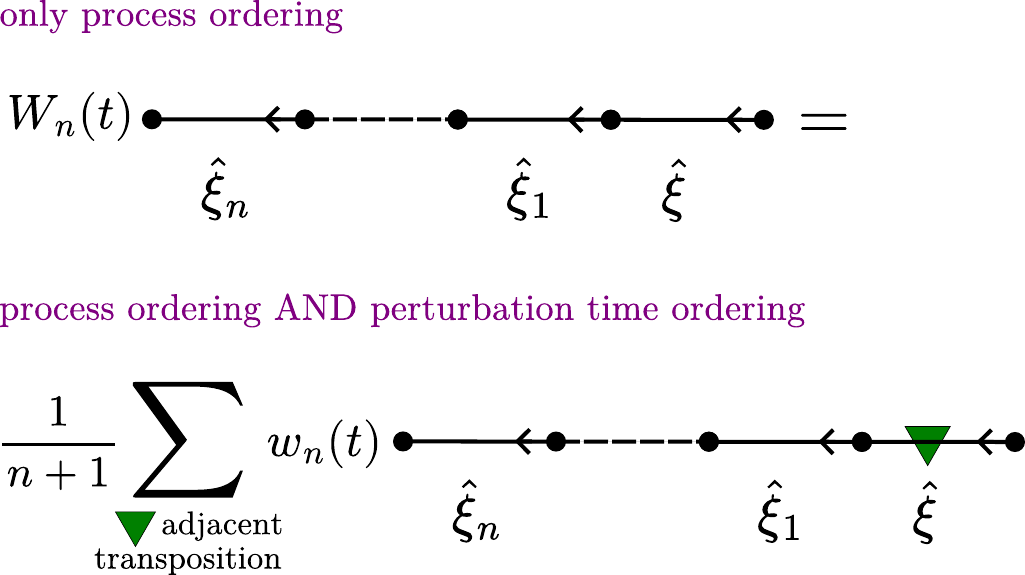}
    \caption{Sketch illustrating rule (R5) for constructing the JLM transition-operator diagram. The weight of an $n$th-order operator $\hat \Xi^{(n)}$ is obtained by summing over all perturbative time orderings, yielding the total weight $\sum_{\text{time orderings}} w_n(t)$. These distinct orderings arise from the $(n+1)$ possible choices of perturbed operator, which are represented diagrammatically by adjacent transpositions of the perturbed-operator label. Since the choice of perturbed operator is unphysical, the different contributions are averaged, leading to the total weight $W_n(t) = \frac{1}{n+1}\sum_{\text{time orderings}} w_n(t)$. In practice, only the canonical diagram -- where the process and perturbative time ordering are identical -- is drawn, with all contributions absorbed into $W_n(t)$. Here, arrows are shown straight for generality, and free-evolution loops are omitted. }
    \label{Figure averaging virtual paths contributions}
\end{figure}

The perturbative time ordering discussed above originates from a distinctive feature of the present JLM perturbation-theory framework: transition operators serve simultaneously as generators of the perturbation and as the dynamical variables being perturbed. In other words, the interaction acts on itself to generate higher-order processes. Consequently, unlike standard perturbation theory constructions, there exists no privileged root identifying an initial state prior to interaction. This reflects the fact that any of the zeroth-order transition operators appearing in Eq.~\eqref{Equation compact general interaction Hamiltonian} may serve either as the generator of the perturbation or as the object being perturbed. At perturbative order $n$, an $n$th-order JLM operator $\hat \Xi^{(n)}$ emerges from the $n+1$ possible choices of which among its $n+1$ constituent transition operators is selected as the zeroth-order object. Within the present JLM diagrammatic formalism, these choices may be sketched by attaching a perturbed-operator label, depicted as a green downward-pointing triangle in Fig.~\ref{Figure averaging virtual paths contributions}. JLM diagrams that differ only by adjacent transpositions of this label -- \textit{i.e.} by the choice of which operator is designated as the perturbed one -- correspond to the same underlying physical process associated with the same $n$th-order JLM transition operator $\hat{\Xi}^{(n)}$, albeit with different weights $w_n(t)$. Namely, adjacent transposition of the perturbed-operator label adds an additional factor $-1$. This sign structure resembles that encountered in fermionic statistics, despite the fact that JLM transition operators are neither bosonic nor fermionic. This fermion-like behavior originates exclusively from the underlying commutator algebra, rather than from intrinsic operator statistics. \\ 
To account for the multiple weights $w_n(t)$ associated with the same $n$th-order physical process, we define the total amplitude $W_n(t)$ of a given $n$th-order transition operator as their average. Since there are exactly $(n+1)$ distinct possible choice of perturbed zeroth-order operator for a $n$th order JLM transition operator $\hat \Xi^{(n)}$, this averaging amounts to dividing by $n+1$, as illustrated in Fig.~\ref{Figure averaging virtual paths contributions}, where the transition arrows are depicted straight for generality. In this way, the choice of which transition operator acts as the perturbation generator is stripped of any physical significance and remains a feature of the perturbative expansion machinery.
We therefore introduce a new diagrammatic rule
\begin{itemize}
    \item (R5) To compute an $n$th-order JLM transition operator $\hat \Xi^{(n)}$, one first constructs the corresponding JLM diagram by following the physical process ordering; this diagram defines the \textit{canonical diagram}. The total weight $W_n(t)$ associated with $\hat \Xi^{(n)}$ is then obtained by accounting for different perturbative time orderings and by averaging over the $(n+1)$ possible choices of operator subjected to the perturbation. In other words, in the $n$-th order correction $\Delta \hat H_{\text{int}}^{(n)}(t)$ to the interaction Hamiltonian, a $n$th-order JLM transition operator $\hat \Xi^{(n)}$ contributes as $W_n(t)\hat \Xi^{(n)}$, with $W_n(t) = \frac{1}{n+1}\sum_{\text{time orderings}} w_n(t)$.
\end{itemize}
\subsubsection{Hermitian conjugate diagram}
\label{Section diagram Hermitian conjugate}
\begin{figure}
    \centering
    \includegraphics[width=\linewidth]{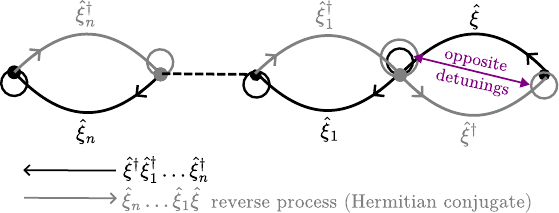}
    \caption{Schematic illustration of how a JLM diagram (of constitutive JLM transition operators indicated in black) and its Hermitian branch (of constitutive JLM transition operators indicated in grey) combine to form a closed loop that returns the system to its initial state. For a given JLM transition operator, the Hermitian conjugate carries the opposite detuning. Here, we highlight this for $\hat \xi$ and $\hat \xi^\dagger$ and the same reasoning applies to all other JLM transition operators.}
    \label{Figure total loop with reverse process}
\end{figure}

For a $n$th-order JLM diagram leading to $\hat \xi_n\dots \hat \xi_1\hat \xi_0$ of weight $w_n(t)$, taking its Hermitian conjugate $\hat \xi_0^\dagger \hat \xi_1^\dagger \dots \hat \xi_n^\dagger$ thus amounts to pertubatively expanding $\hat \xi_0^\dagger$ with $\hat \xi_1^\dagger \dots \hat \xi_n^\dagger$ by applying JLM transition operators on the right. The corresponding weight $w^{\text{reverse}}_n(t)$ can be proven to be exactly equal to the complex conjugate of $w_n(t)$. This comes from the Hermitian conjugate of the different JLM transition operators flipping the cumulative detunings sign and the additional $(-1)^{n}$ derives from the fact that all JLM transitions are now applied on the right of the perturbed operator $\hat \xi^\dagger_0$. Further computation details are provided in Appendix.~\ref{Appendix Computation of the reverse process' weight}. Consequently, each JLM diagram leading to the JLM transition operator $\hat \xi_n \dots \hat \xi_1 \hat \xi_0$ has a corresponding partner diagram representing the reverse physical process, whose weight is the Hermitian conjugate of that of the original diagram. This implies that for $P$ distinct physical processes at a given order, \textit{at most} $P/2$ need to be sketched, as the others can be deduced by taking the Hermitian conjugate. The JLM diagram can be completed to directly incorporate the reverse process, so that the symmetric half of the JLM diagram represents this reversal, restoring the system to its initial state, see Fig.~\ref{Figure total loop with reverse process}.
In practice, however, we display only one branch, with the Hermitian conjugate implicitly understood. Moreover, if the JLM transition operator is itself Hermitian, no corresponding Hermitian-conjugate partner is required and it would overcount said JLM transition operator's contribution. \\
In James’ method, an \textit{a posteriori} symmetrization prescription is required to ensure the Hermitianity of the effective Hamiltonian. In contrast, starting from $\hat H_{\text{int}} = \sum_{m=1}^{M/2}\hat \xi_m + \text{h.c.}$ guarantees Hermiticity at every stage of the expansion. Indeed, we automatically pair each process with its reverse, so that the resulting corrected interaction Hamiltonian is manifestly Hermitian without additional symmetrization.
\subsubsection{Permissible diagrams and circular permutation}
\label{Section Permissible diagrams and circular permutation}
Although the enumeration of all possible transition pathways quickly becomes cumbersome in large Hilbert spaces or at high perturbative order, the matter-overlap constraints of the JLM transition operators drastically reduce the combinatorics. Starting from the interaction Hamiltonian $\hat H_{\mathrm{int}}=\sum_{m=1}^{M/2}\hat\xi_m$, a brute-force construction would involve $M^n$ possible $n$th-order operator strings. However, matter-overlap constraints restrict this number to at most $2(n+1)\binom{M+n}{n+1}$ distinct $n$th-order JLM transition operators. This reduction follows from the fact that a given $n$th-order JLM diagram, when it returns to its initial matter state, corresponds to a set of $(n+1)$ operator strings related by circular permutations of the matter part, while the additional factor $2$ accounts for the Hermitian-conjugate partner. Accordingly, the number of distinct JLM diagrams that need to be drawn at order $n$ is at most $\binom{M+n}{n+1}$. For a fixed matter string, additional distinct operator strings may arise from the bosonic factors attached to it. Operators acting on different photonic modes commute and do not generate new contributions, whereas multiplicities involving creation and annihilation operators acting on the same mode do. If $m$ denotes the number of such multiple positions in a given matter string, this results in at most a factor $2^m$ additional contributions, see Appendix~\ref{Appendix Permissible diagrams}. In practice, this bound is very loose. Indeed, matter-index constraints and detuning or resonance selection eliminate most possible diagrams, so the actual number of nonzero diagrams is much smaller. In Sec.~\ref{Section First-order loop diagrams}, we illustrate this reduction in the first-order case, while the second-order case is discussed further in Appendix~\ref{Appendix Second-order loop diagrams}.
\subsubsection{JLM diagrammatic workflow}
From the previous sections, one can extract a systematic workflow to compute the $n$th-order correction $\Delta \hat H_{\text{int}}^{(n)}(t)$ to the interaction Hamiltonian in the Heisenberg picture. This workflow is the fully general construction; in practice, symmetry, selection rules, and resonance conditions drastically simplify each step. 
\begin{itemize}
\item (S1) Start from the interaction Hamiltonian and identify its building blocks, namely the zeroth-order JLM transition operators. 
\item (S2) Use the combinatorial upper bound discuss in Sec.~\ref{Section Permissible diagrams and circular permutation}, together with physical constraints such as resonance conditions or selection rules to determine how many diagrams are to be drawn. 
\item (S3) Construct the $n$-th order JLM diagrams by concatenating $n+1$ zeroth-order transition operators according to the process ordering. 
\item (S4) A diagram that closes onto its initial matter state, corresponds to $(n+1)$ operator strings related by circular permutations. If there are $m$ multiplicities in assigning creation or annihilation operators on the same bosonic mode, the diagram accounts for up to $2^m$ times distinct operator strings. 
\item (S5) Compute the weight of each $n$th-order contribution by summing over time-ordering contributions Eq.~\eqref{Equation general expression of the time-dependent loop diagram} and including the averaging factor $1/(n+1)$ associated with the choice of perturbed operator, as discussed in Sec.~\ref{Section Perturbator and perturbed object}.
\item (S6) Sum all $n$th-order JLM transition operators with their corresponding weights $W_n(t)$ to obtain $\Delta \hat H_{\text{int}}^{(n)}(t)$. 
\item (S7) Finally, perform the adiabatic elimination by applying the projector $\mathcal P_T$ Eq.~\eqref{Equation projection time-averaging map} to project out the off-resonant contributions over the timescale $T$. 
 \end{itemize}
This workflow provides a general recipe valid at arbitrary perturbative order. In practice, however, only a small subset of diagrams contributes, making the method considerably simpler to implement. We now illustrate it explicitly at first order $n=1$. 
\subsubsection{First-order JLM diagrams}
\label{Section First-order loop diagrams}
\begin{figure}
\label{Figure first-order diagram topologies}
    \centering
    \includegraphics[width=\linewidth]{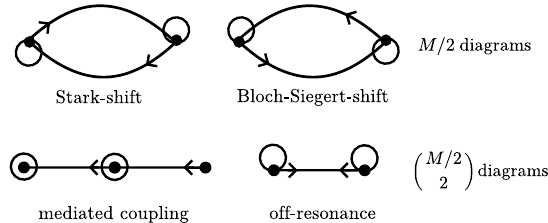}
    \caption{First order ($n=1$) classes of JLM diagrams. The first kind corresponds to terms of the form $\hat \xi_i^\dagger \hat \xi_i$ and $\hat \xi_i\hat \xi_i^\dagger$, representing either Stark or Bloch-Sieger shifts. There are exactly $M/2$ JLM diagrams associated with this first class. (ii) The second class represents mediated coupling or out-of-resonant events. There are at most $\binom{M/2}{2}$ JLM diagrams associated with this second class. Here, some arrows are shown straight for generality. }
    \label{Figure n=1 what loop diagrams should we draw}
\end{figure}

At first order in the perturbative expansion, the time dependence $v_1(t)$ of a first-order JLM diagram weight $w_1(t)$ is given by Eq.~\eqref{Equation general expression of the time-dependent loop diagram}
\begin{equation}
    v_1(t) = \frac{-\left(e^{-i(\Delta_0-i\Theta_0)t}-e^{-i(\Delta_1-i\Theta_1)t} \right)}{\Delta_0-\Delta_1+i(\Theta_1-\Theta_0)}.
\end{equation}
Due to the matter part of the JLM transition operators of the interaction Hamiltonian expressed in its compact form $\hat H_{\text{int}} = \sum_{m=1}^{M/2} \hat \xi_m$, it can straightforwardly be observed that there are three classes of JLM diagrams because of their matter parts: (i) $M/2$ Stark-shift~\cite{QuantumACStarkshiftblais2004cavity} or Bloch-Siegert-shift~\cite{BlochSiegertbloch1940magnetic,BlochSiegertforn2010observation} energy renormalization JLM diagrams corresponding to the operators $\hat \xi_i^\dagger \hat \xi_i$ and $\hat \xi_i\hat \xi_i^\dagger$ (upon cyclic permutation). Performing the circular permutation $\hat \xi_i \leftrightarrow \hat \xi_i^\dagger$ (see Sec.~\ref{Section Permissible diagrams and circular permutation}) yields $\hat \xi_i^\dagger \hat \xi_i$. Their weights are identical up to a sign factor due to the commutator algebra. (ii) $\binom{M/2}{2}$ mediated-coupling or out-of-resonance JLM diagrams defining either $\hat \xi_j\hat \xi_i$ or $\hat \xi_i\hat \xi_j$ (upon cyclic permutation), with $\hat \xi_j \neq \hat \xi_i^\dagger$. Note that depending on the matter-part overlap, $\hat \xi_j\hat \xi_i$ and/or $\hat \xi_j\hat \xi_i$ might be equal to zero. \\
The Bloch-Siegert and out-of-resonance JLM diagrams emerge when taking into account counter-rotating events, as discussed in Sec~\ref{Section Two-level system in a single-mode cavity}. These two classes of first-order JLM diagrams are sketched in Fig.~\ref{Figure n=1 what loop diagrams should we draw}. At first order, there are hence \textit{at most} $M/2+\binom{M/2}{2}$ diagrams to be drawn instead of the $(M/2)^2$ terms needed using the James' method~\cite{Jamesmethodjames2007effective}, which provides a computational advantage for all values of $M$, see Appendix~\ref{Appendix Permissible diagrams}.
The regularized cumulative detuning $\Delta_0-i\Theta_0$ is the regularized detuning $\delta_i-i\theta_i$ of the JLM transition operator at the zeroth-order node while $\Delta_1-i\Theta_1 = \delta_i+\delta_j-i(\theta_i+\theta_j)$ where $\delta_j-i\theta_j$ is the regularized cumulative detuning of the JLM transition operator a first-order. Because of the perturbation time ordering discussed in Sec.~\ref{Subsubsection Process and time ordering}, each two-photon JLM transition operators has a total weight $W_1(t)$ which is the average of the two possible $w_1(t)$ weights depending on which of $\hat \xi_j$ and $\hat \xi_i$ is the perturbed-operator, see Fig.~\ref{Figure order n=1, cyclical topology loop diagram}.
\begin{figure}
    \centering
    \includegraphics[width=0.65\linewidth]{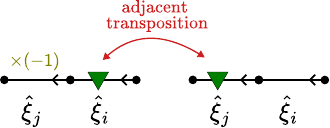}
    \caption{First order $(n=1)$ JLM diagrams adjacent transposition-sign shift. For a given JLM diagram, an adjacent transposition of the perturbed-operator label yields the same two-photon JLM transition operator, but with a different time-dependent weight. The latter changes sign due to the commutator algebra, since $N_L$, the number of operators acting on the left, is either equal to zero or one. Here, arrows are shown straight for generality.}
    \label{Figure order n=1, cyclical topology loop diagram}
\end{figure}
Therefore, for a first-order JLM diagram leading to the JLM transition operator $\hat \xi_j \hat \xi_i$, the total time-dependent part $V_1(t)$ of the weight $W_1(t)$ reads 
\begin{equation}
\begin{split}
\label{Equation total time-dependent weight for n=1}
    V_1(t) &=\frac{1}{2}\bigg( \left[\frac{e^{-i\delta_it}e^{-\theta_it}}{\delta_j-i\theta_j}- \frac{e^{-i\delta_jt}e^{-i\theta_j t}}{\delta_i-i\theta_i}\right]\\
    &+e^{-i(\delta_i+\delta_j)t}e^{-(\theta_i+\theta_j)t}\left[\frac{1}{\delta_i-i\theta_i}-\frac{1}{\delta_j-i\theta_j}\right]\bigg),
\end{split}
\end{equation}
where the additional $-1$ stems from the $\hat \xi_i \leftrightarrow \hat \xi_j$ permutation flipping the sign of the contribution as it entails applying the operator on the right. The weight of the operator $\hat \xi_i\hat \xi_j$ -- if non-zero -- is obtained by cyclic permutation of $\hat \xi_j\hat \xi_i$ is simply the opposite of $W_1(t)$. 
\\ In the discrete case, one has by definition of the dispersive regime $\delta_i \neq 0$ and $\delta_j \neq 0$, which allows one to safely take the limits $\theta_i\rightarrow 0$ and $\theta_j \rightarrow 0$ in Eq.~\eqref{Equation total time-dependent weight for n=1}, yielding 
\begin{equation}
\begin{split}
\label{Equation total time-dependent weight for n=1 and discrete case}
    V_1^{\text{discrete}}(t) &= \frac{1}{2}\bigg( \left[\frac{e^{-i\delta_it}}{\delta_j}- \frac{e^{-i\delta_jt}}{\delta_i}  \right]
    \\
    &+e^{-i(\delta_i+\delta_j)t}\left[\frac{1}{\delta_i}-\frac{1}{\delta_j}\right]\bigg).
\end{split}
\end{equation}
Invoking the adiabatic elimination of fast-evolving one-photon JLM transition operators through the projector $\mathcal P_T$ from Eq.~\eqref{Equation projection time-averaging map}, $V^{\text{discrete}}(t)$ simplifies 
\begin{equation}
\label{Equation total time-dependent weight for n=1 after coarse-grained time}
    V_1^{\text{discrete}}(t) \approx \frac{1}{2}e^{-i(\delta_i+\delta_j)t}\left[\frac{1}{\delta_i}-\frac{1}{\delta_j}\right],
\end{equation}
projecting out the contributions over the time $T$ satisfying $T\gg 1/|\delta_i|$ and $T\gg 1/|\delta_j|$. In the continuous case, one should carefully manipulate the limits $\theta_i,\theta_j \rightarrow 0$ within the frequency continuum, as discussed in Sec.~\ref{Section Three-level systems in a continuum} and Appendix~\ref{Appendix Computation of the effective Hamiltonian for a frequency continuum at first order}. At this stage, one sees that the weight and time dependence Eq.~\eqref{Equation total time-dependent weight for n=1} of any first-order, two-photon JLM transition operator can be computed straightforwardly and systematically from all first-order diagrams. Furthermore, we obtain in Eq.~\eqref{Equation total time-dependent weight for n=1 after coarse-grained time} the \textit{harmonic average} introduced in~\cite{Jamesmethodjames2007effective,Jamesmethodgamel2010time} which in our formalism can thus be interpreted as a direct consequence of the averaging over different time orderings that are topologically equivalent within the JLM diagrams. \\
The second-order JLM diagrams are further detailed in Appendix~\ref{Appendix Second-order loop diagrams} and employed in the companion article~\cite{CompanionarticlePRL} to retrieve the non-trivial three-photon resonance shown by Ma et Law~\cite{Threephotoresonancenma2015three}.
\section{Examples}
\label{Section Examples}
In this section, we illustrate the use of the JLM-transition-operator diagrammatic with several light-matter adiabatic elimination situations up to first order in the perturbation, that is up to second order in the interaction Hamiltonian. We begin with the derivation of the quantum AC Stark shift for a two-level system, then turn to effective qubit-qubit interactions in the Dicke and Tavis-Cummings models. Finally, we derive an effective two-photon interaction Hamiltonian for matter coupled to a frequency continuum.
\subsection{Two-level system in a single-mode cavity}
\label{Section Two-level system in a single-mode cavity}
Consider the two-level system $\{\ket{g},\ket{e};\; \omega_g=0, \omega_e>0\}$ with the ground state taken as the reference of energy. This system, \textit{e.g.} a quantum dot, is placed in a single-mode cavity of frequency $\omega_c$. The interaction Hamiltonian can be expressed as the quantum Rabi Hamiltonian~\cite{RabimodelQuantxie2017quantum,JCmodellarson2021jaynes}
\begin{equation}
\label{Equation Rabi Hamiltonian}
    \hat H_{\text{int}} = g_{g\boldsymbol{\sigma}_c}^e\hat \xi_{g\boldsymbol{\sigma}_c}^e + g_{e\boldsymbol{\sigma}_c}^g\hat \xi^g_{e\boldsymbol{\sigma}_c}+\text{h.c.},
\end{equation}
where $\hat \xi_{g\boldsymbol \sigma_c}^e = \ket{e}\bra{g}\otimes \hat a_c$ and $\hat \xi_{e\sigma_c}^{g} = \ket{g}\bra{e}\otimes \hat a_c$. Possible auxiliary modes to the frequency are accounted for by $\boldsymbol{\sigma}_c$ but are irrelevant here as they do not play a role in the light-matter coupling. The $\boldsymbol \sigma_c$ label is kept regardless in order to follow the transition process, absorption or emission of a photon. Additionally, since there is only one frequency involved, we omit the $\omega_c$ labelling in the JLM transition operators and in their graphical representations. At zeroth order, that is to say up to first order in the interaction Hamiltonian, the JLM transition operators evolve as 
\begin{align}
    \hat \xi_{g\boldsymbol \sigma_c}^{e(0)}(t) = e^{-i\mathcal L_{\text{free}}t}\hat \xi_{g\boldsymbol{\sigma}_c}^{e(0)} &= e^{-i(\omega_c-\omega_e)t}\hat \xi_{g\boldsymbol{\sigma}_c}^{e(0)} \\
    \hat \xi_{e\boldsymbol \sigma_c}^{g(0)}(t)= e^{-i\mathcal L_{\text{free}}t}\hat \xi_{e\boldsymbol{\sigma}_c}^{g(0)} &= e^{-i(\omega_c+\omega_e)t}\hat \xi_{e\boldsymbol{\sigma}_c}^{g(0)}.
\end{align}
The textbook RWA then consists in stating that for coupling terms negligible compared to the $\omega_c$ and $\omega_e$ frequencies, the counter-rotating joint transition operator $\hat \xi_{e\boldsymbol{\sigma}_c}^{g(0)}$ and its Hermitian conjugate average to zero average to zero over a time $T\gg 1/(\omega_c+\omega_e)$. In our framework, RWA then amounts to the projection $\mathcal P_T$ Eq.~\eqref{Equation projection time-averaging map} onto the resonant-transition subspace. Nevertheless, this does not imply that such terms cannot generate resonant contributions at higher perturbative orders, in which case the projection $\mathcal P_T$ must be applied at a later stage of the perturbative expansion. Indeed, let us perturbatively expand the JLM transition operators to first order in the coupling term. \\
According to the Sec.~\ref{Section First-order loop diagrams}, one should draw at most $4/2+\binom{4/2}{2} = 3$ JLM diagrams, see Figure~\ref{Figure 2LE in cavity RWA_only, counter-rotating_only vertices}.
\begin{figure}
\label{First-order JLM diagrams for an atomic two-level system in a single-mode cavity}
    \centering
    \includegraphics[width=1\linewidth]{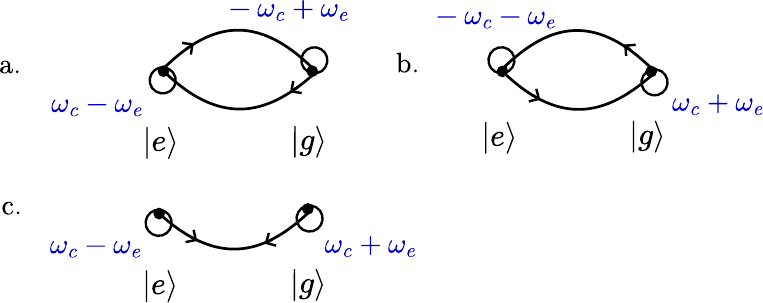}
    \caption{First-order JLM diagrams for an atomic two-level system placed in a single-mode cavity. Three diagrams can be identified: a Stark-shift diagram (panel a.), where interactions are under the RWA and with zero cumulative detuning; a Bloch-Siegert-shift diagram (panel b.), where interactions are out of the RWA and with zero cumulative detuning; an off-resonance diagram (panel c.), where under-the-RWA and out-of-the-RWA interactions are combined and with $2\omega_c$ cumulative detuning. The JLM diagrams panel a. and b. lead to on-resonance contributions whereas the JLM diagram in panel c. results in an out-of-resonance contribution. Detunings (in blue) are indicated for clarity, though this is not necessary in practice.}
    \label{Figure 2LE in cavity RWA_only, counter-rotating_only vertices}
\end{figure}
Interestingly, resonant first-order pathways are composed exclusively of either co-rotating or counter-rotating transitions, whereas off-resonant pathways arise from a mixture of the two. More specifically, here the resonant diagrams exhibit a distinct topology compared to the off-resonant one. Resonant diagrams form loops corresponding to energy renormalization, while the off-resonant diagram is open. From the perspective of the JLM-transition-operator perturbation theory, co-rotating and counter-rotating events play equivalent roles; what ultimately determines resonance is the diagram topology leading to a zero cumulative detuning. \\
The JLM diagrams' weights can be grouped in three time-dependent terms $W_{\text{co}}(t)$, $W_{\text{counter}}(t)$, $W_{\text{mix}}(t)$. They can be expressed using Eq.~\eqref{Equation total time-dependent weight for n=1} as 
\begin{align}
    W_{\text{co}}(t) &= \frac{1}{2}\frac{|g_{g\boldsymbol \sigma_c}^e|^2}{\omega_e-\omega_c}\left[e^{-i(\omega_c-\omega_e)t}+e^{-i(\omega_e-\omega_c)t}-2 \right] \\
    W_{\text{counter}}(t) &= \frac{|g_{g\boldsymbol \sigma_c}^{e}|^2}{\omega_e+\omega_c} \\ 
    W_{\text{mix}}(t) &=\frac{1}{2}\frac{g_{g\boldsymbol \sigma_c}^e g_{e\boldsymbol \sigma_c}^{g}}{\omega_c-\omega_e}e^{-i(\omega_c-\omega_e)t},
\end{align}
where we projected out fast-oscillating terms with $\mathcal P_T$ from Eq.~\eqref{Equation projection time-averaging map} safely assuming that $T\gg 1/(\omega_e+\omega_c)$ and $T\gg 1/\omega_c$ thus dropping the exponential terms evolving at those frequencies. Next, over the coarse-grained time interval $T\gg 1/|\omega_e-\omega_c|$, the exponential $e^{\pm i(\omega_e-\omega_c)t}$ can also be dropped leading to 
\begin{align}
    W_{\text{co}}(t) &= \frac{-|g_{g\boldsymbol \sigma_c}^e|^2}{\omega_e-\omega_c} \\
    W_{\text{counter}}(t) &= \frac{|g_{g\boldsymbol \sigma_c}^e|^2}{\omega_e+\omega_c} \\ 
    W_{\text{mix}}(t) &=0
\end{align}
The JLM diagram where both co-rotating and counter-rotating events occur along the interaction pathways vanishes altogether. As $\omega_e+\omega_c \gg |\omega_e-\omega_c|$, even though the weight $ W_{\text{counter}}(t)$ is not equal to zero, it is still negligible compared to the weight $W_{\text{co}}(t)$. The Bloch-Siegert shift~\cite{BlochSiegertbloch1940magnetic,BlochSiegertforn2010observation} associated with $W_{\text{counter}}(t)$ can nevertheless become significant in certain physical systems. \\ Dropping the Bloch-Siegert shift and grouping the second order contributions together in the total effective Hamiltonian (up to an overall energy shift) and using the atomic subspace identity operator $\mathbb 1_\alpha = \ket{e}\bra{e}+\ket{g}\bra{g}$
\begin{equation}
\label{Equation two-level system in a single-mode cavity AC Stark shift}
\Delta \hat H_{\text{int}}^{(1)} = \left(\omega_c+\frac{|g_{g\boldsymbol \sigma_c}^e|^2}{\omega_e-\omega_c}\hat \sigma_z \right)\hat n_c +\frac{1}{2}\left(\omega_e+\frac{|g_{g\boldsymbol \sigma_c}^e|^2}{\omega_e-\omega_c} \right)\hat \sigma_z
\end{equation}
where $\hat n_c = \hat a_c^\dagger \hat a_c$ and $\hat \sigma_z = \ket{e}\bra{e} - \ket{g}\bra{g}$ are the number and atomic system Pauli-Z operator, respectively. The obtained first-order JLM transition operators are Hermitian, no Hermitian partner should be added. The second-order Hamiltonian Eq.~\eqref{Equation two-level system in a single-mode cavity AC Stark shift} features the so-called CQED quantum photon-number dependent-AC Stark shift~\cite{QuantumACStarkshiftblais2004cavity}. It emerges because an off-resonant photon mode can be virtually absorbed and re-emitted by the atomic system, as shown by the loop diagram in Fig.~\ref{Figure 2LE in cavity RWA_only, counter-rotating_only vertices}. Each energy-conserving $\ket{g}\bra{g}\otimes\hat a_c^\dagger\hat a_c $ or $\ket{e}\bra{e}\otimes\hat a_c \hat a_c^\dagger$ round-trip process dresses the atomic levels and produces a net level shift of order $|g_{g\boldsymbol \sigma_c}^e|^2/(\omega_e-\omega_c)$ per photon along with an additional vacuum-induced shift sometimes referred to as the Lamb shift~\cite{QuantumACStarkshiftblais2004cavity,CircuitQEDblais2021circuit}. The first-order correction to the Hamiltonian \label{Equation two-level system in a single-mode cavity AC Stark shift} is also discussed in the companion article~\cite{CompanionarticlePRL} wherein the Bloch-Siegert correction is additionally computed explicitly along with the three-photon interaction correction at second-order when $\omega_c \approx \omega_e/3$. 
\subsection{Multiple qubits coupled to a single oscillator}
\label{Subsection Multiple qubits coupled to a single oscillator}
\begin{figure*}
    \centering
    \includegraphics[width=\linewidth]{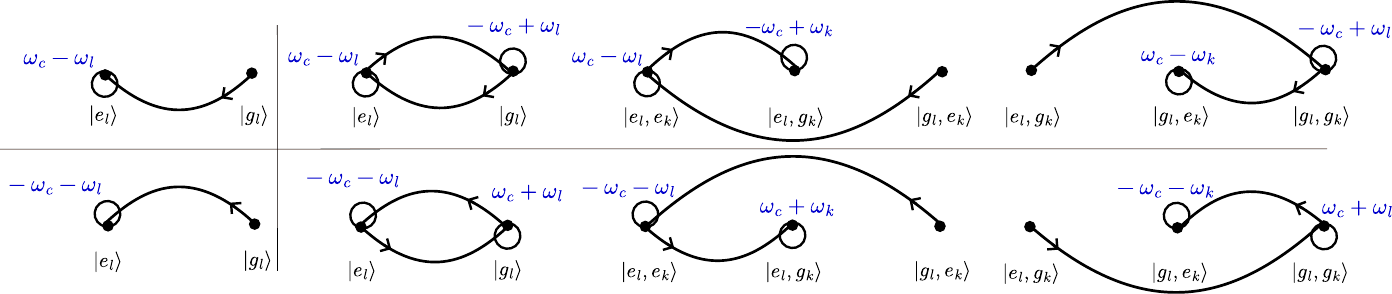}
    \caption{Zeroth (left column) and first-order (right column) JLM diagrams for the Tavis-Cummings model (top row) and the additional counter-rotating contributions for the Dicke model (bottom row). At first order, we only depict the six resonant diagrams; the off-resonant diagrams involving two emission or two absorption events are omitted. For the JLM diagram corresponding to the transition $\ket{g_l,e_k}\rightarrow \ket{e_l,e_k} \rightarrow \ket{e_l,g_k}$ via photon absorption and emission, the associated JLM transition operator is $\hat \sigma_-^k \hat \sigma_+^l \otimes \hat a_c^\dagger \hat a_c$. The same reasoning holds for the other JLM diagrams. As in Fig.~\ref{First-order JLM diagrams for an atomic two-level system in a single-mode cavity}, since only one photon frequency is involved in the light-matter transitions, we omit the $\omega_c$ in the JLM diagrams. Detunings (in blue) are indicated for clarity, though this is not necessary in practice.}
    \label{Figure Dicke and TC firts-order diagrams}
\end{figure*}

The interaction Hamiltonian $\hat H_{\text{int}}$ in Eq.~\eqref{Equation general interaction Hamiltonian} describes a $N$-level system coupled to bosonic degrees of freedom. The interaction Hamiltonian may equivalently describe $N$ two-level (qubit) systems coupled to single-mode cavity of frequency $\omega_c$; this is the Dicke Hamiltonian~\cite{dicke1954coherence,JCmodellarson2021jaynes} 
\begin{equation}
\label{Equation Dicke Hamiltonian}
\begin{split}
    \hat H_{\text{Dicke}} &= \omega_c\hat a_c^\dagger \hat a_c + \sum_{l=1}^N\frac{\omega_l}{2}\hat \sigma_z^l + \sum_{l=1}^Ng_l\left(\hat \sigma_+^l\hat a_c + \hat \sigma_-^l \hat a_c + \text{h.c}\right),
\end{split}
\end{equation}
where, for the $l$-th qubit, $\omega_l$ is its frequency, $g_l$ the coupling term between the cavity and the qubit, $\hat \sigma_z^l$ the Pauli-Z operator defined by $\hat \sigma_z^l = \ket{e_l}\bra{e_l}-\ket{g_l}\bra{g_l}$, $\hat \sigma_+^l = \ket{e_l}\bra{g_l}$, and $\hat \sigma_-^l = \ket{g_l}\bra{e_l}$, the raising and lowering operator, respectively. As explained in Appendix.~\ref{Appendix Equation interaction Hamiltonian multiple qubits}, a $N$-qubit system can be equivalently represented as $2^N$-level atomic system, in which the $4N$ transition operators from Eq.~\eqref{Equation Dicke Hamiltonian} decompose into to $2N\times2^N$ transition operators in the $2^N$-level-system description. As a result, according to Sec.~\ref{Section First-order loop diagrams}, at most $(N\times2^N)+\binom{N\times2^N}{2}$ JLM diagrams must be considered. At first order, however, the zeroth-order JLM diagrams couple only pairwise transitions. Consequently, the interaction part of the Dicke Hamiltonian in Eq.~\eqref{Equation Dicke Hamiltonian} can be viewed as an interaction Hamiltonian composed of four terms and their Hermitian conjugates associated with the $l$th and $k$th qubits. Within this set of four operators and their Hermitian conjugates, the pairwise structure of the zeroth-order diagrams implies that the resulting combinations can involve either two transitions belonging to the same qubit or transitions belonging to two different qubits. In Fig.~\ref{Figure Dicke and TC firts-order diagrams}, we represent the JLM diagrams at first order of the dispersive perturbative expansion for the Dicke model Eq.~\eqref{Equation Dicke Hamiltonian} and Tavis-Cummings model which amounts to the Dicke model without considering counter-rotating contributions $\hat \sigma_-^l\hat a_c+\text{h.c}$. 
The JLM diagrams in Fig.~\ref{Figure Dicke and TC firts-order diagrams} show three types of diagrams: (i) the ones where the $l$th-qubit system is closed and does not couple to another $k$th-qubit system. This leads to Stark- or Bloch-Siegert-shift diagrams depending on whether the constitutive transitions are within or out of the RWA; (ii) processes in which a photon is first absorbed, driving the transition $\ket{g_l}\rightarrow\ket{e_l}$, respectively $\ket{e_l}\rightarrow\ket{g_l}$, followed by photon emission driving the transition $\ket{e_k}\rightarrow\ket{g_k}$, respectively $\ket{g_k}\rightarrow\ket{e_k}$; these correspond to co-rotating and counter-rotating transitions, respectively; (iii) processes in which a photon is first emitted, driving the transition $\ket{e_l}\rightarrow\ket{g_l}$, respectively $\ket{g_l}\rightarrow\ket{e_l}$, followed by photon absorption driving the transition $\ket{g_k}\rightarrow\ket{e_k}$, respectively $\ket{e_k}\rightarrow\ket{g_k}$; these again correspond to co-rotating and counter-rotating transitions, respectively. As in the previous section, we compute the first-order correction to the interaction Hamiltonian $\Delta \hat H_{\text{TC}}^{(1)}(t)$ for the Tavis-Cummings model 
\begin{equation}
\label{Equation TC Hamiltonian first-order correction}
\begin{split}
    &\Delta \hat H_{\text{TC}}^{(1)}(t) = \sum_{l=1}^N\frac{g_l^2}{\delta_l}\hat \sigma_z^l\left(\hat a_c^\dagger \hat a_c+\frac{1}{2}\right) \\
    &+2\sum_{l>k}\frac{g_lg_k}{\overline{\delta}_{lk}}\left(e^{-i(\omega_k-\omega_l)t}\hat \sigma_-^k\hat \sigma_+^l+e^{-i(\omega_l-\omega_k)t}\hat \sigma_+^k\hat \sigma_-^l\right),
\end{split}
\end{equation}
where we define the harmonic average
\begin{equation}
    \frac{1}{\overline \delta_{lk}} = \frac{1}{2}\left(\frac{1}{\delta_l}+\frac{1}{\delta_k}\right),
\end{equation}
and the one-photon detunings $\delta_i=\omega_i-\omega_c$. Note that we obtained Eq.~\eqref{Equation TC Hamiltonian first-order correction} by using the bosonic commutation relation $[\hat a_c,\hat a_c^\dagger] = \mathbb 1_{\omega}$, $\mathbb 1_{\omega}$ being the identity operator, and by projecting out fast-oscillating contributions with $\mathcal P_T$ over a time $T\gg 1/|\omega_c-\omega_k|$. The effective correction to the interaction Hamiltonian in Eq.~\eqref{Equation TC Hamiltonian first-order correction} shows that the cavity mediates an isotropic XY-type qubit-qubit interaction~\cite{Dispersiveregimezueco2009qubit,SWayyash2025dispersive} analogous to the isotropic XY-type spin-spin interaction, whose interaction Hamiltonian is expressed as $\hat H_{\text{XY}} =-J/2\sum_{\langle k, l\rangle}\left(\hat S_-^k\hat S_+^l+\hat S_+^k\hat S_-^l\right)$, where $J$ is the coupling term and $\hat S_\pm^l$ the raising and lowering spin operators, respectively. Our expression further establishes a direct connection to the classical isotropic XY-type spin-spin model~\cite{Condensedmatterchaikin1995principles} of Hamiltonian $H_{\text{XY}}=-J\sum_{\langle k, l\rangle}\cos(\theta_k-\theta_l)$, where $\theta_k-\theta_l$ is the angle mismatch between neighboring spins. Indeed, in $\Delta \hat H_{\text{TC}}^{(1)}(t)$ Eq.~\eqref{Equation TC Hamiltonian first-order correction} the coupling term is $J=- 2g_lg_k/\overline{\delta}_{lk}$ and the phase factor $(\omega_k-\omega_l)t$ -- originating from energy mismatch between qubits -- is analogous to the angle mismatch in the XY model. This correspondence between energy detuning and angle mismatch is obscured in the Schrieffer-Wolff derivations of~\cite{Dispersiveregimezueco2009qubit,SWayyash2025dispersive}. Indeed, their rotating-frame choices absorb the phase $e^{i(\omega_k-\omega_l)t}$ into the generator of the unitary transformation, thus yielding effective Hamiltonians where the spectral structure is no longer explicit. In the present Heisenberg-picture formulation, no such frame is chosen and the phase is retained as a direct consequence of the perturbative expansion in terms of transition operators. The first-order correction to the interaction Hamiltonian $\Delta \hat H_{\text{Dicke,non-RWA}}^{(1)}(t) = \Delta \hat H_{\text{Dicke}}^{(1)}(t) -\Delta \hat H_{\text{TC}}^{(1)}(t)$ for the Dicke model due to the counter-rotating contributions, can similarly be calculated as
\begin{equation}
    \begin{split}
        &\Delta \hat H_{\text{Dicke,non-RWA }}^{(1)}(t) = \sum_{l=1}^N\frac{g_l^2}{\Sigma_l}\hat \sigma_z^l\left(\hat a_c^\dagger \hat a_c+\frac{1}{2}\right) \\
        &-2\sum_{l>k}\frac{g_lg_k}{\overline \Sigma_{lk}}\bigg(e^{-i(\omega_k-\omega_l)t}\hat \sigma_-^k\hat \sigma_+^l\otimes \hat a_c\hat a_c^\dagger \\
        &+ e^{-i(\omega_l-\omega_k)t}\hat \sigma_+^k\hat \sigma_-^l\otimes\hat a_c^\dagger \hat a_c+\text{h.c}\bigg)
    \end{split},
\end{equation}
where $\Sigma_l = \omega_l+\omega_c$ and 
\begin{equation}
    \frac{1}{\Sigma_{lk}} = \frac{1}{2}\left(\frac{1}{\Sigma_k}+\frac{1}{\Sigma_l}\right).
\end{equation}
The two corrections to the interaction Hamiltonian, $\Delta \hat H_{\text{TC}}^{(1)}(t)$ and $\Delta \hat H_{\text{Dicke,non-RWA}}^{(1)}(t)$ are written in a unique Heisenberg picture and are therefore not tied to a particular transformation, unlike in~\cite{Dispersiveregimezueco2009qubit,SWayyash2025dispersive}. As a result, the transition from the Dicke to the Tavis-Cummings model follows directly by simply removing the counter-rotating contributions. This contrasts with the results of~\cite{Dispersiveregimezueco2009qubit,SWayyash2025dispersive}, where the dispersive expansion of the Dicke Hamiltonian -- exhibiting an effective Ising-type qubit-qubit interaction -- does not smoothly reduce to the dispersive expansion of the Tavis-Cummings Hamiltonian once the counter-rotating terms are discarded (see Eq.~(20) and Eq.~(23) of~\cite{Dispersiveregimezueco2009qubit} and Eq.~(14) and Eq.~(15) of~\cite{SWayyash2025dispersive}). This discrepancy arises because in~\cite{Dispersiveregimezueco2009qubit,SWayyash2025dispersive}, two different Schrieffer-Wolff transformations were used for the Tavis-Cummings and Dicke models.

\subsection{Three-level systems in a continuum: $\Xi$, $\Lambda$ and $V$ configuration}
\label{Section Three-level systems in a continuum}
\begin{figure}
    \centering
    \includegraphics[width=\linewidth]{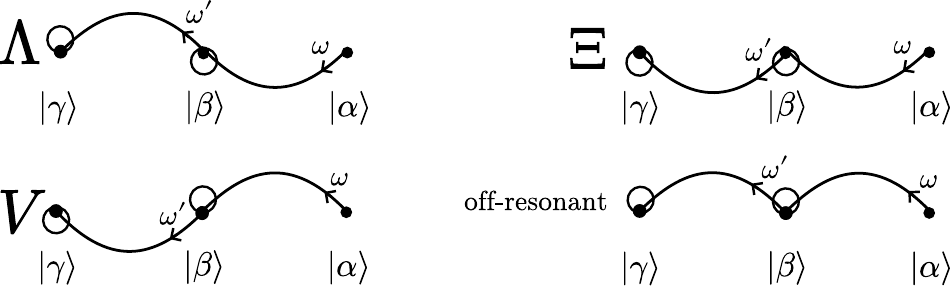}
    \caption{First-order mediated-coupling JLM diagrams for a three-level system placed coupled to a frequency continuum. There are four distinct JLM diagrams. First of all, JLM diagrams that lead to off-resonant contributions regardless of the three-level configuration because we assumed that $\omega_\gamma > \omega_\alpha$. The remaining three classes correspond to JLM diagrams whose nodes all satisfy the RWA for one of the three canonical three-level configurations if the three-level system matches one of the three possible three-level configuration: the $\Xi$ configuration ($\omega_\alpha < \omega_\beta < \omega_\gamma$), the $\Lambda$ configuration ($\omega_\alpha < \omega_\gamma < \omega_\beta$) and the $V$ configuration ($\omega_\beta < \omega_\alpha < \omega_\gamma$). }
    \label{Figure Three-level system loop first-order without Stark}
\end{figure}

\begin{figure}
    \centering
    \includegraphics[width=0.8\linewidth]{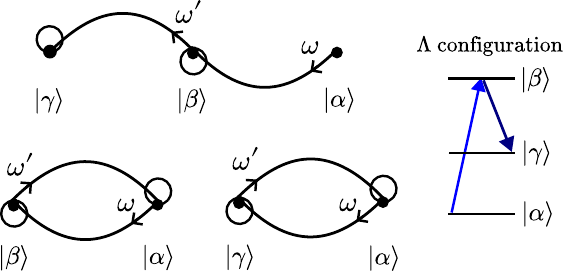}
    \caption{Sketch of the three JLM diagrams diagrams required to capture the first-order JLM-transition-operator perturbation expansion. The two loop JLM diagrams correspond to Stark shifts leading to diagonal terms in $\ket{\alpha}\bra{\alpha}\otimes \hat a_{\boldsymbol \sigma_i}^\dagger(\omega')\hat a_{\boldsymbol \sigma_i}(\omega)$ and $\ket{\gamma}\bra{\gamma}\otimes \hat a_{\boldsymbol \sigma_j}^\dagger(\omega')\hat a_{\boldsymbol \sigma_j}(\omega)$. The diagonal terms proportional to $\ket{\beta}\bra{\beta}$ are obtained by circular permutations of the JLM transition operators in the Stark-shift loop diagram; it amounts to adding an extra $-1$. There is one mediated-coupling JLM-diagram translating the effective two-photon coupling driving a transition from the state $\ket{\alpha}$ to the state $\ket{\gamma}$.}
    \label{Figure n=1 loop diagram for Lambda system in RWA}
\end{figure}

Consider a three-level system $\{\ket{\alpha},\ket{\beta}, \ket{\gamma}\}$ of natural frequencies $\omega_\alpha$, $\omega_\beta$ and $\omega_\gamma$ coupled to a frequency continuum. At first order in the interaction Hamiltonian, only the $\ket{\alpha}\leftrightarrow\ket{\beta}$ and $\ket{\beta} \leftrightarrow \ket{\gamma}$ transitions are available. The $\ket{\gamma}$ is assumed to have a higher energy than the state $\ket{\alpha}$. No assumptions are made about the energy of the state $\ket{\beta}$ for now. The interaction Hamiltonian reads 
\begin{equation}
\begin{split}
    \hat H_{\text{int}} &= \int_{\mathbb R^+}d\omega\;\left(g_{\alpha \boldsymbol \sigma_i}^\beta(\omega)\hat \xi_{\alpha \boldsymbol \sigma_i}^\beta(\omega)+g_{\beta \boldsymbol \sigma_i}^\alpha (\omega)\hat \xi_{\beta \boldsymbol \sigma_i}^\alpha(\omega) \right) \\
    &+ \int_{\mathbb R^+}d\omega\;\left(g_{\beta \boldsymbol \sigma_j}^\gamma(\omega)\hat \xi_{\beta \boldsymbol \sigma_j}^\gamma(\omega)+g_{\gamma \boldsymbol \sigma_j}^\beta (\omega)\hat \xi_{\gamma \boldsymbol \sigma_j}^\beta(\omega) \right)+\text{h.c},
\end{split}
\end{equation}
where the auxiliary degrees of freedom $\boldsymbol \sigma_i$ and $\boldsymbol \sigma_j$ may label the polarization, the direction of propagation, etc. Let us now perform a first-order perturbative expansion of the JLM transition operators. Note that at this stage, since no surmises have been considered regarding the $\ket{\beta}$ energy level, all these JLM transition operators could be either co-rotating or counter-rotating contributions. This will prove relevant shortly as we show how the JLM diagrams resulting from the pertubative expansion encompass all possible three-level consideration: $\Xi$ where $\omega_\alpha \leq \omega_\beta \leq \omega_\gamma$, $\Lambda$ where $\omega_\alpha \leq \omega_\gamma \leq \omega_\beta$ and $V$ where $\omega_\beta \leq \omega_\alpha \leq \omega_\gamma$. First of all, let us analyze the possible Stark and Bloch-Siegert shifts. At first order, a JLM diagram resulting in JLM transition operators with $\hat a_{\boldsymbol \sigma_k}(\omega) \hat a_{\boldsymbol \sigma_k}(\omega')$ (and equivalently the h.c) for $k=i,j$ are always out of resonance and can thus be ignored. On the other hand, JLM transition operators with $\hat a_{\boldsymbol \sigma_k}(\omega)\hat a_{\boldsymbol \sigma_k}^\dagger(\omega)$ (and equivalently the h.c) evolve at a frequency $|\omega-\omega'|$ and can therefore be kept and looked for. Nonetheless, we can in first approximation ignore these Stark-shift and Bloch-Siegert-shift pathways and focus on transitions between two different matter states. The corresponding JLM diagrams leading to off-diagonal matter terms -- \text{i.e.} terms in $\ket{j}\bra{i}$ with $j\neq i$ -- are displayed in Fig.~\ref{Figure Three-level system loop first-order without Stark}. 
Remarkably, sketching all possible JLM diagrams without minding whether the transitions are co-rotating or counter-rotating events, naturally leads to all the three-level energy-level configurations $\Lambda$, $V$ and $\Xi$ by identifying JLM diagrams with co-rotating transitions only. In other words, deciding which JLM diagrams to keep reduces to identifying the system’s energy-level configuration. Some JLM diagrams are always off-resonant because of the initial assumption that $\omega_\gamma > \omega_\alpha$, see Fig.~\ref{Figure Three-level system loop first-order without Stark}. The JLM diagrams' weights can be computed directly using Eq.~\eqref{Equation total time-dependent weight for n=1}. For instance, focusing on the $\Lambda$ configuration in the RWA, only three JLM diagrams are needed, see Fig.~\ref{Figure n=1 loop diagram for Lambda system in RWA}.
One can compute the interaction Hamiltonian correction $\Delta \hat H^{(1)}_{\text{med.}}(t)$ due to the first-order JLM transition operators $\ket{\gamma}\bra{\alpha}\otimes \hat a_{\boldsymbol \sigma_j}^\dagger(\omega')\hat a_{\boldsymbol \sigma_i}(\omega)$ as 
\begin{equation}
\label{Equation Hamiltonian correction due to mediated coupling three-level system continuum}
    \begin{split}
        \Delta \hat H^{(1)}_{\text{med.}}(t) &= \ket{\gamma}\bra{\alpha}\otimes \mathcal P\bigg( \int_{\left(\mathbb R^+\right)^2}d\omega'd\omega\;g_{\beta}^{\gamma}(\omega')g_{\alpha}^\beta(\omega)\\
        &\times\frac{e^{-i\left(\omega-\omega'-(\omega_\gamma-\omega_\beta)\right)t}}{\overline \delta(\omega',\omega)}\hat a_{\boldsymbol \sigma_j}^\dagger(\omega')\hat a_{\boldsymbol \sigma_i}(\omega)\bigg) +\text{h.c},
    \end{split}
\end{equation}
where $\overline \delta(\omega',\omega)$ denotes the harmonic average
\begin{equation}
\label{Equation harmonic average for the continuum in a three-level system}
    \frac{1}{\overline \delta(\omega',\omega)} = \frac{1}{2}\left(\frac{1}{\omega-\omega_\beta}-\frac{1}{\omega_\gamma-\omega_\beta-\omega'}\right),
\end{equation}
and the Stark-shift contribution $\Delta \hat H_{\text{Stark,$\alpha$}}^{(1)}(t)$ to the state $\ket{\alpha}$
\begin{equation}
\label{Equation Hamiltonian correction due to alpha Stark shift coupling three-level system continuum}
\begin{split}
    &\Delta \hat H_{\text{Stark,$\alpha$}}^{(1)}(t) = \frac{1}{2}\ket{\alpha}\bra{\alpha}\otimes \mathcal P \bigg(\int_{(\mathbb{R}^+)^2}d\omega'd\omega \; \\
&\Big[
F_i(\omega,\omega')\hat a_{\boldsymbol \sigma_i}^\dagger(\omega')\hat a_{\boldsymbol \sigma_i}(\omega)
+ F_i^*(\omega',\omega)\hat a_{\boldsymbol \sigma_i}^\dagger(\omega)\hat a_{\boldsymbol \sigma_i}(\omega')
\Big]\bigg), 
\end{split}
\end{equation}
where we consider the kernel 
\begin{equation}
    F_i(\omega,\omega') = g_{\beta}^\gamma(\omega')g_{\alpha}^\beta(\omega)e^{-i(\omega-\omega')t}\left[\frac{1}{\omega-\omega_\beta}+\frac{1}{\omega'-\omega_\beta} \right]
\end{equation}
in case $[\hat a_{\boldsymbol \sigma_i}(\omega),\hat a_{\boldsymbol \sigma_i}^\dagger(\omega')]\neq 0$, in which case $\hat a^\dagger_{\boldsymbol \sigma_i}(\omega')\hat a_{\boldsymbol \sigma_i}(\omega)$ is not Hermitian and its Hermitian should consequently be accounted for. $\mathcal P$ labels the Cauchy principal value and is to be understood as applied individually where the singularities emerge. The computations details are given in Appendix~\ref{Appendix Computation of the effective Hamiltonian for a frequency continuum at first order}, where the other Stark shifts leading to diagonal terms in $\ket{\beta}\bra{\beta}$ and $\ket{\gamma}\bra{\gamma}$ are obtained as well. Eq.~\eqref{Equation Hamiltonian correction due to alpha Stark shift coupling three-level system continuum} indicates that, for matter coupled to a frequency continuum, the Stark-shift contribution naturally decomposes into two distinct parts. Photons resonant at the same frequency lead to energy renormalization as in Eq.~\eqref{Equation two-level system in a single-mode cavity AC Stark shift}, whereas photons at different frequencies mediate genuine interaction terms. \\
The principal values in Eq.~\eqref{Equation Hamiltonian correction due to mediated coupling three-level system continuum} can be controlled with a frequency cutoff close to the resonant frequencies. The frequency cutoff can be introduced \textit{ad hoc} -- for example via an exponential decay of the coupling at high frequencies~\cite{UVcutoffleppakangas2018quantum} --, derived microscopically by going beyond the dipolar approximation in Eq.~\eqref{Equation general interaction Hamiltonian}, as in~\cite{UVcutoffparra2018quantum} where boundary conditions at the coupling point enforce mode renormalization, or obtained from the intrinsic cutoff set by the waveguide density of states, a common example being 3D microwave guide mode from aluminium or copper. \\
The Hamiltonian correction in Eq.~\eqref{Equation Hamiltonian correction due to mediated coupling three-level system continuum} was derived by applying the adiabatic elimination through $\mathcal P_T$ from Eq.~\eqref{Equation projection time-averaging map}, thereby neglecting one-photon fast-oscillating terms. This may be quantified with the following practical rule of thumb. Denote $\mathcal R_T = \{\omega, |\delta(\omega)| \not \gg 1/T\}$ the resonant region of the frequency continuum and $\mathcal R_T^c$ its complement, that is the off-resonant region. Define the ratio $R_T = \int_{\mathcal R_T}d\omega\; |g(\omega)|^2/\int_{\mathcal R_T^c}d\omega\; |g(\omega)|^2$ weighing the fraction of the frequency continuum at resonance, where $\delta(\omega)$ is the one-photon detuning at a given frequency $\omega$. If $R_T \ll 1$, the frequency continuum can safely be assumed to be roughly off-resonant. The validity of the adiabatic elimination thus depends on two main factors. First, coupling locality: since the interaction strength varies with frequency, only modes with significant coupling play a role. If the one-photon transition frequencies lie in regions where the coupling is weak or near its cutoff, they can be safely neglected compared to the modes fulfilling the two-photon resonance condition. Second, density of states: even weakly coupled modes may contribute if many are clustered near resonance. Conversely, when most modes are far from resonance, their individual contributions are weak and they collectively have little effect on one-photon transitions. \\
Note that in the specific case where all of the continuum frequencies are far-detuned from the one-photon resonance, \textit{i.e.} there is a lower bound $|\delta_{\text{lower}}|$ large with respect to the coupling terms, then exactly all frequency-dependent transitions oscillate fast and can safely be neglected for interaction times large compared to $1/|\delta_{\text{lower}}|$ and the singularities in Eq.~\eqref{Equation Hamiltonian correction due to mediated coupling three-level system continuum} are avoided. This simpler configuration was considered in our previous work~\cite{Adiabaticeliminationmeguebel2025generation}, where we showed that a four-level quantum dot embedded in a waveguide can be employed as a photon-photon frequency entangling gate bypassing significant drawbacks from traditional nonlinear-material-based or linear optics post-selection methods. The present treatment applies to a broader class of waveguide QED settings and is more systematic at arbitrary perturbation order, while also being more transparent. Indeed, it provides a direct derivation together with a diagrammatic representation, rather than relying on the involved system of differential equations employed in~\cite{Adiabaticeliminationmeguebel2025generation}.
\section{Discussion: comparison with other methods}
In this section, we summarize how the present transition-centric framework and its diagrammatic formulation compare with other methods to compute effective Hamiltonians. \\
A key feature of our approach is that it operates directly at the level of transition operators, which are eigenoperators of the free Liouvillian. As a result, the spectral structure of the interaction – and in particular the transition detunings governing the validity of the dispersive regime – is made explicit from the outset, without requiring any prior and non-unique choice of frame transformation as in state-centric approaches~\cite{AdiabaticeliminationLambdasystembrion2007adiabatic,Adiabaticeliminationbeyondpaulisch2014beyond,adiabaticeliminationatomicphysicsalexanian1995unitary,SchriefferWolffbravyi2011schrieffer,Dispersiveregimezueco2009qubit,SWayyash2025dispersive}. Moreover, the explicit determination of the transformation generator in the Schrieffer-Wollf can be highly non-trivial at higher perturbative orders or in complex multilevel settings. In contrast, our formalism avoids this difficulty altogether by remaining within the uniquely defined Heisenberg picture. \\ \\
Our method also bridges two perturbative viewpoints that are often treated separately: projection-based approaches, such as Nakajima-Zwanzig~\cite{Generalopenquantumsystembreuer2002theory,Adiabaticeliminationprojectorappraochgonzalez2024tutorial} or resolvent-based methods~\cite{AdiabaticeliminationLambdasystembrion2007adiabatic}; and time-averaging schemes, such as James' method. In the present JLM transition framework, the elimination of rapidly oscillating transition operators is formulated as the adiabatic elimination of a non-resonant subspace, whereas the relevant subspace is selected by a resonance criterion over a given time. \\
In light of this, the weights in Eq.~\eqref{Equation general expression of the time-dependent loop diagram} resolve the multi-scale dynamics underlying the generation of $n$th-order JLM transition operators. In contrast to James' method~\cite{Jamesmethodjames2007effective,Jamesmethodgamel2010time}, which applies a uniform low-pass filtering of fast dynamics, our formulation retains distinct throughout the perturbative construction. In particular, counter-rotating transition operators are kept in the intermediate steps and treated on equal footing with co-rotating contributions in the diagrammatics. Although they may ultimately be eliminated, these terms can play a crucial role in mediating higher-order resonant processes. \\ \\
Another important feature of working at the transition level is that no truncation of the underlying Hilbert space is required in order to capture higher-order perturbative effects and processes involving virtual intermediate transitions are retained explicitly, unlike for instance the Ma and Law~\cite{Threephotoresonancenma2015three} approach to deriving a three-photon resonance in a far-detuned cavity. The latter is covered in the companion article~\cite{CompanionarticlePRL}. Retaining transition pathways explicitly is especially valuable in waveguide QED where the relevant field modes form a continuum. \\
Furthermore, the diagrammatic formulation, although it can become intricate at high order, remains systematic at arbitrary perturbative order. Whereas Schrieffer-Wolff calculations quickly become cumbersome because of the proliferation of nested commutators and generator choices, the JLM diagrams provide a transparent graphical encoding of operator ordering, selection rules, and detuning denominators, and naturally extend to arbitrary order. \\ \\
The present JLM diagrammatics is conceptually related to that used in nonlinear spectroscopy~\cite{CloseloopMukamelmarx2008nonlinear,CloseloopMukamelroslyak2009unified,DoublesidedFeynmandiagramsmukamel2010ultrafast,DoublesidedFeynmandiagramsRevModPhys.88.045008}, where perturbation theory is carried out in Liouville space, \textit{i.e.} the space of operators acting on the Hilbert space. While nonlinear spectroscopy emphasizes expansions of the density matrix, our approach operates directly at the level of light-matter transition operators. As discussed in Appendix~\ref{Appendix Transition state-dressed state correspondence}, these operators form the ensemble from which the density matrix is constructed, so the underlying perturbative structures are formally analogous. Correspondingly, nonlinear-spectroscopy diagrams~\cite{CloseloopMukamelmarx2008nonlinear,CloseloopMukamelroslyak2009unified,DoublesidedFeynmandiagramsmukamel2010ultrafast,DoublesidedFeynmandiagramsRevModPhys.88.045008} are used to compute optical response functions, whereas our diagrammatics is designed to extract effective Hamiltonians via adiabatic elimination. Additionally, although the graphical structure also bears a superficial resemblance to Feynman diagrams~\cite{QDTzee2010quantum}, the physical content is different. In quantum field theory, diagrams encode contributions to probability amplitudes, whereas here they represent contributions to a coarse-grained generator of dynamics. The associated denominators, time orderings, and combinatorial factors therefore reflect cumulative detunings, Dyson ordering, and operator selection rules, rather than scattering amplitudes, despite the shared perturbative origin. \\ \\
Finally, a very recent work~\cite{WMIhuang2026theory} developed a degenerate Floquet~\cite{Floquetguerin2003control} perturbation theory for periodically driven quantum systems, deriving effective Hamiltonians and introducing a diagrammatic representation. Despite this conceptual overlap, our approach differs fundamentally from~\cite{WMIhuang2026theory} in several respects. The framework of~\cite{WMIhuang2026theory} is state-centric, taking quantum states and quasi-energies as its primary objects. Moreover, it treats the electromagnetic field as a classical periodic drive rather than a quantized mode, making extensions to regimes where field quantization is essential – such as few-photon cavity or waveguide QED settings, paramount to photonic quantum information, – nontrivial. By contrast, our transition-centric formalism applies uniformly to both discrete and continuum spectra within a fully quantized framework in which matter and light are handled jointly.
\section{Conclusion}
In this work, we have developed a fully quantized, transition-operator-oriented perturbation theory that treats light and matter on equal footing. The framework recasts perturbation theory at the level of JLM transition operators, yielding effective Hamiltonians directly in terms of the remaining slow dynamical processes. By retaining the frequency-resolved structure of light–matter interactions, our formalism captures higher-order processes with explicit dependence on all participating frequencies, thereby extending standard dispersive treatments beyond few-mode settings. Unlike previous methods, our approach accommodates continua of field modes, as encountered in waveguide QED. At order $n$, the method generates $(n+1)$ transition operators whose contributions can be organized diagrammatically, providing direct insight into the underlying physical pathways. Although developed for light–matter systems, the formalism applies more generally to spin-boson models~\cite{leggett1987dynamics,ripoll2022quantum} with similar interaction structures, provided the relevant operators form eigenoperators of the free Liouvillian.\\ 
The main technical contributions of this work are fourfold: (i) a perturbative expansion built on transition operators, making the frequency structure of the interaction immediate and avoiding the non-unique frame transformations required by state-centric approaches; (ii) a transition-level formulation of adiabatic elimination that bridges projection-based and time-averaging perturbative viewpoints, (iii) a framework that requires no truncation of the underlying Hilbert space, preserving virtual intermediate transitions at every perturbative order; and (iv) a diagrammatic calculus that organizes operator ordering, selection rules, and detuning denominators in a transparent and systematic way to arbitrary order. These features make the approach particularly well suited to multimode and continuum problems, as well as to situations where counter-rotating and higher-order effects are relevant. A limitation of the diagrammatic approach is the growth in combinatorial complexity at high perturbative order. Nevertheless, we have shown that constraints arising from diagram topology and transition-operator algebra significantly reduce the number of relevant contributions in practice. \\ 
More broadly, the JLM-diagram framework connects ideas from diagrammatic many-body theory, nonlinear spectroscopy, and adiabatic-elimination techniques in quantum optics. It provides a unified, operator-level language for analyzing light–matter interactions and offers a practical route toward engineering effective high-order Hamiltonians for quantum technologies~\cite{bazavan2403squeezing}. \\
Extensions of this work include incorporating both intrinsic dissipation and effective losses induced by the elimination of fast transition sectors, as well as generalizing the formalism to explicitly time-dependent interactions, for instance incorporating insights from~\cite{WMIhuang2026theory} to our framework using Floquet theory~\cite{Floquetguerin2003control}.
\section*{ACKNOWLEDGMENT}
M. Meguebel acknowledges support from the Program QuanTEdu-France n° ANR-22-CMAS-0001 France 2030. L. Garbe acknowledges funding from the Munich Quantum Valley, which is supported by the Bavarian state government with funds from the Hightech Agenda Bayern Plus.

\newpage
\onecolumngrid
\appendix

\section{Transition state-dressed state correspondence} 
\label{Appendix Transition state-dressed state correspondence}
In Sec.~\ref{Subsubsection Time-dependent perturbation theory in the Heisenberg-Dirac picture}, we argued that the detuning of a JLM transition operator $\hat \xi_{i\boldsymbol \sigma}^j(\omega)$, as defined by the free Liouvillian, is the fundamental quantity characterizing the corresponding light-matter transition, in analogy with energy for states. We want to go a step further by introducing the notion of \textit{transition state}, motivated by our desire to describe the light-matter system primarily in terms of relative states (\textit{i.e.}, transitions) rather than bare states. \\
Consider a general light-matter multimode dressed state reading 
\begin{equation}
\label{Equation multimode dressed state}
    \ket{\psi} = \sum_{i=1}^N \sum_{n=0}^{+\infty}\sumint_{\{\boldsymbol \sigma,\omega\}_n}\alpha_{i,\{\boldsymbol \sigma, \omega\}_n}\ket{i}\otimes \ket{\{\boldsymbol \sigma,\omega\}_n},
\end{equation}
where $\ket{\{\boldsymbol \sigma, \omega\}_n}$ denotes the $n$-photon state with the multipartite configuration $\{\boldsymbol \sigma,\omega\}_n$, with the $k$th-photon having the auxiliaray degrees of freedom $\boldsymbol \sigma_k$ and frequency $\omega_k$: $\ket{\{\boldsymbol \sigma,\omega\}_n} = \otimes_{k=1}^n \ket{\boldsymbol \sigma_k, \omega_k}$. Together, the JLM states $\ket{i}\otimes \ket{\{\boldsymbol \sigma,\omega\}_n}$ encode all the information about the light-matter vector state $\ket{\psi}$. \\
The density matrix $\hat \rho = \ket{\psi}\bra{\psi}$ associated with the vector state $\ket{\psi}$ thus reads 
\begin{equation}
\label{Equation multimode dressed state density matrix}
    \hat \rho = \sum_{i,j=1}^N\sum_{n,m=0}^{+\infty}\sumint_{\substack{\{\boldsymbol \sigma,\omega\}_n \\ \{\boldsymbol \sigma',\omega'\}_m}}\alpha_{j,\{\boldsymbol \sigma', \omega'\}_m}\alpha^*_{i,\{\boldsymbol \sigma, \omega\}_n}\ket{j}\bra{i}\otimes \ket{\{\boldsymbol \sigma',\omega'\}_m}\bra{\{\boldsymbol \sigma,\omega\}_n} .
\end{equation}
Defining the general JLM transition operator $\hat \xi_{i \{\boldsymbol \sigma,\omega\}_n}^{j\{\boldsymbol \sigma',\omega'\}_m}$ to be the transition operator translating the matter transition $\ket{i}\rightarrow \ket{j}$ by means of annihilation of the photonic configuration $\{\boldsymbol \sigma,\omega\}_n$ and the creation of the $\{\boldsymbol \sigma',\omega'\}_m$ photonic configuration -- \textit{i.e.} $\hat \xi_{i \{\boldsymbol \sigma,\omega\}_n}^{j\{\boldsymbol \sigma',\omega'\}_m} = \ket{\{\boldsymbol \sigma',\omega'\}_m}\bra{\{\boldsymbol \sigma,\omega\}_n} $ -- the density matrix can be expressed as 
\begin{equation}
\label{Equation multimode dressed state density matrix with JLM transition operators}
 \hat \rho = \sum_{i,j=1}^N\sum_{n,m=0}^{+\infty}\sumint_{\substack{\{\boldsymbol \sigma,\omega\}_n \\ \{\boldsymbol \sigma',\omega'\}_m}}\alpha_{j,\{\boldsymbol \sigma', \omega'\}_m}\alpha^*_{i,\{\boldsymbol \sigma, \omega\}_n}\hat \xi_{i \{\boldsymbol \sigma,\omega\}_n}^{j\{\boldsymbol \sigma',\omega'\}_m}.
\end{equation}
The $\hat \xi_{i \{\boldsymbol \sigma,\omega\}_n}^{j\{\boldsymbol \sigma',\omega'\}_m}$ operators connect to the $\hat \xi_{i\boldsymbol \sigma}^{j}(\omega)$ operators: $ \hat \xi_{i\boldsymbol \sigma}^{j}(\omega) = \sum_{n_{\boldsymbol \sigma}(\omega) \geq 1}\sqrt{n_{\boldsymbol \sigma}(\omega)}\hat \xi_{i\{n_{\boldsymbol \sigma}(\omega)-1\}}^{j\{n_{\boldsymbol{\sigma}}(\omega)\}}$, where $\hat \xi_{i\{n_{\boldsymbol \sigma}(\omega)-1\}}^{j\{n_{\boldsymbol \sigma}(\omega)\}}$ is a shorthand notation for the operator driving the $\ket{i}\rightarrow \ket{j}$ transition by destroying configurations where there are at least one photon in mode $(\boldsymbol \sigma,\omega)$ and creating a configuration $\{n_{\boldsymbol \sigma}(\omega)-1\}$ identical to $\{n_{\boldsymbol{\sigma}}(\omega)\}$ but where the mode $(\boldsymbol \sigma,\omega)$ has been deprived of one photon. \\
According to Eq.~\eqref{Equation multimode dressed state density matrix with JLM transition operators}, the JLM transition operators can hence be seen as the basis elements of the light-matter Liouville space of density matrices. Accordingly, the $\hat \xi_{i \{\boldsymbol \sigma,\omega\}_n}^{j\{\boldsymbol \sigma',\omega'\}_m}$ can be referred to as (multimode) \textit{transition states}. In a sense, transition states are the Liouville-space equivalent to the JLM states $\ket{i}\otimes \ket{\{\boldsymbol \sigma,\omega\}_n}$ in the vector state Hilbert space. Namely, the diagonal elements, \textit{i.e.} the populations, correspond to the weight of the transition states $\hat \xi_{i \{\boldsymbol \sigma,\omega\}_n}^{i\{\boldsymbol \sigma,\omega\}_n}$ in Eq.~\eqref{Equation multimode dressed state density matrix with JLM transition operators}. The off-diagonal elements of $\hat \rho$ on the other hand, \textit{i.e.} the coherences, match all the other transition states. This definition translates the idea that light-matter density matrices can be thought of as a collection of light-matter transitions. In that regard, $\hat \rho$ like $\ket{\psi}$ Eq.~\eqref{Equation multimode dressed state} can be labelled as a (multimode) \textit{dressed transition state} in that it is expressed as a superposition of different transition states. The present JLM operator-based perturbation theory thus places transition states at the forefront. 
\section{Joint light-matter transition operators' weights calculation}
In this section, we derive the JLM diagrams' weights at arbitrary order $n$, and more specifically their time-dependence $v_n(t)$ (putting aside the coupling terms).
\subsection{Computation with nested convolution products}
\label{Appendix Joint light-matter transition operators' weights calculation}
We first carry out the calculation by means of nested convolution products arising from the perturbative expansion. Recall the perturbative expansion Eq.~\eqref{Equation perturbation theory expansion in Heisenberg-interaction picture}
\begin{align}
\label{Appendix Equation perturbation theory expansion in Heisenberg-interaction picture}
    \hat \xi_{i\boldsymbol\sigma}^{j(0)}(\omega,t) &= e^{-i\mathcal L_{\text{free}}(t-t_0)}\hat \xi_{i\boldsymbol\sigma}^{j}(\omega) \\ 
    \hat \xi_{i\boldsymbol\sigma}^{j(1)}(\omega,t) &= -i\int_{t_0}^td\tau\; e^{-i\mathcal L_{\text{free}}(t-\tau)}\mathcal L_{\text{int}}e^{-i\mathcal L_{\text{free}}(\tau-t_0)}\hat \xi_{i\boldsymbol\sigma}^{j}(\omega) \\ 
    \hat \xi_{i\boldsymbol\sigma}^{j(2)}(\omega,t) &= (-i)^2\int_{t_0}^t\int_{t_0}^\tau d\tau d\tau'\; e^{-i\mathcal L_{\text{free}}(t-\tau)}\mathcal L_{\text{int}}e^{-i\mathcal L_{\text{free}}(\tau-\tau')}\mathcal L_{\text{int}}e^{-i\mathcal L_{\text{free}}(\tau'-t_0)}\hat \xi_{i\boldsymbol\sigma}^{j}(\omega) \\ 
    &\;\;\vdots \notag.
\end{align}
A $n$th-order JLM diagram thus have a weight $w_n(t)$ in the $n$th-order correction $\Delta \hat H_{\text{int}}^{(n)}(t)$ to the interaction Hamiltonian. $w_n(t)$ that can be expressed as 
\begin{equation}
    w_n(t) = (-i)^n\int_{t_0}^{t}d\tau_{n} \dots \int_{t_0}^{\tau_2}d\tau_1 \sumint_{\boldsymbol \sigma_n}\sumint_{\omega_n}g(\boldsymbol \sigma_n, \omega_n)\dots \sumint_{\boldsymbol \sigma_0}\sumint_{\omega_0}g(\boldsymbol \sigma_0, \omega_0)e^{-i\Delta_n(t-\tau_{n})}\dots e^{-i\Delta_0(\tau_1-t_0)},
\end{equation}
where we have used the fact that the JLM transition operators are eigenoperators of the free-Hamiltonian commutator free Liouvillian. $g(\boldsymbol \sigma_i,\omega_i)$ and $\Delta_i$ denote, respectively, the coupling term and the detuning involved in the JLM diagram pathway at $i$th order. The detunings are cumulative, \textit{i.e.} $\Delta_i = \sum_{l=0}^i \delta_l$ where $\delta_l$ is detuning of JLM transition order driving the $l$th order perturbation. The $i$th order detuning then depends on the detunings at preceding orders, that is to say on the $\omega_{j\leq i}$ frequencies contained in the $\sumint$s. For full rigor, we attach to the individual detunings $\delta_l$ a regularization term $-i\theta_l$ with $\theta_l > 0$ to ensure the integrability of the terms involved in the expansion when considering long-time regimes. At the very end of the computation, we send $\theta_l \rightarrow 0$ so that no extra \textit{ad hoc} physical terms are added. This regularization is needed when dealing with continua, for instance as in waveguide QED. \\
$w_n(t)$ then reads
\begin{equation}
     w_n(t) = (-i)^n\int_{t_0}^{t}d\tau_{n} \dots \int_{t_0}^{\tau_2}d\tau_1 \sumint_{\boldsymbol \sigma_n}\sumint_{\omega_n}g(\boldsymbol \sigma_n, \omega_n)\dots \sumint_{\boldsymbol \sigma_0}\sumint_{\omega_0}g(\boldsymbol \sigma_0, \omega_0)e^{-i(\Delta_n-i\Theta_n)(t-\tau_{n})}\dots e^{-i(\Delta_0-i\Theta_0)(\tau_1-t_0)},
\end{equation}
where similarly to the cumulative detunings, the $\Theta_i$s denote the cumulative regularization terms. We further assume that the coupling functions are integrable, for instance requiring them to be compactly supported. With the latter and the extra regularization terms, one may reorganize $w_n(t)$ by applying Fubini-Lebesgue theorem~\cite{Mathematicalappel2002mathematiques}.
\begin{equation}
    w_n(t) = \sumint_{\boldsymbol \sigma_n}\sumint_{\omega_n}g(\boldsymbol \sigma_n, \omega_n)\dots \sumint_{\boldsymbol \sigma_0}\sumint_{\omega_0}g(\boldsymbol \sigma_0, \omega_0) v_n(t),
\end{equation}
with the weight time-dependence $v_n(t)$ 
\begin{equation}
\label{Appendix Equation general expression of the loop time-dependence}
    v_n(t) = (-i)^n\int_{0}^t d\tau_{n}\dots \int_{0}^{\tau_2}d\tau_1\; e^{-i(\Delta_n-i\Theta_n)(t-\tau_{n})}\dots e^{-i(\Delta_0-i\Theta_0)\tau_1},
\end{equation}
where we have set $t_0=0$ for simplicity and without loss of generality. For example, at second order 
\begin{equation}
    v_2(t) = (-i)^2\int_0^t d\tau_2 \int_0^{\tau_2} d\tau_1 e^{-i(\Delta_2-i\Theta_2)(t-\tau_2)}e^{ -i(\Delta_1-i\Theta_1)(\tau_2-\tau_1)}e^{-i(\Delta_0-i\Theta_0)\tau_1}.
\end{equation}
Note that, although not explicitly indicated, $v_n(t)$ depends on the $\boldsymbol \sigma_i$s and $\omega_i$s and is thus positioned within the product of $\sumint$s. Eq.~\eqref{Appendix Equation general expression of the loop time-dependence} is a nested integral with complex exponentials. It can be expressed as successive convolution products $\left(f_n*(f_{n-1}*(f_{n-2}*\dots\right)(t)$ with $f_n : \tau \mapsto e^{-i(\Delta_n-i\Theta_n)\tau}$ which can be dealt with using a Laplace transform $\mathcal L_a$. At second order for instance, this Laplace transform can be cast as
\begin{equation}
     F_n(s) = \mathcal L_a[v_n(t)]= (-i)^2\frac{1}{s+i\Delta_2+\Theta_2}\frac{1}{s+i\Delta_1+\Theta_1}\frac{1}{s+i\Delta_0+\Theta_0},
\end{equation}
and more generally at the $n$th order 
\begin{equation}
   F_n(s)=(-i)^n\prod_{k=0}^n\frac{1}{s+i\Delta_k+\Theta_k}.
\end{equation}
We now wish to express $F(s)$ as a sum to separate the contributing orders. We do so by performing a partial fraction decomposition 
\begin{equation}
\label{Equation Laplace transform partial fraction decomposition}
    F_n(s) = (-i)^n\prod_{k=0}^n\frac{1}{s+i\Delta_k+\Theta_k} = \sum_{l=1}^r\sum_{p=1}^{m_l}\frac{C_{lp}}{(s+i\Delta_l+\Theta_l)^p},
\end{equation}
where the $(\Delta_l-i\Theta_l)$s are the $r$ distinct regularized cumulative detunings of multiplicity $m_l$. The cumulative regularization terms $\Theta_l$ are equal to $\sum_{i=0}^{l}\theta_l$ where $\theta_l$ are the individual regularization terms attached to the individual detunings. Since the $\theta_l$s are strictly positive, one cannot have $\Theta_l = \Theta_k$ for $k\neq l$. Subsequently, the regularized cumulative detunings all have multiplicity $m_l=1$. Therefore, Eq.~\eqref{Equation Laplace transform partial fraction decomposition} reduces to a single sum 
\begin{equation}
    F_n(s) = \sum_{l=0}^n\frac{C_l}{(s+i\Delta_l+\Theta_l)}, \;\; C_l = \prod_{\substack{k=0 \\ k\neq l}}^n\frac{1}{\Delta_l-\Delta_k+i(\Theta_k-\Theta_l)},
\end{equation}
where the coefficients $C_l$ have been computed using the residue of $F(s)$ about the different regularized cumulative detunings~\cite{MathematicalQkreyszig2021advanced}. With the previous expression of the Laplace transform, one can go back to the time domain using the Laplace transform linearity and standard Laplace-transform calculus.
\begin{equation}
\label{Appendix equation simplified no degeneracies form of the JLM transition operator weight}
   v_n(t)= \sum_{l=0}^n e^{-i\Delta_l t}e^{-\Theta_l t} \prod_{\substack{k=0 \\ k\neq l}}^n\frac{1}{\Delta_l-\Delta_k+i(\Theta_k-\Theta_l)}
\end{equation}
Adding the extra $(-1)^{N_L}$ factors to by virtue of rule (R4) of Sec.~\ref{Section JLM diagram method}, with $N_L$ the number of operators applied on the left, one retrieves Eq.~\eqref{Equation general expression of the time-dependent loop diagram} from the main text. 
\subsection{Computation with the resolvent}
\label{Appendix Computation with the resolvent}
In this subsection, we demonstrate how the time weights $v_n(t)$ in Eq.~\eqref{Appendix equation simplified no degeneracies form of the JLM transition operator weight} may equivalently be retrieved by manipulating the resolvent $\mathcal G(s)$ of the Liouvillian $\mathcal L$ 
\begin{equation}
\label{Appendix Equation general expression of the resolvent of the Liouvillian}
    \mathcal G(s) = \frac{1}{s+i\mathcal L}, \; \Re(s) > 0.
\end{equation}
$\mathcal G(s)$ is obtained by considering the evolution superoperator in Liouville space $\mathcal U(t,t_0) = e^{-i\mathcal L(t-t_0)}$ define as 
\begin{equation}
\label{Appendix Equation time-evolution of the evolution superoperator in Liouville space}
    \frac{d\;\mathcal U(t,t_0)}{dt}=-i\mathcal L\;\mathcal U(t,t_0),\;\; \mathcal U(t_0,t_0) = \mathcal I.
\end{equation}
Considering $t_0=0$ without loss of generality and taking the Laplace transform of Eq.~\eqref{Appendix Equation time-evolution of the evolution superoperator in Liouville space} yields the expression Eq.~\eqref{Appendix Equation general expression of the resolvent of the Liouvillian} of the resolvent $\mathcal G(s)$.\\
As discussed in the main text, we may expand $\mathcal G(s)$ as a Neumann series about $\mathcal L_{\text{int}}\left(s+\mathcal L_{\text{free}}\right)^{-1}$
\begin{equation}
     \mathcal G(s) = \sum_{n=0}^{+\infty}\mathcal G^{(n)}(s),
\end{equation}
where the $n$-th order term $\mathcal G^{(n)}(s)$ reads 
\begin{equation}
    \mathcal G^{(n)}(s)= (-i)^n\left(s+i\mathcal L_{\text{free}}\right)^{-1}\left[\mathcal L_{\text{int}}\left(s+i\mathcal L_{\text{free}}\right)^{-1} \right]^{n}.
\end{equation}
Because all JLM transition operators are eigenoperators of the free Liouvillian $\mathcal L_{\text{free}}$, applying $\mathcal G^{(n)}(s)$ to a zeroth-order JLM transition operator $\hat \xi^{(0)}$ yields for a given interaction path
\begin{equation}
     \mathcal G^{(n)}(s)\hat \xi^{(0)} = \left(\sumint_{\boldsymbol \sigma_n}\sumint_{\omega_n}g(\boldsymbol \sigma_n, \omega_n)\dots \sumint_{\boldsymbol \sigma_1}\sumint_{\omega_1}g(\boldsymbol \sigma_1, \omega_1) F(s)\right)\hat \Xi^{(n)},
\end{equation}
with 
\begin{equation}
     F_n(s)=(-i)^n\prod_{k=0}^n\frac{1}{s+i\Delta_k+\Theta_k},
\end{equation}
where $\Delta_k-i\Theta_k$ is the regularized cumulative detuning after the $k$-th action of the interaction Liouvillian and $\hat \Xi^{(n)}$ is the resulting $n$-th order JLM transition operator upon applying $\mathcal G^{(n)}(s)$ on the zeroth-order JLM transition operator $\hat \xi^{(0)}$ and which depends \textit{a priori} on all the $(\boldsymbol \sigma_i,\omega_i)$s. $F_n(s)$ is therefore again the Laplace transform of the $v_n(t)$ 

As in the previous subsection, a partial fraction decomposition on $F(s)$ can be carried out to separate the contributing orders. Toggling back to the real $t$-space gives the expression of $v_n(t)$ Eq.~\eqref{Appendix equation simplified no degeneracies form of the JLM transition operator weight} previously calculated.
\subsection{Computation of the reverse process' weight}
\label{Appendix Computation of the reverse process' weight}
This section provides a computation of the time weight $v^{\text{reverse}}_n(t)$ of the reverse process of a given $n$-th order JLM transition operator of time weight $v_n(t)$, which reads
\begin{equation}
    \label{Appendix Equation general expression of the time-dependent loop diagram}
    v_n(t) = (-1)^{N_L} \sum_{l=0}^{n} e^{-i\Delta_l t}e^{-\Theta_l t} \prod_{\substack{k=0 \\ k\neq l}}^n\frac{1}{\Delta_l-\Delta_k+i(\Theta_k-\Theta_l)}
\end{equation}
More specifically, we show that $ v^\text{reverse}_n(t) = v_n^*(t)$, entailing that for $P$ possible $n$th-order JLM transition operators emerging from the adiabatic elimination procedure, one only needs to compute $P/2$ as the remaining $P/2$ can be inferred by taking the Hermitian conjugates of the first $P/2$ terms. $v_n^*(t)$ can be immediately computed from Eq.~\eqref{Appendix Equation general expression of the time-dependent loop diagram} as 
\begin{equation}
    v_n^*(t) = (-1)^{N_L}\sum_{l=0}^ne^{i\Delta_l t}e^{-\Theta_l}\prod_{\substack{k=0 \\ k\neq l}}^n \frac{1}{\Delta_l-\Delta_k +i(\Theta_l-\Theta_k)}.
\end{equation}
Factoring out the $(-1)^n$ term from the product gives 
\begin{equation}
\label{Appendix complex conjugate Equation general expression of the time-dependent loop diagram}
    v_n^*(t) =(-1)^{N_L}(-1)^n \sum_{l=0}^ne^{i\Delta_l t}e^{-\Theta_l}\prod_{\substack{k=0 \\ k\neq l}}^n \frac{1}{\Delta_k-\Delta_l +i(\Theta_k-\Theta_l)}.
\end{equation}
Eq.~\eqref{Appendix complex conjugate Equation general expression of the time-dependent loop diagram} corresponds precisely to the time weight $w_n^{\text{reverse}}(t)$ of the reverse process of time weight $v_n(t)$. Indeed, as argued in Sec.~\ref{Section diagram Hermitian conjugate}, for a given $n$th-order JLM diagram, the reverse process can be obtained by taking the Hermitian conjugate of the JLM transition operators by reading the JLM diagram in the opposite direction. The former flips all the cumulative detunings signs which adds an extra $(-1)^n$. \\
When coupling terms are taken into account, it follows that $W^{\text{reverse}}_n(t) = W_n(t)$, that is to say that an $n$th-order JLM transition operator and its Hermitian conjugate reverse process have weights that are complex conjugates of the other.
\section{Computation of the effective Hamiltonian for a frequency continuum at first order: the $\Lambda$ three-level system}
\label{Appendix Computation of the effective Hamiltonian for a frequency continuum at first order}
In this section, we provide computation details of the effective Hamiltonian for a frequency continuum at order $n=1$ of the JLM-transition-operator adiabatic elimination. We tackle the specific example of the $\Lambda$ three-level system discussed in Sec.~\ref{Section Three-level systems in a continuum}. The generalization to other systems is immediate. Consider the unperturbed interaction Hamiltonian 
\begin{equation}
    \hat H_{\text{int}} = \int_{\mathbb R^+}d\omega\;g_{\alpha \boldsymbol \sigma_i}^\beta(\omega)\hat \xi_{\alpha \boldsymbol \sigma_i}^\beta(\omega)+\int_{\mathbb R^+}d\omega\;g_{\beta \boldsymbol \sigma_j}^\gamma(\omega)\hat \xi_{\beta \boldsymbol \sigma_j}^\gamma(\omega) + \text{h.c},
\end{equation}
where we disregarded the counter-rotating events for simplicity. The free Hamiltonian can be cast as 
\begin{equation}
    \hat H_{\text{free}} = \omega_{\beta}\ket{\beta}\bra{\beta} + \omega_{\gamma}\ket{\gamma}\bra{\gamma} + \sum_{k=i,j}\int_{\mathbb R^+} d\omega\; \omega \hat a_{\boldsymbol \sigma_k}^\dagger(\omega)\hat a_{\boldsymbol \sigma_k}(\omega),
\end{equation}
where the identity operators have been dropped for simplicity. As argued in Sec.~\ref{Section First-order loop diagrams} and Sec.~\ref{Section Three-level systems in a continuum}, since the interaction Hamiltonian possesses $M/2 = 2$ JLM transition operators (and their $M/2 = 2$ Hermitian partners), one should draw at most three JLM diagrams to correctly capture the perturbative expansion at first order, see Fig.~\ref{Figure n=1 loop diagram for Lambda system in RWA}. The first-order JLM transition operators have weights of time-dependence that can be computed employing Eq.~\eqref{Equation total time-dependent weight for n=1}, which we recall 
\begin{equation}
\label{Appendix Reminder weight-time-dependence at first order}
     V_1(t) = \left[\frac{e^{-i\delta_it}e^{-\theta_it}}{\delta_j-i\theta_j}- \frac{e^{-i\delta_jt}e^{-i\theta_j t}}{\delta_i-i\theta_i} \right] +e^{-i(\delta_i+\delta_j)t}e^{-(\theta_i+\theta_j)t}\left[\frac{1}{\delta_i-i\theta_i}-\frac{1}{\delta_j-i\theta_j}\right].
\end{equation}
The total weight is simply the product of $\mathcal W_1^{\text{total}}(t)$ with the coupling terms. Let us apply formula Eq.~\eqref{Appendix Reminder weight-time-dependence at first order} for the two classes of JLM diagrams: the mediated coupling and the Stark shifts. 
\subsection{Mediated-coupling correction}
The correction to the interaction Hamiltonian due to the mediated coupling $\ket{\gamma}\bra{\alpha}\otimes \hat a_{\boldsymbol \sigma_j}^\dagger(\omega')\hat a_{\boldsymbol \sigma_i}(\omega)$ reads 
\begin{equation}
    \Delta \hat H_{\text{med.}}^{(1)}(t) = \frac{1}{2}\int_{\mathbb R^+}d\omega\; g_{\alpha}^{\beta}(\omega)\int_{\mathbb R^+}d\omega'\; g_{\beta}^\gamma(\omega') W_1^{\text{total,med.}}(t)\ket{\gamma}\bra{\alpha}\otimes\hat a_{\boldsymbol \sigma_j}^\dagger(\omega')\hat a_{\boldsymbol \sigma_i}(\omega)+\text{h.c},
\end{equation}
where the $1/2$ factor stems from rule (R5)~\ref{Section JLM diagram method}. Using Eq.~\eqref{Appendix Reminder weight-time-dependence at first order}, 
\begin{equation}
\begin{split}
    \Delta \hat H_{\text{med.}}^{(1)}(t) &= \frac{1}{2}\int_{\mathbb R^+}d\omega\; g_{\alpha}^{\beta}(\omega)\int_{\mathbb R^+}d\omega'\; g_{\beta}^\gamma(\omega') \bigg(\left[\frac{e^{-i\delta_i(\omega)t}e^{-\theta_it}}{\delta_j(\omega')-i\theta_j}- \frac{e^{-i\delta_j(\omega')t}e^{-i\theta_j t}}{\delta_i(\omega)-i\theta_i}  \right] \\
    &+e^{-i(\delta_i(\omega)+\delta_j(\omega'))t}e^{-(\theta_i+\theta_j)t}\left[\frac{1}{\delta_i(\omega)-i\theta_i}-\frac{1}{\delta_j(\omega')-i\theta_j}\right] \bigg)\ket{\gamma}\bra{\alpha}\otimes\hat a_{\boldsymbol \sigma_j}^\dagger(\omega')\hat a_{\boldsymbol \sigma_i}(\omega)+\text{h.c},
\end{split}
\end{equation}
where $\delta_i(\omega) = \omega-\omega_\beta$ and $\delta_j(\omega') = \omega_\beta-\omega_\gamma -\omega'$ and $\theta_i$ and $\theta_j$ are the regularization terms associated with the two zeroth-order transitions. Reorganizing the previous expression yields
\begin{equation}
    \begin{split}
        \Delta \hat H_{\text{med.}}^{(1)}(t) &= \frac{1}{2}\ket{\gamma}\bra{\alpha}\otimes\bigg(\int_{\mathbb R^+}d\omega'\;\frac{g_{\beta}^{\gamma}(\omega')}{\delta_j(\omega')-i\theta_j} \hat a_{\boldsymbol \sigma_j}^\dagger(\omega')\int_{\mathbb R^+}d\omega\; g_{\alpha}^{\beta}(\omega)e^{-i\delta_i(\omega)t}e^{-\theta_it} \hat a_{\boldsymbol \sigma_i}(\omega) \\
        &-\int_{\mathbb R^+}d\omega'\;g_{\beta}^{\gamma}(\omega')e^{-i\delta_j(\omega')t}e^{-\theta_j t} \hat a_{\boldsymbol \sigma_j}^\dagger(\omega')\int_{\mathbb R^+}d\omega\; \frac{g_{\alpha}^\beta(\omega)}{\delta_i(\omega)-i\theta_i}\hat a_{\boldsymbol \sigma_i}(\omega) \\
        &+ \int_{\mathbb R^+}d\omega'\;g_{\beta}^\gamma(\omega')e^{-i\delta_j(\omega')t}e^{-\theta_jt}\hat a_{\boldsymbol \sigma_j}^\dagger(\omega')\int_{\mathbb R^+}d\omega\; \frac{g_{\alpha}^\beta(\omega)e^{-i\delta_i(\omega)t}e^{-\theta_it}}{\delta_i(\omega)-i\theta_i}\hat a_{\boldsymbol \sigma_i}(\omega) \\
        &-\int_{\mathbb R^+}d\omega'\; \frac{g_{\beta}^\gamma (\omega')e^{-i\delta_j(\omega')t}e^{-\theta_jt}}{\delta_j(\omega')-i\theta_j}\hat a_{\boldsymbol \sigma_j}^\dagger(\omega')\int_{\mathbb R^+}d\omega\; g_{\alpha}^\beta(\omega)e^{-i\delta_i(\omega)t}e^{-\theta_it}\hat a_{\boldsymbol \sigma_i}(\omega)\bigg)+\text{h.c}.
    \end{split}
\end{equation}
The limits when we take the regularization terms $\theta_i,\theta_j$ to zero can be handled by using the Sokhotski-Plemelj theorem (see~\cite{Generalopenquantumsystembreuer2002theory} page 145, Eq.~(3.202) and~\cite{Mathematicalphysicsblanchard2015mathematical}) where $\mathcal P$ denotes the Cauchy principal value
\begin{equation}
    \begin{split}
         \Delta \hat H_{\text{med.}}^{(1)}(t) &= \frac{1}{2}\ket{\gamma}\bra{\alpha}\otimes\bigg( \left[\mathcal P\int_{\mathbb R^+}d\omega'\;\frac{g_{\beta}^\gamma(\omega')}{\delta_j(\omega')}\hat a_{\boldsymbol \sigma_j}^\dagger(\omega') +i\pi g_{\beta}^\gamma(\omega_\gamma-\omega_\beta)\hat a_{\boldsymbol \sigma_j}^\dagger(\omega_\gamma-\omega_\beta)\right]\int_{\mathbb R^+}d\omega\; g_{\alpha}^\beta(\omega)e^{-i\delta_i(\omega)t}\hat a_{\boldsymbol \sigma_i}(\omega) \\
         &- \int_{\mathbb R^+}d\omega'\;g_{\beta}^{\gamma}(\omega')e^{-i\delta_j(\omega')t} \hat a_{\boldsymbol \sigma_j}^\dagger(\omega')\left[\mathcal P\int_{\mathbb R^+}d\omega\; \frac{g_{\alpha}^\beta(\omega)}{\delta_i(\omega)}\hat a_{\boldsymbol \sigma_i}(\omega)+i\pi g_{\alpha}^\beta(\omega_\beta)\hat a_{\boldsymbol \sigma_i}(\omega_\beta)\right] \\
         &+\int_{\mathbb R^+}d\omega'\;g_{\beta}^\gamma(\omega')e^{-i\delta_j(\omega')t}\hat a_{\boldsymbol \sigma_j}^\dagger(\omega')\left[\mathcal P\int_{\mathbb R^+}\frac{g_{\alpha}^\beta(\omega)e^{-i\delta_i(\omega)t}}{\delta_i(\omega)}\hat a_{\boldsymbol \sigma_i}(\omega)+i\pi g_{\alpha}^\beta(\omega_\beta)\hat a_{\boldsymbol \sigma_i}(\omega_\beta) \right] \\
         &-\left[\mathcal P\int_{\mathbb R^+}d\omega'\;\frac{g_{\beta}^\gamma(\omega')e^{-i\delta_j(\omega')t}}{\delta_j(\omega')}\hat a_{\boldsymbol \sigma_j}^\dagger(\omega') +i\pi g_{\beta}^{\gamma}(\omega_\gamma-\omega_\beta)\hat a_{\boldsymbol \sigma_j}^\dagger(\omega_\gamma-\omega_\beta) \right]\int_{\mathbb R^+}d\omega\; g_{\alpha}^\beta(\omega)e^{-i\delta_i(\omega)t}\hat a_{\boldsymbol \sigma_i}(\omega)\bigg) + \text{h.c},
    \end{split}
\end{equation}
that is 
\begin{equation}
    \begin{split}
          \Delta \hat H_{\text{med.}}^{(1)}(t) &= \frac{1}{2}\ket{\gamma}\bra{\alpha}\otimes\bigg(\mathcal P\int_{\mathbb R^+}d\omega'\;\frac{g_{\beta}^\gamma(\omega')}{\delta_j(\omega')}\hat a_{\boldsymbol \sigma_j}^\dagger(\omega') \int_{\mathbb R^+}d\omega\; g_{\alpha}^\beta(\omega)e^{-i\delta_i(\omega)t}\hat a_{\boldsymbol \sigma_i}(\omega)\\
          &-\int_{\mathbb R^+}d\omega'\;g_{\beta}^{\gamma}(\omega')e^{-i\delta_j(\omega')t} \hat a_{\boldsymbol \sigma_j}^\dagger(\omega')\;\mathcal P\int_{\mathbb R^+}d\omega\; \frac{g_{\alpha}^\beta(\omega)}{\delta_i(\omega)}\hat a_{\boldsymbol \sigma_i}(\omega) \\
          &+\int_{\mathbb R^+}d\omega'\;g_{\beta}^\gamma(\omega')e^{-i\delta_j(\omega')t}\hat a_{\boldsymbol \sigma_j}^\dagger(\omega')\;\mathcal P\int_{\mathbb R^+}\frac{g_{\alpha}^\beta(\omega)e^{-i\delta_i(\omega)t}}{\delta_i(\omega)}\hat a_{\boldsymbol \sigma_i}(\omega) \\
          &-\mathcal P\int_{\mathbb R^+}d\omega'\;\frac{g_{\beta}^\gamma(\omega')e^{-i\delta_j(\omega')t}}{\delta_j(\omega')}\hat a_{\boldsymbol \sigma_j}^\dagger(\omega')\int_{\mathbb R^+}d\omega\; g_{\alpha}^\beta(\omega)e^{-i\delta_i(\omega)t}\hat a_{\boldsymbol \sigma_i}(\omega)\bigg)+\text{h.c},
    \end{split}
\end{equation}
where the terms in $i\pi$ cancel out. The singularities in $\Delta \hat H_{\text{med.}}^{(1)}(t)$ can be mitigated by introducing a frequency cutoff near the resonant frequencies. This cutoff can be derived microscopically by going beyond the dipole approximation, may be implemented \textit{ad hoc}, or naturally determined by the intrinsic cutoff set by the waveguide’s density of states. As discussed in the main text, the frequency continuum coupled to the matter can be regarded as \textit{predominantly} off-resonant quantified by the ratio $R_T = \int_{\mathcal R_T}d\omega\; |g(\omega)|^2/\int_{\mathcal R_T^c}d\omega\; |g(\omega)|^2$ which measures the fraction of the continuum lying near resonance, where $|\delta(\omega)|$ denotes the one-photon detuning at frequency $\omega$. Then, the condition $R_T \ll 1$ ensures that the continuum is mostly off-resonant and the contributions involving only single-photon transition detunings can be safely neglected. 
\subsection{Stark-shift correction}
Employing the same procedure for the Stark-shift contributions lead to the total correction to the Hamiltonian at first order $\Delta \hat H^{(1)}_{\text{int}}(t)$
\begin{equation}
\begin{split}
\Delta \hat H_{\text{int}}^{(1)}(t) 
&= \frac{1}{2}\ket{\gamma}\bra{\alpha}\otimes \mathcal P\left(\int_{(\mathbb{R}^+)^2}d\omega'd\omega \;
g_{\beta}^\gamma(\omega')g_{\alpha}^\beta(\omega)
e^{-i\left(\omega-\omega'-\omega_\gamma\right)t}
\left[\frac{1}{\delta_i(\omega)}-\frac{1}{\delta_j(\omega')}\right]
\hat a_{\boldsymbol \sigma_j}^\dagger(\omega')\hat a_{\boldsymbol \sigma_i}(\omega)\right)+\text{h.c} \\
&+\frac{1}{2}\ket{\alpha}\bra{\alpha}\otimes \mathcal P \left(\int_{(\mathbb{R}^+)^2}d\omega'd\omega \;
\Big[
F_i(\omega,\omega')\hat a_{\boldsymbol \sigma_i}^\dagger(\omega')\hat a_{\boldsymbol \sigma_i}(\omega)
+ F_i^*(\omega',\omega)\hat a_{\boldsymbol \sigma_i}^\dagger(\omega)\hat a_{\boldsymbol \sigma_i}(\omega')
\Big]\right) \\
&-\frac{1}{2}\ket{\gamma}\bra{\gamma} \otimes \mathcal P \left(\int_{(\mathbb{R}^+)^2}d\omega'd\omega \;
\Big[
F_j(\omega,\omega')\hat a_{\boldsymbol \sigma_j}^\dagger(\omega')\hat a_{\boldsymbol \sigma_j}(\omega)
+ F_j^*(\omega',\omega)\hat a_{\boldsymbol \sigma_j}^\dagger(\omega)\hat a_{\boldsymbol \sigma_j}(\omega')
\Big]\right) \\
&-\frac{1}{2}\ket{\beta}\bra{\beta}\otimes \Bigg[
\mathcal P \left(\int_{(\mathbb{R}^+)^2}d\omega'd\omega \;
\Big[
F_i(\omega,\omega')\hat a_{\boldsymbol \sigma_i}^\dagger(\omega')\hat a_{\boldsymbol \sigma_i}(\omega)
+ F_i^*(\omega',\omega)\hat a_{\boldsymbol \sigma_i}^\dagger(\omega)\hat a_{\boldsymbol \sigma_i}(\omega')
\Big]\right) \\
& +
\mathcal P \left(\int_{(\mathbb{R}^+)^2}d\omega'd\omega \;
\Big[
F_j(\omega,\omega')\hat a_{\boldsymbol \sigma_j}^\dagger(\omega')\hat a_{\boldsymbol \sigma_j}(\omega)
+ F_j^*(\omega',\omega)\hat a_{\boldsymbol \sigma_j}^\dagger(\omega)\hat a_{\boldsymbol \sigma_j}(\omega')
\Big]\right)
\Bigg],
\end{split}
\end{equation}
where we introduced the kernels 
\begin{align}
    F_i(\omega,\omega') &= g_{\beta}^\gamma(\omega')g_{\alpha}^\beta(\omega)e^{-i(\omega-\omega')t}\left[\frac{1}{\delta_i(\omega)}+\frac{1}{\delta_i(\omega')} \right] \\
    F_j(\omega,\omega') &= g_{\beta}^{\gamma}(\omega')g_{\gamma}^{\beta}(\omega)e^{-i(\omega-\omega')t}\left[\frac{1}{\delta_j(\omega)}+\frac{1}{\delta_j(\omega')} \right]
\end{align}
in case $[\hat a_{\boldsymbol \sigma_k}(\omega),\hat a_{\boldsymbol \sigma_k}^\dagger(\omega')] \neq 0$, for $k=i,j$, in which case $\hat a^\dagger(\omega')\hat a(\omega)$ is not Hermitian and its Hermitian conjugate should hence be added. This implies that when matter is coupled to a frequency continuum, the Stark-shift contribution splits into two parts. Photons at the same frequency give rise to an energy renormalization, while photons at different frequencies generate genuine interaction terms.

\section{Permissible diagrams}
\label{Appendix Permissible diagrams}
\subsection{Combinatorial upper bound}
\label{Appendix subsection Combinatorial upper bound}
This section is dedicated to evaluating the number of JLM diagrams needed to encapsulate the whole system's dynamics. Consider the interaction Hamiltonian Eq.~\eqref{Equation general interaction Hamiltonian} in a more compact form Eq.~\eqref{Equation compact general interaction Hamiltonian} with a number $M$ of JLM transition operators $\hat H_{\text{int}} = \sum_{m=1}^{M/2}\hat \xi_m + \text{h.c}$ where $\hat \xi_m$ are the zeroth-order JLM transition operators. 
\subsubsection{Multilevel atomic system}
\label{Appendix subsubsection Combinatorial upper bound Multilevel atomic system}
\begin{figure}
    \centering
    \includegraphics[width=0.85\linewidth]{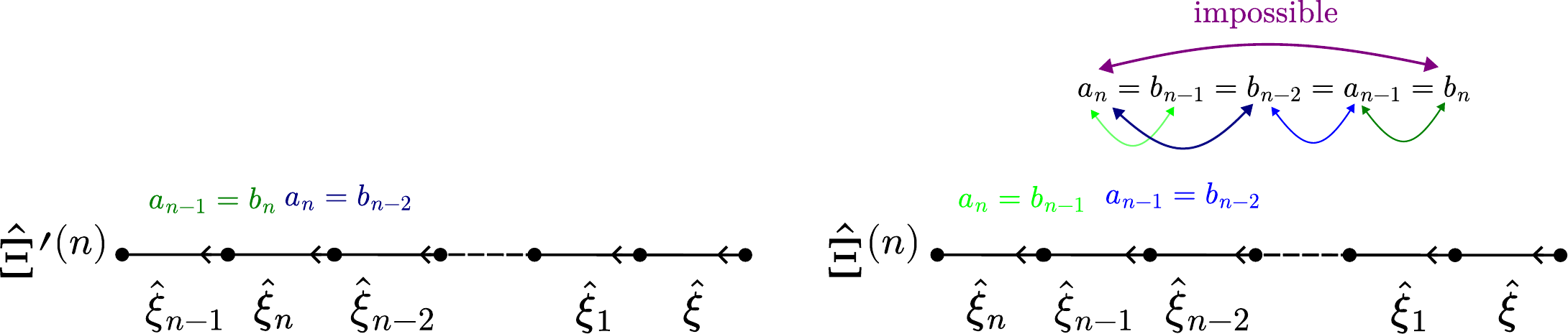}
    \caption{Pictorial representation of the proof of Eq.~\eqref{Appendix formal claim nth order string}. In the $n$-th order JLM transition operator $\hat \Xi'^{(n)}$, the products $\hat \xi_{n-1}\hat \xi_n$ and $\hat \xi_n\hat \xi_{n-2}$ imply, respectively, $a_{n-1}=b_n$ and $a_n=b_{n-2}$. In the $n$-th order JLM transition operator $\hat \Xi^{(n)}$, the products $\hat \xi_n\hat \xi_{n-1}$ and $\hat \xi_{n-1}\hat \xi_{n-2}$ entail, respectively, the additional conditions $a_n=b_{n-1}$ and $a_{n-1}=b_{n-2}$.}
    \label{Figure appendix diagrammatic illustration of atomic-overlap permissible diagrams}
\end{figure}

Let us first consider the situation of a multilevel atomic systems. The $\hat \xi_m$'s have matter parts $\ket{b_m}\bra{a_m}$ with $b_m \neq a_m$ and $\braket{b_m|a_l} \neq 0 \Rightarrow b_m=a_l$ for $l\neq m$ by definition of the JLM transition operators; corresponding to transitions from one atomic state to another. Each $\hat \xi_m$ has a partner operator which is its Hermitian conjugate, so $M$ is an even number. \\ At $n$th order, JLM transition operators are strings of $n+1$ zeroth-order JLM transition operators. Denote $\mathcal S_n$ the set of all $n$th-order JLM transition operators, \textit{i.e.} the set of all $(n+1)$-long strings of zeroth-order JLM transition operators. Let $\hat \Xi^{(n)} = \hat \xi_n\dots \hat \xi_1\hat \xi_0$ be such a $(n+1)$-long string. In principle, there are $M^{n+1}$ such a $(n+1)$-long string. For instance, at $n=1$ order with $M=4$ (such as in the quantum Rabi Hamiltonian Eq.~\eqref{Equation Rabi Hamiltonian}) there would be 16 terms to determine. Fortunately, because of the JLM transition operators algebra, and in particular their matter part, one drastically narrows down the amount of terms that need to be computed. \\ 
Indeed, we claim if $\hat \Xi^{(n)} \neq 0$, then there exists no other $n$-th order JLM transition operator $\hat \Xi'^{(n)}$ composed of operators drawn from the alphabet $\mathcal A_n = \{\hat \xi_n,\dots,\hat \xi_1,\hat \xi_0\}$ that is not equal to zero and does correspond to a cyclic permutation of the terms composing $\hat \Xi^{(n)} = \hat \xi_n\dots \hat \xi_0$. Formally
\begin{equation}
    \label{Appendix formal claim nth order string}
\hat \Xi^{(n)} = \hat \xi_n \dots \hat \xi_0 \neq 0 \Rightarrow  \nexists \; \hat \Xi'^{(n)} \in \mathcal S_n(\mathcal A_n) \left(\hat \Xi'^{(n)} \neq 0\; \wedge \hat \Xi'^{(n)} \notin \mathcal C(\hat \Xi^{(n)})\right),
\end{equation}
where $\mathcal C(\hat \Xi^{(n)})$ denotes the set of cyclic permutations of the \textit{matter parts} of $\hat \Xi^{(n)}$ and $\mathcal S_n(\mathcal A_n)$ denotes elements of $\mathcal S_n$ drawn from the alphabet $\mathcal A_n$. $\mathcal C(\hat \Xi^{(n)})$ acts on matter indices only; bosonic operators attached to those matter indices may differ and can produce distinct, non-cyclic operator-strings. Instead of proving Eq.~\eqref{Appendix formal claim nth order string} as stated, we demonstrate its contrapositive
\begin{equation}
    \label{Appendix contrapositive formal claim nth order string}
\exists \;\hat \Xi'^{(n)} \in \mathcal S_n(\mathcal A_n) \left(\hat \Xi'^{(n)} \neq 0 \wedge \hat \Xi'^{(n)} \notin \mathcal C(\hat \Xi^{(n)})\right) \Rightarrow \hat \Xi^{(n)} = 0.
\end{equation}
Consider such $\hat \Xi'^{(n)} \in \mathcal S_n(\mathcal A_n)\left(\hat \Xi'^{(n)} \neq 0 \wedge \hat \Xi'^{(n)} \notin \mathcal C(\hat \Xi^{(n)})\right)$, for instance $\hat \Xi'^{(n)} = \hat \xi_{n-1}\hat \xi_n\hat \xi_{n-2}\dots \hat \xi_0$. $\hat \Xi'^{(n)} \neq 0$ implies by overlapping of the matter parts of the zeroth-order JLM transition operator that $a_{n-1} = b_n$ and $a_n = b_{n-2}$. Given that $\hat \Xi^{(n)} = \hat \xi_n \dots \hat \xi_0$, one has $\hat \Xi^{(n)}\propto \braket{a_n|b_{n-1}}\braket{a_{n-1}|b_{n-2}}$, that is to say $\hat \Xi^{(n)} \propto \braket{b_{n-2}|b_{n-1}}\braket{a_{n-1}|b_{n-2}}$ because $a_{n-1}=b_n$ and $a_n=b_{n-2}$. $\hat \Xi^{(n)}\neq 0$ requires $b_{n-2}=b_{n-1}=a_{n-1}$. However, by construction of the JLM transition operators, $a_{n-1}\neq b_{n-1}$. Therefore, one necessarily has $\hat \Xi^{(n)}= 0$. This proves the contrapositive Eq.~\eqref{Appendix contrapositive formal claim nth order string}. In Fig.~\ref{Figure appendix diagrammatic illustration of atomic-overlap permissible diagrams}, we pictorially illustrate the matter-overlap constraint used in the demonstration.
Therefore, at the perturbative order $n$, instead of evaluating all $M^{n+1}$ possible operator strings of length $(n+1)$, it suffices to consider the distinc alphabets $\mathcal A_n$, since all nonzero operator strings built from a given $\mathcal A_n$ are related by circular permutations of their matter parts, while all other sequences are excluded by the matter-part constraints. Given that the interaction Hamiltonian $\hat H_{\text{int}} = \sum_{m=1}^{M/2}\hat \xi_m+\text{h.c}$ contains $M$ zeroth-order JLM transition operators, the number of such alphabets with repetition allowed is $\binom{M+n}{n+1}$. Accordingly, from Eq.~\eqref{Appendix formal claim nth order string}, one needs to enumerate \textit{at most} $2(n+1)\binom{M+n}{n+1}$ JLM transition operators, instead of $M^{n+1}$. The factor $(n+1)$ accounts for the cyclic permutations of a given $(n+1)$-long string $\hat \Xi^{(n)}$, while the additional factor 2 accounts for the Hermitian-conjugate contributions. \\
Each $n$th-order JLM transition operator can equivalently be represented as a JLM diagram, whose topology further reduces the computational overhead. First, as discussed in Sec.~\ref{Section diagram Hermitian conjugate}, each non-Hermitian JLM diagram has a partner corresponding to the reverse physical process, whose weight is simply the Hermitian conjugate of the original one. Hence, out of the $2(n+1)\binom{M+n}{n+1}$ possible JLM diagrams, only $(n+1)\binom{M+n}{n+1}$ need to be sketched when Hermitian-conjugate partners are counted separately. Second, due to their diagrammatic structure, JLM diagrams naturally incorporate the $(n+1)$ cyclic permutations of the underlying operator strings. More specifically, a given $n$th-order JLM diagram accounts for $(n+1)$ circularly permuted operator strings when its matter part forms a closed loop, \textit{i.e.} returns to its initial state. In other words, instead of drawing $(n+1)\binom{M+n}{n+1}$ JLM diagrams, \textit{at most} $\binom{M+n}{n+1}$ are required to capture the full $n$th-order perturbative regime. In contrast, with the interaction Hamiltonian $\hat H_{\text{int}}=\sum_{m=1}^{M/2}\hat \xi_m+\text{h.c}$, the James' method~\cite{Jamesmethodjames2007effective,Jamesmethodgamel2010time,Jamesmethodshao2017generalized,Jamesmethodrosado2023comment} requires the computation of \textit{at least} $(M/2)^{n+1}$ terms at $n$th order (see Eq.~$(3.10)$~\cite{Jamesmethodjames2007effective}, Eq.~$(30)$~\cite{Jamesmethodgamel2010time}, Eq.~$(15)$~\cite{Jamesmethodshao2017generalized} and Eq.~$(19)$~\cite{Jamesmethodrosado2023comment}). \\ \\
Notwithstanding this result, the counting above only used the matter parts of the zeroth-order JLM transition operators, not their bosonic components. Bosonic operators of a fixed matter-string produce additional distinct operator-strings only when they involve the same bosonic mode with different creation or annihilation assignments. The reason these bosonic swaps do not proliferate uncontrolledly is the bosonic commutation relations. Indeed, operators acting on different modes commute Eq.~\eqref{Equation bosonic commutation relation},
$[\hat a_{\boldsymbol \sigma}(\omega),\hat a_{\boldsymbol{\sigma}'}^\dagger(\omega')] = \delta(\boldsymbol \sigma - \boldsymbol \sigma')\delta(\omega-\omega')$, so swapping bosonic factors attached to different modes does not change the overall operator and therefore does not produce a new, distinct operator-string. Consequently, many naively distinct permutations of bosonic operators collapse to the same algebraic operator and are absorbed into a single diagram. By contrast, reordering creation and annihilation operators acting on the same bosonic mode cannot, in general, be eliminated by commutation and thus produces genuinely distinct operator strings. Applying the canonical bosonic commutation relations, such reorderings generate additional operator contributions with structures different from the original string. These extra terms correspond to lower-photon processes and must therefore be treated separately in the diagrammatic expansion. This is why only multiplicities involving creation and annihilation operators acting on the same bosonic mode give rise to a multiplicity of operator strings associated with a given matter string, whereas permutations involving different modes do not. In turn, the only multiplicity that must be accounted for per matter string beyond cyclic permutations and Hermitian conjugation arises from the binary choice at positions where the \textit{same} mode admits both $\hat a_{\boldsymbol \sigma}(\omega)$ and $\hat a_{\boldsymbol \sigma}^\dagger(\omega)$. Swaps between different modes are redundant, while same-mode assignments $\hat a_{\boldsymbol \sigma}(\omega) \leftrightarrow \hat a^\dagger_{\boldsymbol \sigma}(\omega)$ constitute the source of multiplicity. For a given matter-string of length $L=n+1$ let $m$ be the number of positions in that string at which the same bosonic mode admits both $\hat a_{\boldsymbol \sigma}(\omega)$ and $\hat a^\dagger_{\boldsymbol \sigma}(\omega)$. Each such position produces a creation-annihilation binary choice, hence the bosonic realizations of that matter-string adds at most $2^{m}$ multiplicity factor to the previous $2(n+1)\binom{M+n}{n+1}$ bound. This bound is intentionally a loose worst-case bound applied for every order $n$ of the perturbative expansion. In practice resonance or RWA conditions reduce the effective number of nonzero and distinct contributions, often by orders of magnitude. \\
In summary, (i) the matter-overlap argument guarantees uniqueness of the matter-string up to cyclic permutation and therefore collapses the naive $M^{n+1}$ count to the multiset count $\binom{M+n}{n+1}$, (ii) bosonic multiplicities associated with the same mode introduce at most a factor $2^{m}$ per matter-string; (iii) Hermitian-conjugate partners and cyclic permutations are handled by the explicit $2(n+1)$ factor shown. Importantly, because different-mode bosonic operators commute, a single diagram with a fixed matter-string can correspond to many permutations of bosonic factors which do not change the operator and these permutations are absorbed into the same diagram, whereas only same-mode $\hat a_{\boldsymbol \sigma}(\omega)$ or $\hat a^\dagger_{\boldsymbol \sigma}(\omega)$ choices yield additional inequivalent operator-strings counted by the $2^{m}$ factor.
\subsubsection{Multiple qubits} 
\label{Appendix subsubsection Combinatorial upper bound Multilevel atomic system}
For an $N$-qubit system, the interaction Hamiltonian Eq.~\eqref{Equation compact general interaction Hamiltonian} reads 
\begin{equation}
\label{Appendix Equation interaction Hamiltonian multiple qubits}
    \hat H_{\text{int}} = \sum_{l=1}^N\left(\hat \xi^{l}_{\text{RWA}}+\hat \xi^{l}_{\text{non-RWA}} + \text{h.c.}\right),
\end{equation}
where $\hat \xi^l_{\text{RWA}}$ and $\hat \xi^l_{\text{non-RWA}}$ correspond, respectively, to co-rotating and counter-rotating transitions $\ket{g_l}\rightarrow \ket{e_l}$, where $\{\ket{g_l},\ket{e_l}\}$ is the $l$-th qubit system. Eq.~\eqref{Appendix Equation interaction Hamiltonian multiple qubits} amounts to the Dicke Hamiltonian Eq.~\eqref{Equation Dicke Hamiltonian} when the associated transitions are driven by a single-mode cavity. Because the different qubtis live in distinct Hilbert spaces, the atomic-overlap argument used for multilevel atomic systems must be adapted with greater care, although it remains applicable. Indeed, an $N$-qubit system may be embedded into $2^N$-dimensional tensor-product basis and therefore viewed as a $2^N$-level atomic system. For example, a two-qubit system with local basis $\{\ket{g_l}, \ket{e_l}\}_{l=1,2}$ can be represented by the four-level basis $\{\ket{i}\}_{i\in \llbracket 2, 4 \rrbracket}$, 
\begin{equation}
\ket{1}=\ket{g_1,g_2}, \;\; \ket{2} = \ket{g_1,e_2},\;\; \ket{3}=\ket{e_1,g_2},\;\; \ket{4} = \ket{e_1,e_2}.
\end{equation}
In this representation, the overlap condition $\braket{i|j} = \delta_{ij}$ is recovered, so the reasoning developed for multilevel atomic systems can be applied directly. More generally, each local qubit transition operator such as $\ket{g_l}\bra{e_l}$, lifts to $\ket{g_l}\bra{e_l}\otimes \mathbb 1_{\neq l}$, which decomposes into $2^{N-1}$ elementary transitions in the full $2^N$-dimensional basis. Consequently, the interaction Hamiltonian Eq.~\eqref{Appendix Equation interaction Hamiltonian multiple qubits}, which contains $4N$ transition operators in the $N$-qubit description, decomposes into $2N\times2^N$ elementary transition operators in the $2^N$-level atomic representation. Accordingly, the previous diagram upper bound $\binom{M+n}{n+1}$ becomes $\binom{2N\times2^N + n}{n+1}$ for an $N$-qubit system. Fortunately, as shown in Sec.~\ref{Subsection Multiple qubits coupled to a single oscillator}, additional overlap constraints and symmetries drastically reduce the number of distinct diagrams required in practice. 
\subsection{JLM diagrams multiplicities}
In this subsection, we illustrate, for a multilevel atomic system, the matter-cyclic condition and the role of bosonic-operator permutations for first- and second-order JLM diagrams. For simplicity and without loss of generality, the bosonic modes are defined for a discrete set, not a continuum.
\subsubsection{First order}
\begin{figure}
    \centering
    \includegraphics[width=0.6\linewidth]{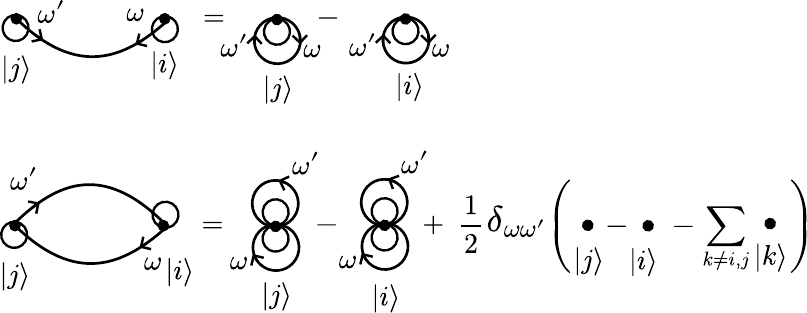}
    \caption{Equivalence of JLM diagrams at first order when there are multiplicities due to their reading orientation.}
    \label{Appendix figure JLM diagrams multiplicities first order}
\end{figure}

Consider a first-order JLM transition operator written $\hat \Xi^{(1)} = \hat \xi_j \hat \xi_i$ as a $2$-long string constructed from the zeroth-order JLM transition operators $\hat \xi_i$ and $\hat \xi_j$ of respective matter operator $\ket{b_i}\bra{a_i}$ and $\ket{b_j}\bra{a_j}$, with the matter-overlap condition $b_i=a_j$. The circular permutation $\hat \xi_i \hat \xi_j$ is non-zero provided the additional condition $b_j = a_i$. In other words, a given JLM diagram contributes two first-order JLM transition operators (and two additional ones from the Hermitian conjugate) provided that the matter state returns to itself, that is $\ket{a_i} = \ket{b_j} \equiv \ket{i}$, and because of the matter-overlap condition $\ket{b_i} = \ket{a_j} \equiv \ket{j}$. Moreover, there are no multiple same-mode positions because $\hat \xi_i \neq \hat \xi_j$. Two strings $\hat \xi_i \hat \xi_j$ and $\hat \xi_j \hat \xi_i$ both non-zero, that is with $b_i=a_j$ and $b_j=a_i$, have according to the expression of $V_1(t)$ Eq.~\eqref{Equation total time-dependent weight for n=1} exactly opposite weight $W_1(t)$. \\ Because of that, the two possible permutations may be grouped together into a single term. Consider first the first-order JLM transition operator $\hat \Xi^{(1)} = \hat \xi_j\hat \xi_i$ and $ \mathcal C\hat\Xi^{(1)}$ the operator obtained by cyclic permutations of $\hat \xi_i$ and $\hat \xi_j$, where $\hat \xi_j = \ket{i}\bra{j}\otimes \hat a_{\boldsymbol \sigma'}(\omega')$ and $\hat \xi_i = \ket{j}\bra{i}\otimes \hat a_{\boldsymbol \sigma}(\omega)$. Define the operator $\hat \sigma_z^{ij} = \ket{j}\bra{j}-\ket{i}\bra{i}$, which amounts to a Pauli-Z matrix in the $\{\ket{i},\ket{j}\}$ subspace. Using the identity operator on the Hilbert space of matter $\mathbb{1}_{\alpha} = \sum_{k=1}^{N}\ket{k}\bra{k}$, 
\begin{align}
    \hat \Xi^{(1)} &= \left(\frac{\mathbb {1}_\alpha-\hat \sigma_z^{ij}}{2}-\frac{1}{2}\sum_{k\neq i,j}\ket{k}\bra{k}\right)\otimes \hat a_{\boldsymbol \sigma'}(\omega')\hat a_{\boldsymbol \sigma}(\omega) \\
    \mathcal C \hat \Xi^{(1)} &= \left(\frac{\mathbb {1}_\alpha+\hat \sigma_z^{ij}}{2}-\frac{1}{2}\sum_{k\neq i,j}\ket{k}\bra{k}\right)\otimes\hat a_{\boldsymbol \sigma'}(\omega')\hat a_{\boldsymbol \sigma}(\omega),
\end{align}
where we have invoked $[\hat a_{\boldsymbol \sigma'}(\omega'),\hat a_{\boldsymbol \sigma}(\omega)]=0$. If $\mathcal C \hat \Xi^{(1)}$ has weight $W_1(t)$, then $\hat \Xi_1$ and has the opposite weight $-W_1(t)$, leading to the first-order correction $\Delta \hat H_{\text{int}}^{(1)}(t)$ to the interaction Hamiltonian 
\begin{equation}
    \Delta \hat H_{\text{int}}^{(1)}(t) = W_1(t)\;\hat \sigma_z^{ij}\otimes \hat a_{\boldsymbol \sigma'}(\omega')\hat a_{\boldsymbol \sigma}(\omega)+\text{h.c},
\end{equation}
where we added the Hermitian conjugate because $\hat \Xi^{(1)}$ and $\mathcal C\hat \Xi^{(1)}$ are not Hermitian. \\
Take now the case where $\hat \xi_j = \ket{i}\bra{j}\otimes \hat a_{\boldsymbol \sigma'}^\dagger(\omega')$ and $\hat \xi_i = \ket{j}\bra{i}\otimes \hat a_{\boldsymbol \sigma}(\omega)$. $\hat \Xi^{(1)}$ and $\mathcal C \hat \Xi^{(1)}$ read 
\begin{align}
    \hat \Xi^{(1)} &= \left(\frac{\mathbb {1}_\alpha-\hat \sigma_z^{ij}}{2}-\frac{1}{2}\sum_{k\neq i,j}\ket{k}\bra{k} \right)\otimes \hat a_{\boldsymbol \sigma'}^\dagger(\omega') \hat a_{\boldsymbol \sigma}(\omega) \\
    \mathcal C \hat \Xi^{(1)} &= \left(\frac{\mathbb {1}_\alpha+\hat \sigma_z^{ij}}{2}-\frac{1}{2}\sum_{k\neq i,j}\ket{k}\bra{k}\right)\otimes\left(\hat a_{\boldsymbol \sigma'}^\dagger(\omega') \hat a_{\boldsymbol \sigma}(\omega)+\delta_{\boldsymbol \sigma \boldsymbol \sigma'}\delta_{\omega\omega'}\right),
\end{align}
where we used the canonical commutation relation $[\hat a_{\boldsymbol \sigma}(\omega),\hat a_{\boldsymbol{\sigma}'}^\dagger(\omega')] = \delta_{\boldsymbol \sigma \boldsymbol \sigma'}\delta_{\omega\omega'}$ for a discrete set of bosonic modes. Again, if $\mathcal C\hat \Xi_1$ has weight $W_1(t)$, then the first-order correction $\Delta\hat H_{\text{int}}^{(1)}(t)$ reads 
\begin{equation}
\begin{split}
    \Delta \hat H_{\text{int}}^{(1)}(t) &= W_1(t)\left(\hat \sigma_z^{ij}\otimes \hat a_{\boldsymbol \sigma'}^\dagger(\omega') \hat a_{\boldsymbol \sigma}(\omega)+ \delta_{\boldsymbol \sigma \boldsymbol \sigma'}\delta_{\omega\omega'}\left(\frac{\hat \sigma_z^{ij}}{2}-\frac{1}{2}\sum_{k\neq i,j}\ket{k}\bra{k}\right)\right)+\left(1-\delta_{\boldsymbol \sigma \boldsymbol \sigma'}\delta_{\omega\omega'}\right)\text{h.c},
\end{split}
\end{equation}
where we discarded the term global energy shift carried by $\mathbb 1_{\alpha}$ and where the Hermitian conjugate should be added if $\hat \Xi^{(1)}$ and $\mathcal C\hat \Xi^{(1)}$ are not Hermitian, that is whenever $[\hat a_{\boldsymbol \sigma}(\omega),\hat a_{\boldsymbol{\sigma}'}^\dagger(\omega')]= 0$. The additional vacuum-induced shift induced by the commutation relation contribution $\delta_{\boldsymbol \sigma \boldsymbol \sigma'}\delta_{\omega\omega'}$ acts both on the subspace $\{\ket{i},\ket{j}\}$ involved in the light-matter interaction and on the the non-interacting matter states $\{\ket{k}\}_{k\neq i,j}$. The results from this section may be represented with JLM diagrams, see Fig.~\ref{Appendix figure JLM diagrams multiplicities first order}.
\subsubsection{Second order}
\begin{figure}
    \centering
    \includegraphics[width=0.8\linewidth]{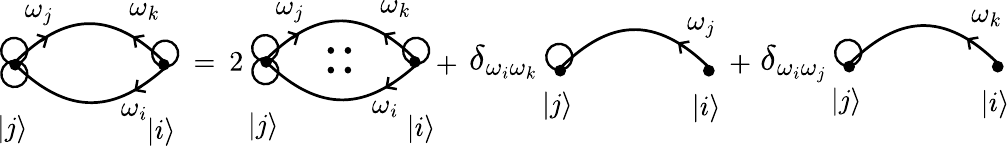}
    \caption{Equivalence of JLM diagrams at second order when there are multiplicities due to their reading orientation.}
    \label{Appendix figure JLM diagrams multiplicities second order}
\end{figure}

Consider a second-order JLM transition operator written $\hat \Xi^{(2)} = \hat \xi_k \hat \xi_j \hat \xi_i$ as $3$-long string composed of the zeroth-order JLM transition operators $\hat \xi_i$, $\hat \xi_j$ and $\hat \xi_j$ of respective matter operator $\ket{b_l}\bra{a_l}$, with $l=i,j,k$ with the overlap condition $b_i = a_j$ and $b_j=a_k$. Consider a first circular permutation $\hat \xi_i\hat \xi_k \hat \xi_j$. For it to be non-zero, one additionally requires $a_i=b_k$. Next, for the second circular permutation $\hat \xi_j\hat \xi_i \hat \xi_k$ to be non-zero, the matter-overlap constraint imposes again the new condition $a_i=b_k$. Therefore, circular permutations of the string $\hat \Xi^{(2)} = \hat \xi_k \hat \xi_j \hat \xi_i$ are non-zero when $a_i=b_k$, that is to say that, the matter state returns to itself, as for the first-order case. \\
Let us now examine how the placement of bosonic operators encodes different transitions and leads to distinct simplifications arising from the bosonic commutation relations. For $\hat \Xi^{(2)}$ to be non-zero, $\hat \xi_j \neq \hat \xi_i$ and $\hat \xi_j \neq \hat \xi_k$, but one may have the same matter parts for $\hat \xi_i$ and $\hat \xi_k$, that is $a_i = a_k$ and $b_i = b_k$, that is to say no matter part circular permutation but a possible multiple same-mode positions. Consider for instance the second-order JLM transition operator $\hat \Xi^{(2)}=\hat \xi_k\hat \xi_j\hat \xi_i$ and $\mathcal S_{ik}\hat \Xi^{(2)}$ the operator obtained by swapping the bosonic operators of $\hat \xi_i$ and $\hat \xi_k$, where $\hat \xi_k = \ket{j}\bra{i}\otimes \hat a_{\boldsymbol \sigma_k}^\dagger(\omega_k)$, $\hat \xi_i=\ket{j}\bra{i}\otimes \hat a_{\boldsymbol \sigma_i}(\omega_i)$. $\hat \Xi^{(2)}$ and $\mathcal S_{ik}\hat \Xi^{(2)}$ are represented by the same JLM diagram, see Fig.~\ref{Appendix figure JLM diagrams multiplicities second order}. Using the commutation relations 
\begin{align}
    \hat \Xi^{(2)} &= \ket{j}\bra{i}\otimes\hat a_{\boldsymbol \sigma_k}^\dagger(\omega_k)\hat a_{\boldsymbol \sigma_j}^\dagger(\omega_j)\hat a_{\boldsymbol \sigma_i}(\omega_i) \\
    \mathcal S_{ik}\hat \Xi^{(2)} &= \hat \Xi^{(2)}+ \ket{j}\bra{i}\otimes \left(\delta_{\boldsymbol \sigma_i\boldsymbol \sigma_k}\delta_{\omega_i\omega_k}\hat a_{\boldsymbol \sigma_j}^\dagger(\omega_j)+\delta_{\boldsymbol \sigma_i\boldsymbol \sigma_j}\delta_{\omega_i\omega_j}\hat a_{\boldsymbol \sigma_k}^\dagger(\omega_k)\right).
\end{align}
$\mathcal S_{ik}\hat \Xi^{(2)}$ corresponds to swapping the first and third operators in $\hat \Xi^{(2)}$, so $\hat \Xi^{(2)}$ and $\mathcal S_{ik}\hat \Xi^{(2)}$ have the same weight $W_2(t)$ and the second-order correction $\Delta \hat H_{\text{int}}^{(2)}(t)$ associated with the second-order JLM diagram reads 
\begin{equation}
   \Delta \hat H_{\text{int}}^{(2)}(t) = W_2(t)\left(2\ket{j}\bra{i}\hat a_{\boldsymbol \sigma_k}^\dagger(\omega_k)\hat a_{\boldsymbol \sigma_j}^\dagger(\omega_j)\hat a_{\boldsymbol \sigma_i}(\omega_i)+\ket{j}\bra{i}\otimes \left(\delta_{\boldsymbol \sigma_i\boldsymbol \sigma_k}\delta_{\omega_i\omega_k}\hat a_{\boldsymbol \sigma_j}^\dagger(\omega_j)+\delta_{\boldsymbol \sigma_i\boldsymbol \sigma_j}\delta_{\omega_i\omega_j}\hat a_{\boldsymbol \sigma_k}^\dagger(\omega_k)\right)\right),
\end{equation}
that is that vacuum fluctuations contribute to one-photon transitions at second order of the perturbation theory. The results presented in this section admit a diagrammatic representation in terms of JLM diagrams, see Fig.~\ref{Appendix figure JLM diagrams multiplicities second order}. The symbol $::$ denotes normal ordering of bosonic operators~\cite{Quantumopticsjohn2014quantum}, whereby all annihilation operators are arranged to the right of the creation operators.
\subsection{Number of JLM diagrams}
This subsection is dedicated to calculating an upper limit on the number of JLM diagrams to be drawn at first and second order of the perturbation expansion. 
\subsubsection{First order}
\begin{figure}
\label{Appendix Figure number of JLM VS number of James}
\centering
\includegraphics[width=0.4\linewidth]{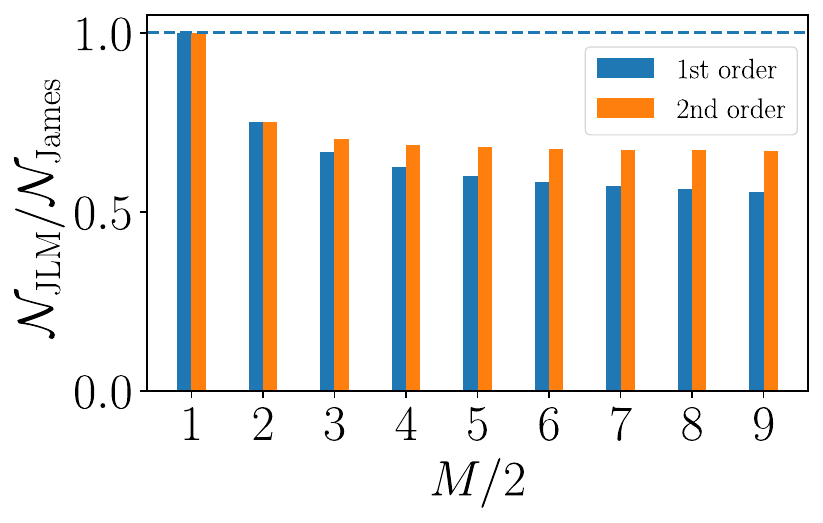}
\caption{Comparison of the maximum number of diagrams, $\frac{1}{2}\binom{M}{2}$ and $\frac{M(M^2+2)}{12}$, required versus the minimum number of terms, $(M/2)^2$ and $(M/2)^3$, that need to computed with the James' method at first (see Eq.~$(3.10)$~\cite{Jamesmethodjames2007effective}, Eq.~$(30)$~\cite{Jamesmethodgamel2010time}) and second order (see Eq.~$(15)$~\cite{Jamesmethodshao2017generalized} and Eq.~$(19)$~\cite{Jamesmethodrosado2023comment}), respectively. $M$ denotes the number of JLM transition operators in the interaction Hamiltonian ($M/2$ plus the Hermitian conjugate partner). In practice, many terms computed from James' method vanish, but all must be evaluated before this can be determined, whereas the JLM-diagram bound already incorporates matter-overlap constraints by the very definition of JLM diagrams.}
\label{Appendix comparison number loop VS number James}
\end{figure}

At first order $n=1$ of the perturbation expansion, the previous section demonstrated that \textit{at most} $\binom{M+1}{2}$ JLM diagrams -- which correspond to \textit{at most} $4\binom{M+1}{2}$ first-order JLM transition operators -- were needed to entirely capture the perturbative expansion of the $M$ zeroth-order JLM transition operators. As discussed in Sec.~\ref{Section First-order loop diagrams}, further analyses actually lead to additional reduction in the number of diagrams that need to be drawn: $\frac{1}{2}\binom{M}{2}$ \textit{at most}, where there are $M/2$ Stark-shift (or Bloch-Siegert if counter-rotating contributions) diagrams and $\binom{M/2}{2}$ mediated-coupling diagrams. We recall that the $1/2$ factor stands for the Hermitian-conjugate partner diagrams. Furthermore, the Stark-shift contributions $\hat{\xi}_i \hat{\xi}_i^\dagger$ and $\hat{\xi}_i^\dagger \hat{\xi}_i$ carry opposite weights. In addition, if the mediated coupling $\hat{\xi}_j \hat{\xi}_i \neq 0$, then it necessarily follows that $\hat \xi_i\hat{\xi}_j = 0$. The comparison between the maximum number of JLM diagrams needed to address all possible first-order JLM transition operators versus that of the $(M/2)^2$ terms that should be computed using James' method is shown in Fig.~\ref{Appendix comparison number loop VS number James}, showcasing the computational advantage of our method compared to James'. Nevertheless, Fig.~\ref{Appendix comparison number loop VS number James} shows that the computational overhead of our method relative to James' is lower for interaction Hamiltonians with few terms. As the number of interaction increases, the computational efficiency of our approach improves, although the relative advantage becomes less pronounced. 
\subsubsection{Second order}
\label{Appendix Second-order number of diagrams}
First consider the special case where the entire diagram is built only from a single operator and its Hermitian partner, $\hat\xi_i$ and $\hat\xi_i^\dagger$. Matter-part matching then allows only the three-photon combination $\hat\xi_i\,\hat\xi_i^\dagger\,\hat\xi_i$ (and its Hermitian conjugate). There are $M/2$ such diagrams (accounting for $M$ second-order JLM transition operators). \\ \\
Next consider sequences that mix $\hat\xi_i$ (or $\hat\xi_i^\dagger$) with operators $\hat\xi_j$ and $\hat\xi_j^\dagger$ whose matter part is different from that of $\hat\xi_i$. One can construct either $\hat \xi_j\hat \xi_i^\dagger\hat \xi_i$, $\hat \xi_j\hat \xi_i\hat \xi_i^\dagger$, $\hat \xi_i^\dagger \hat \xi_i \hat \xi_j$ or $\hat \xi_i \hat \xi_i^\dagger \hat \xi_j$, of which only one of the four sequences is allowed, since the others are mutually incompatible, either because they result in a vanishing overlap between the matter states or because they would together entail $\hat \xi_j = \hat \xi_i$ or $\hat \xi_j = \hat \xi_i^\dagger$, which have already been addressed in the previous case. Counting these thus gives $\frac{1}{2}\times\frac{M}{2}(M-2)=\frac{M(M-2)}{4}$ distinct JLM diagrams representatives, where the prefactor $1/2$ accounts for Hermitian-conjugate pairing. \\ \\
Finally, allow three distinct matter parts $\hat\xi_i,\hat\xi_j,\hat\xi_k$ (each possibly daggered). Grouping by cyclic rotation and Hermitian conjugation yields at most $\frac{1}{2}\binom{M}{3}$ diagram representatives from this class. \\ \\
Summing the three contributions gives an upper bound on the number of diagrams to evaluate at $n=2$, that is \textit{at most}
\begin{equation}
\mathcal N_{\text{JLM}} = \frac{M}{2} + \frac{M(M-2)}{4} + \frac{1}{2}\binom{M}{3}
= \frac{M\bigl(M^2+2\bigr)}{12}
\end{equation}
to draw in order to capture the perturbation-expansion physics. Here we have already used the fact that permutations of bosonic operators acting on different modes commute and therefore do not generate additional diagram representatives; only same-mode creation and annihilation multiplicities are absorbed into a given diagram.
Naturally, problem-specific matter-overlap selection rules in the RWA typically reduce this number in practice; as exhibited in the examination of the three-photon resonance where only one relevant second-order JLM diagram is needed instead of six. The number of JLM diagrams required at second-order is always smaller than the number of terms that are to be computed in James' method, that is at least $(M/2)^3$, see Fig.~\ref{Appendix Figure number of JLM VS number of James}.
\section{Second-order JLM transition operator weights}
\label{Appendix Second-order loop diagrams}
As in the first-order case, we now examine the topology and weights of second-order JLM diagrams. Each second-order diagram involves three JLM transition operators which we denote $ \hat\xi_i=\ket{e_i}\bra{g_i}\otimes\cdots,\; \hat\xi_j=\ket{e_j}\bra{g_j}\otimes\cdots,\; \hat\xi_k=\ket{e_k}\bra{g_k}\otimes\cdots $, with $ \braket{g_l|g_{l'}}=\braket{e_l|e_{l'}}=\delta_{ll'}$ and $\braket{g_l|e_l'}=0$. The bosonic factors inside the JLM operators do not affect the matter-part overlap condition and are therefore omitted; we abbreviate each operator by its matter part and write $ \hat\xi_l $ for $ l\in\{i,j,k\} $ in accordance with the interaction Hamiltonian's compact expression $\hat H_{\text{int}} = \sum_{m=1}^{M/2}\hat \xi_m + \text{h.c}$ where $\hat \xi_m$ are the zeroth-order JLM transition operators along with their coupling terms.
\subsection{Weight computation}
\label{Appendix Second-order loop diagrams Weight computation}
In this subsection, we compute the weight $V_2(t)$ of the second-order three-photon JLM diagrams. As before, $V_2(t)$ denotes the time-dependent contribution of $W_2(t)$ which also accounts for the coupling terms involved in the corresponding diagram. Recalling the general canonical time weight Eq.~\eqref{Equation general expression of the time-dependent loop diagram} at $n$th order of a given $n$th order JLM transition Hamiltonian 
\begin{equation}
     v_n(t) = (-1)^n \sum_{l=0}^n e^{-i\Delta_l t}e^{-\Theta_l t} \prod_{\substack{k=0 \\ k\neq l}}^n\frac{1}{\Delta_l-\Delta_k+i(\Theta_k-\Theta_l)},
\end{equation}
we can apply it to write its form for a given second-order JLM diagram leading to the JLM transition operator $\hat \xi_k \hat \xi_j \hat \xi_i$
\begin{equation}
\label{Equation appendix n=2 time-weight without detuning degeneracies}
\begin{split}
    v_2^{\text{\tiny{canonical}}}(t) &= \frac{e^{-i(\Delta_0-i\Theta_0)t}}{\left(\Delta_0-\Delta_1+i(\Theta_1-\Theta_0)\right)\left(\Delta_0-\Delta_2+i(\Theta_2-\Theta_0)\right)} \\
    &+ \frac{e^{-(i\Delta_1-i\Theta_1) t}}{\left(\Delta_1-\Delta_0 +i(\Theta_0-\Theta_1)\right)\left(\Delta_1-\Delta_2+i(\Theta_2-\Theta_1)\right)} \\
    &+\frac{e^{-i(\Delta_2-i\Theta_2)t}}{\left(\Delta_2-\Delta_0+i(\Theta_0-\Theta_2)\right)\left(\Delta_2-\Delta_1+i(\Theta_1-\Theta_2)\right)},
\end{split}
\end{equation}
where the $\Delta_l$ and $\Theta_l$ are the cumulative detunings and regularization terms, respectively. For instance, if the JLM diagram leading to $\hat \xi_k \hat \xi_j \hat \xi_i$ is constructed such as $\hat \xi_i$ is the zeroth-order term (placed at the perturbative square-vertex) and the diagram is thus read clockwise 
\begin{equation}
\label{Appendix canonical weight at second order}
    v_2^{\text{\tiny{canonical}}}(t) = \frac{e^{-i(\delta_i-i\theta_i)t}}{\left(\delta_j-i\theta_j\right)(\delta_j+\delta_k-i(\theta_j+\theta_k))}
    - \frac{e^{-i\left(\delta_i+\delta_j -i(\theta_i+\theta_j)\right)t}}{\left(\delta_j-i\theta_j\right)\left(\delta_k-i\theta_k\right)}
    +\frac{e^{-i\left(\delta_i+\delta_j+\delta_k-i(\theta_i+\theta_j+\theta_k)\right)t}}{\left(\delta_j+\delta_k-i(\theta_j+\theta_k)\right)\left(\delta_k-i\theta_k\right)}.
\end{equation}
As previously discussed however, adjacent transpositions of the JLM diagram perturbative square-vertex give the same JLM transition operator $\hat \xi_k \hat \xi_j \hat \xi_i$. This amounts to saying $\hat \xi_i$, $\hat \xi_j$ and $\hat \xi_k$ can all play the role of perturbed operator. \\
If $\hat \xi_k$ acts as the perturbed operator, two adjacent transpositions have been performed so based on the rule (7) of JLM-diagram construction, the sign does not change and the weight is now equal to 
\begin{equation}
\label{Appendix two adjacent transpositions weight at second order}
    v_2^{\text{\tiny{two adj. transp.}}}(t) = \frac{e^{-i(\delta_k-i\theta_k)t}}{\left(\delta_j-i\theta_j\right)(\delta_j+\delta_i-i(\theta_j+\theta_i))}
    - \frac{e^{-i\left(\delta_k+\delta_j -i(\theta_k+\theta_j)\right)t}}{\left(\delta_j-i\theta_j\right)\left(\delta_i-i\theta_i\right)}
    +\frac{e^{-i\left(\delta_i+\delta_j+\delta_k-i(\theta_i+\theta_j+\theta_k)\right)t}}{\left(\delta_j+\delta_i-i(\theta_j+\theta_i)\right)\left(\delta_i-i\theta_i\right)},
\end{equation}
which is equivalent to the permutation $i \leftrightarrow k$. Lastly, $\hat \xi_j$ may play the role of the zeroth-order, perturbed JLM transition operator. In this case, we need to do one adjacent transposition thus adding an extra $(-1)$ factor. Besides, having $\hat \xi_j$ as the zeroth-order term entails $\hat \xi_i$ acting on its right and $\hat \xi_k$. Since operators acting on different strands are not bound to process ordering, they can act as both the first and the second order term. In other words, two additional terms should be computed, together yielding
\begin{equation}
\label{Appendix one adjacent transpositions weight at second order}
    v_2^{\text{\tiny{one adj. transp.}}}(t) = \frac{-e^{-i\left(\delta_j -i\theta_j\right)t}+e^{-i\left(\delta_i+\delta_j -i(\theta_i+\theta_j)\right)t}+e^{-i\left(\delta_k+\delta_j-i(\theta_k+\theta_j)\right)t}-e^{-i\left(\delta_i+\delta_j+\delta_k -i(\theta_i+\theta_j+\theta_k)\right)t}}{\left(\delta_i-i\theta_i\right)\left(\delta_k-i\theta_k\right)}.
\end{equation}
The total second-order JLM transition operator's time-weight is the sum of all adjacent-transposition contributions
\begin{equation}
\label{Appendix Equation total second order time weight}
    V_2(t) = v_2^{\text{\tiny{canonical}}}(t) +v_2^{\text{\tiny{two adj. transp.}}}(t) + v_2^{\text{\tiny{one adj. transp.}}}(t).
\end{equation}
\subsection{Dealing with cumulative-detuning degeneracies}
\label{Appendix Dealing with cumulative-detuning degeneracies}
As opposed to the first-order computation covered in Sec.~\ref{Section First-order loop diagrams}, poles may emerge if $\delta_j=-\delta_k$, $\delta_j = -\delta_i$ or $\delta_k=\delta_i=-\delta_j$, which amounts to having cumulative-detuning degeneracies $\Delta_0 = \Delta_2$. We now show that the divergences cancel. Let us assume for instance that $\delta_j = -\delta_k$ and analyze $v_2^{\text{\tiny{canonical}}}(t)$ where a pole may rise as we send the regularization terms to zero in $v_2^{\text{\tiny{canonical}}}(t)$, which in turn would signal a divergence. If $\delta_j=-\delta_k$, then
\begin{equation}
    v_2^{\text{\tiny{canonical}}}(t) = \frac{e^{-i(\delta_i-i\theta_i)t}}{-i(\theta_i+\theta_j)(\delta_j-i\theta_j)}-\frac{e^{-i(\delta_i+\delta_j-i(\theta_i+\theta_j))t}}{(\delta_j-i\theta_j)(-\delta_j-i\theta_k)}+\frac{e^{-i(\delta_i-i(\theta_i+\theta_j+\theta_k))t}}{-i(\theta_j+\theta_k)(-\delta_j-i\theta_k)}. 
\end{equation}
Grouping the two terms in $v_2^{\text{\tiny{canonical}}}(t)$ which may lead to a divergence 
\begin{equation}
    v_2^{\text{\tiny{canonical}}}(t) = \frac{e^{-i(\delta_i+\delta_j-i(\theta_i+\theta_j))t}}{(\delta_j-i\theta_j)(-\delta_j-i\theta_k)} + e^{-i(\delta_i-i\theta_i)t}\left[\frac{1}{-i(\theta_i+\theta_j)(\delta_j-i\theta_j)}+\frac{e^{-(\theta_j+\theta_k)t}}{-i(\theta_j+\theta_k)(-\delta_j-i\theta_k)} \right].
\end{equation}
Taking the limit $\theta_i, \theta_k \rightarrow 0$ 
\begin{equation}
    v_2^{\text{\tiny{canonical}}}(t) \underset{\theta_i,\theta_k\rightarrow 0}{\longrightarrow} -\frac{e^{-i(\delta_i+\delta_j)t}}{\delta_j^2}+e^{-i\delta_it} \left[\frac{1}{-i\theta_j(\delta_j-i\theta_j)}+\frac{e^{-i\theta_jt}}{i\theta_j\delta_j} \right],
\end{equation}
and then $\theta_j \rightarrow 0$
\begin{equation}
    v_2^{\text{\tiny{canonical}}}(t) \underset{\theta_i,\theta_k, \theta_j \rightarrow 0}{\longrightarrow} -\frac{e^{-i(\delta_i+\delta_j)t}}{\delta_j^2},
\end{equation}
Hence, the two problematic terms in $v_2^{\text{\tiny{canonical}}}(t)$ Eq.~\eqref{Appendix canonical weight at second order} cancel out when $\delta_j=-\delta_k$, and Eq.~\eqref{Appendix canonical weight at second order}, Eq.~\eqref{Appendix two adjacent transpositions weight at second order} and Eq.~\eqref{Appendix one adjacent transpositions weight at second order} may be rewritten as 
\begin{equation}
    v_{2,\delta_j=-\delta_k}^{\text{\tiny{canonical}}}(t) = -\frac{e^{-i(\delta_i+\delta_j-i(\theta_i+\theta_j))t}}{(\delta_j-i\theta_j)(-\delta_j-i\theta_k)}, 
\end{equation}
\begin{equation}
    v_{2, \delta_j=-\delta_k}^{\text{\tiny{two adj. transp.}}}(t) = \frac{e^{-i(-\delta_j-i\theta_k)t}}{\left(\delta_j-i\theta_j\right)(\delta_j+\delta_i-i(\theta_k+\theta_j))}
    - \frac{e^{-(\theta_k+\theta_j)t}}{\left(\delta_j-i\theta_j\right)\left(\delta_i-i\theta_i\right)}
    +\frac{e^{-i\left(\delta_i-i(\theta_i+\theta_j+\theta_k)\right)t}}{\left(\delta_j+\delta_i-i(\theta_j+\theta_i)\right)\left(\delta_i-i\theta_i\right)},
\end{equation}
\begin{equation}
     v_{2,\delta_j=-\delta_k}^{\text{\tiny{one adj. transp.}}}(t) = \frac{-e^{-i\left(\delta_j -i\theta_j\right)t}+e^{-i\left(\delta_i+\delta_j -i(\theta_i-\theta_j)\right)t}+e^{-(\theta_k+\theta_j)t}-e^{-i\left(\delta_i -i(\theta_i+\theta_j+\theta_k)\right)t}}{\left(\delta_i-i\theta_i\right)\left(-\delta_j-i\theta_k\right)},
\end{equation}
and the total weight 
\begin{equation}
    V_{2,\delta_j=-\delta_k}(t) =  v_{2,\delta_j=-\delta_k}^{\text{\tiny{canonical}}}(t) + v_{2, \delta_j=-\delta_k}^{\text{\tiny{two adj. transp.}}}(t) +v_{2,\delta_j=-\delta_k}^{\text{\tiny{one adj. transp.}}}(t).
\end{equation}
The reasoning is identical in the case where $\delta_j=-\delta_i$ and $\delta_j\neq-\delta_k$. Should $\delta_j=-\delta_i=-\delta_k$, the previous expression now becomes
\begin{equation}
    v_{2,\delta_j=-\delta_k=-\delta_i}^{\text{\tiny{canonical}}}(t) = -\frac{e^{-(\theta_i+\theta_j)t}}{(\delta_j-i\theta_j)(-\delta_j-i\theta_k)}, 
\end{equation}
\begin{equation}
    v_{2, \delta_j=-\delta_k=-\delta_i}^{\text{\tiny{two adj. transp.}}}(t) =
    - \frac{e^{-(\theta_k+\theta_j)t}}{\left(\delta_j-i\theta_j\right)\left(-\delta_j-i\theta_i\right)},
\end{equation}
\begin{equation}
     v_{2,\delta_j=-\delta_k=-\delta_i}^{\text{\tiny{one adj. transp.}}}(t) = \frac{-e^{-i\left(\delta_j -i\theta_j\right)t}+e^{-(\theta_i-\theta_j)t}+e^{-(\theta_k+\theta_j)t}-e^{-i\left(-\delta_j -i(\theta_i+\theta_j+\theta_k)\right)t}}{\left(-\delta_j-i\theta_i\right)\left(-\delta_j-i\theta_k\right)}.
\end{equation}
$v_{2,\delta_j=-\delta_k=-\delta_i}^{\text{\tiny{canonical}}}(t)$ and $v_{2, \delta_j=-\delta_k=-\delta_i}^{\text{\tiny{two adj. transp.}}}(t)$ will cancel out with the terms in $e^{-i(\theta_i-\theta_j)t}$ and $e^{-(\theta_k+\theta_j)t}$ in the $\theta_i,\theta_j,\theta_k \rightarrow 0$ limit, in such a way that we may rewrite 
\begin{equation}
    v_{2,\delta_j=-\delta_k=-\delta_i}^{\text{\tiny{canonical}}}(t) = 0, 
\end{equation}
\begin{equation}
    v_{2, \delta_j=-\delta_k=-\delta_i}^{\text{\tiny{two adj. transp.}}}(t) = 0,
\end{equation}
\begin{equation}
     v_{2,\delta_j=-\delta_k=-\delta_i}^{\text{\tiny{one adj. transp.}}}(t) =\frac{-e^{-i\left(\delta_j-i\theta_j\right)t}-e^{-i\left(-\delta_j -i(\theta_i+\theta_j+\theta_k)\right)t}}{\left(-\delta_j-i\theta_i\right)\left(-\delta_j-i\theta_k\right)},
\end{equation}
and therefore 
\begin{equation}
    V_{2,\delta_j=-\delta_k=-\delta_i}(t) = v_{2,\delta_j=-\delta_k=-\delta_i}^{\text{\tiny{one adj. transp.}}}(t).
\end{equation}
The terms proportional to $e^{-(\theta_k+\theta_j)t}$ appearing in $v_{2, \delta_j=-\delta_k}^{\text{\tiny{two adj. transp.}}}(t)$ and $v_{2,\delta_j=-\delta_k}^{\text{\tiny{one adj. transp.}}}(t)$ generate a constant contribution $-2/(\delta_i\delta_j)$ in the limit $\theta_i,\theta_j,\theta_k\rightarrow 0$. This term does not vanish under time-averaged adiabatic elimination. Consequently, the associated second-order JLM transition operator does not produce genuine dynamical transitions but instead acts as a renormalization of the free Hamiltonian $H_{\text{free}}$, \textit{i.e.} as an energy shift, even though it may originates from operators that couple distinct matter states. This constant contribution arises from the two-photon resonance condition $\delta_j+\delta_k=0$. Such a resonance already induces an energy renormalization at first order, from which the term proportional to $2/(\delta_i\delta_j)$ ultimately rises from. In the dispersive regime, operator weights decrease with perturbative order. Therefore, this accidental second-order energy shift is parametrically smaller than the first-order shift responsible for it and may consistently be neglected. Moreover, since the corresponding second-order transition operator remains off-resonant, its contribution to the energy shift is necessarily weak. The same reasoning applies to the terms proportional to $e^{-(\theta_i-\theta_j)t}$ in $v_{2,\delta_j=-\delta_k=-\delta_i}^{\text{\tiny{canonical}}}(t)$, $v_{2, \delta_j=-\delta_k=-\delta_i}^{\text{\tiny{two adj. transp.}}}(t)$ and $v_{2,\delta_j=-\delta_k=-\delta_i}^{\text{\tiny{one adj. transp.}}}(t)$. \\
In summary, only second-order JLM transition operators satisfying the three-photon resonance condition $\delta_i+\delta_j+\delta_k \approx 0$ must be retained. All other second-order contributions either (i) average to zero under adiabatic elimination or (ii) reduce -- through accidental two-photon resonance conditions $\delta_i+\delta_j=0$ or $\delta_i+\delta_j+\delta_k =0$ -- to effective energy renormalizations that can be absorbed into the free Hamiltonian and neglected in the dispersive regime in first approximation, since the corresponding first-order shifts dominate parametrically. \\
Therefore, in the discrete case, we obtain
\begin{equation}
\label{Equation appendix n=2 total time-weight with no degeneracies}
\begin{split}
    V_2(t) =& \frac{e^{-i\delta_i t}}{\delta_j(\delta_j+\delta_k)}+\frac{e^{-i\delta_k t}}{\delta_j(\delta_i+\delta_j)}-\frac{e^{-i\delta_j t}}{\delta_i\delta_k} +\frac{e^{-i(\delta_i+\delta_j)t}}{\delta_k}\left(\frac{1}{\delta_i}-\frac{1}{\delta_j} \right) + \frac{e^{-i(\delta_k+\delta_j)t}}{\delta_i}\left(\frac{1}{\delta_k}-\frac{1}{\delta_j} \right) \\ 
    &+e^{-i(\delta_i+\delta_j+\delta_k)t}\left(\frac{1}{\delta_k(\delta_j+\delta_k)}+\frac{1}{\delta_i(\delta_i+\delta_j)} -\frac{1}{\delta_i\delta_k}\right).
\end{split}
\end{equation}
Taking the limits $\theta_i,\theta_j,\theta_k \rightarrow 0$ before performing the $\sumint$ is permissible when the frequency spectrum is discrete. In contrast, for a continuous frequency distribution, these limits must be taken inside the integrals. The previous expressions has thus been calculated considering no accidental first-order resonances $\delta_i+\delta_j = 0$ or $\delta_j+\delta_k = 0$. Further assuming that only the three-photon processes are close to resonance, Eq.~\eqref{Equation appendix n=2 total time-weight with no degeneracies} simplifies to 
\begin{equation}
    V_2(t) = e^{-i(\delta_i+\delta_j+\delta_k)t}\left(\frac{1}{\delta_k(\delta_k+\delta_j)}+\frac{1}{\delta_i(\delta_i+\delta_j)} -\frac{1}{\delta_i\delta_k}\right),
\end{equation}

\bibliography{biblio}

\end{document}